\documentclass[11pt,a4paper]{article}

\pdfoutput=1

\usepackage{jheppub}

\usepackage{graphicx}
\usepackage{bm}
\usepackage{amssymb}
\usepackage{amsmath}
\usepackage{physics}
\usepackage{color}
\usepackage{slashed}
\usepackage{hyperref}
\usepackage{multirow,array}
\usepackage{subcaption}
\usepackage{enumitem}
\usepackage[normalem]{ulem}

\allowdisplaybreaks[1]

\newcommand{\GeV}{{\text{GeV}}}

\newcommand{\BR}{\text{BR}}
\newcommand{\SM}{\mathrm{SM}}
\newcommand{\ie}{\textit{i.e.}}

\makeatletter
\gdef\@fpheader{}
\makeatother

\preprint{OU-HET 1330}

\title{Electroweak corrections to the SM-like Higgs boson decays and the Higgs trilinear coupling in the real Higgs triplet model}

\author[a]{Masashi Aiko,}
\author[b]{Shinya Kanemura,}
\author[c]{Mariko Kikuchi,}
\author[d]{Kodai Sakurai,}
\author[b]{Sora Taniguchi}
\author[e]{and Kei Yagyu}

\affiliation[a]{National Institute of Technology, Miyakonojo College, Miyakonojo, Miyazaki 885-8567, Japan}
\affiliation[b]{Department of Physics, University of Osaka, Toyonaka, Osaka 560-0043, Japan}
\affiliation[c]{Department of Physics, Saga University, 1 Honjomachi, Saga 840-8502, Japan}
\affiliation[d]{National Institute of Technology, Tsuruoka College, Tsuruoka, Yamagata 997-0842, Japan}
\affiliation[e]{Department of Physics, Tokyo University of Science, 1-3, Kagurazaka, Shinjuku-ku, Tokyo 162-8601, Japan}

\emailAdd{m-aiko@miyakonojo.kosen-ac.jp}
\emailAdd{kanemu@het.phys.sci.osaka-u.ac.jp}
\emailAdd{mkikuchi@cc.saga-u.ac.jp}
\emailAdd{kodai.sakurai@tsuruoka-nct.ac.jp}
\emailAdd{taniguchi@het.phys.sci.osaka-u.ac.jp}
\emailAdd{yagyu@rs.tus.ac.jp}

\abstract{
We study new physics effects on the decays of the standard model-like Higgs boson ($h$) and the Higgs trilinear coupling including the full one-loop corrections in the real Higgs triplet model (RHTM).
We adopt the on-shell renormalization scheme, in which the scalar mixing angles are renormalized in a gauge-independent way using the pinch technique.
We found that sizable one-loop effects can arise in the decay rates through the Higgs wave-function (external-leg) corrections, while the tree-level effects are rather small due to the constraint on the vacuum expectation value of the triplet field from the electroweak precision measurements.
In addition, the $h\to\gamma\gamma$ decay rate and the $hhh$ coupling are sensitive to loop contributions.
The RHTM may be distinguishable from other extended Higgs models by studying the correlations among the $h\to\tau\tau$, $h\to ZZ^{*}$, and $h\to\gamma\gamma$ decay modes, and such characteristic predictions can be tested at the High-Luminosity LHC or future Higgs factories.
}

\begin{document}

\maketitle

\section{Introduction}
\label{sec: introduction}

The standard model (SM) of particle physics successfully explains various experimental data at the electroweak (EW) scale.
However, several phenomena cannot be explained within the SM, such as neutrino oscillations, the existence of dark matter, the baryon asymmetry of the Universe, and inflation.
These phenomena provide strong motivation for physics beyond the SM (BSM).

The discovery of the 125~GeV Higgs boson established the existence of a Higgs field associated with EW symmetry breaking.
Although the SM provides a minimal realization of the Higgs mechanism, there is no fundamental principle that determines the Higgs sector, and its structure remains unknown.
In many BSM models, the Higgs sector is extended from the SM by introducing additional scalar fields with various representations.

The real Higgs triplet model (RHTM)~\cite{Ross:1975fq} is a simple extension of the SM with an additional real isospin-triplet scalar field.
Extensions of the SM with scalar triplets are qualitatively different from those with singlet or doublet scalars. This is because the vacuum expectation values (VEVs) of triplet fields shift the tree-level value of the EW rho parameter from unity\footnote{In the Georgi-Machacek model~\cite{Georgi:1985nv, Chanowitz:1985ug}, which includes real and complex triplet scalar fields, the EW rho parameter is unity at tree level due to the vacuum alignment.}.
In addition, the extended Higgs sector of the RHTM can affect the nature of the EW phase transition.
EW phase transitions in real triplet extensions have been studied in Refs.~\cite{Patel:2012pi, Niemi:2018asa, Niemi:2020hto}.
Gravitational-wave signatures associated with first-order phase transitions have also been studied in Refs.~\cite{Chala:2018opy, Bandyopadhyay:2021ipw, Bandyopadhyay:2025jlg}.

The collider phenomenology of the additional Higgs bosons in the RHTM has been studied extensively.
Direct searches at the LHC can probe the additional neutral and charged Higgs bosons mainly through pair and associated production, while the diphoton decay of the SM-like Higgs boson provides an indirect constraint on the charged Higgs boson mass~\cite{Wang:2013jba, Chabab:2018ert, Bell:2020gug, Ashanujjaman:2023etj, Butterworth:2023rnw, Ashanujjaman:2024pky, Ashanujjaman:2024lnr, Crivellin:2024uhc, Maharathy:2026ooo}.
For a charged Higgs boson lighter than the top quark, searches for exotic top-quark decays provide additional constraints~\cite{Ashanujjaman:2025una}.
The prospects for direct searches at future electron-positron colliders are discussed in Ref.~\cite{Maharathy:2025yyo}.

The signatures of the RHTM can appear in the SM-like Higgs boson observables.
At tree level, the couplings of the SM-like Higgs boson are modified from their SM values through scalar mixing effects.
In addition, the additional Higgs bosons affect these couplings through radiative corrections.
The precision of the Higgs coupling measurements is expected to reach a few-percent level at the High-Luminosity LHC (HL-LHC)~\cite{Cepeda:2019klc,CMS:2025hfp} and a per-mille level at future Higgs factories, such as the International Linear Collider~\cite{Baer:2013cma,Fujii:2017vwa,Asai:2017pwp,Fujii:2019zll}, the Circular Electron-Positron Collider~\cite{CEPC-SPPCStudyGroup:2015csa,Ai:2025cpj}, and the Future Circular Collider~\cite{Gomez-Ceballos:2013zzn,FCC:2025lpp}.
To fully exploit the precision of these measurements, theoretical predictions with comparable precision are essential, requiring the inclusion of radiative corrections.

New physics effects in the SM-like Higgs boson observables have been widely studied, especially in singlet and doublet extensions.
In the two-Higgs doublet model (2HDM), one-loop corrections to the Higgs boson couplings~\cite{Kanemura:2004mg, Kanemura:2014dja, Kanemura:2015mxa, Arhrib:2015hoa, Kanemura:2016sos}, decays of the SM-like Higgs boson~\cite{Kanemura:2019kjg, Kanemura:2018yai, Altenkamp:2017ldc}, and decays of heavy Higgs bosons~\cite{Aiko:2021can, Kanemura:2022ldq, Aiko:2022gmz, Krause:2019qwe, Akeroyd:1998uw, Akeroyd:2000xa, Krause:2016oke, Krause:2016xku} have been investigated. 
Similar studies have also been performed in the Higgs singlet model (HSM)~\cite{Bojarski:2015kra, Kanemura:2015fra, Kanemura:2016lkz, Altenkamp:2018bcs, Egle:2023pbm}.
Two-loop corrections have been studied~\cite{Braathen:2019pxr, Braathen:2019zoh, Braathen:2020vwo, Sturm:2022hak, Aiko:2023nqj, Degrassi:2024qsf, Degrassi:2025pqt}.
As a public tool for the numerical evaluation of higher-order corrections, the authors of this paper have developed the \texttt{H-COUP} program~\cite{Kanemura:2017gbi, Kanemura:2019slf, Aiko:2023xui}.
Other public tools, such as \texttt{2HDECAY}~\cite{Krause:2018wmo}, \texttt{ewN2HDECAY}~\cite{Engeln:2018mbg}, \texttt{EWsHDECAY}~\cite{Egle:2023pbm}, \texttt{Prophecy4f}~\cite{Denner:2019fcr}, and \texttt{FlexibleDecay}~\cite{Athron:2021kve}, are also available.
\texttt{2HDECAY}, \texttt{ewN2HDECAY}, and \texttt{EWsHDECAY} calculate the full NLO corrections to the two-body decays of all Higgs bosons for a given extended Higgs model.
\texttt{Prophecy4f} computes full NLO corrections for specific decay modes, namely the four-body decays of CP-even Higgs bosons into four fermions, in the HSM and 2HDMs.
\texttt{FlexibleDecay} incorporates various BSM models and computes higher-order corrections to Higgs boson decays without including BSM loop effects.
The main difference between \texttt{H-COUP} and these programs is that \texttt{H-COUP} evaluates the full one-loop electroweak corrections to all major decay modes of both the SM-like Higgs boson and additional Higgs bosons in various extended Higgs models.
These calculations are performed using a common on-shell renormalization scheme~\cite{Kanemura:2004mg, Kanemura:2017wtm}, in which the scalar mixing angles are renormalized in a gauge-independent way using the pinch technique~\cite{Papavassiliou:1994pr, Papavassiliou:1989zd, Cornwall:1981zr, Binosi:2009qm, Degrassi:1992ue, Cornwall:1989gv}\footnote{Other gauge-invariant renormalization schemes for extended Higgs sectors are discussed in e.g., Refs.~\cite{Krause:2016oke,Denner:2018opp,Dittmaier:2022ivi,Kanemura:2024ium}.}
\texttt{H-COUP} enables systematic comparisons of predictions among different models, such as four types of 2HDMs with the softly broken $Z_{2}$ symmetry, the inert doublet model (IDM), and the HSM.

In the RHTM, the renormalization procedure in the SM must be extended to accommodate the additional input parameter associated with the triplet VEV.
In Ref.~\cite{Blank:1997qa}, a renormalization scheme for the electroweak sector based on the effective weak mixing angle of the electron was proposed for the RHTM.
This scheme was used to evaluate the electroweak precision observables (EWPOs) at one-loop level~\cite{Chen:2005jx, Chen:2006pb} and was applied to the complex Higgs triplet model (CHTM) in Ref.~\cite{Kanemura:2012rs}.
The EWPOs were also studied using a renormalization scheme with the $W$-boson mass as an input~\cite{Passarino:1989py, Passarino:1990nu}, as well as the star scheme~\cite{Lynn:1990zk}.
The renormalization of the triplet VEV in the $\overline{\mathrm{MS}}$ scheme was studied in Refs.~\cite{Chankowski:2006hs, Chivukula:2007koj, Chen:2008jg}, and this approach was used in a two-loop calculation of the Higgs di-photon decay rate~\cite{Degrassi:2023eii}.
A different approach based on scalar mixing angles was developed for the CHTM, and the one-loop corrections to Higgs boson couplings were studied~\cite{Aoki:2012yt, Aoki:2012jj}.
This scheme was formulated in a gauge-invariant way in the CHTM, and its impact on Higgs boson decays was investigated in Ref.~\cite{Aiko:2026qzi}.

In this paper, we present a renormalization scheme for the RHTM based on the scalar mixing angles and investigate electroweak one-loop corrections to the decays of the SM-like Higgs boson.
The counterterms for the scalar mixing angles are determined
by the self-energies, and the gauge dependence is eliminated by the pinch technique~\cite{Papavassiliou:1994pr,Papavassiliou:1989zd,Cornwall:1981zr,Binosi:2009qm,Degrassi:1992ue,Cornwall:1989gv}.
In particular, the charged scalar sector involves additional diagrams
arising from the tree-level $H^\pm W^\mp Z$ coupling, which are absent
in the 2HDMs~\cite{Krause:2016oke, Kanemura:2017wtm, Krause:2017mal}.
We analytically confirm the cancellation of the gauge dependence
in the general $R_\xi$ gauge, and provide the explicit expressions for the pinch terms in the 't Hooft--Feynman gauge.

The decay rates of the SM-like Higgs boson are calculated including NLO EW and higher-order QCD corrections.
We find that sizable one-loop effects can arise in the decay rates and the Higgs trilinear coupling, while the tree-level effects are rather small due to the small triplet VEV.
The resulting patterns of deviations in decay rates of the SM-like Higgs boson can help distinguish the RHTM from other extended Higgs models.

This paper is organized as follows.
In Sec.~\ref{sec: RHTM}, we define the RHTM Lagrangian.
In Sec.~\ref{sec: Renormalization}, we discuss the renormalization of the RHTM.
The numerical study of the Higgs boson decays is presented in Sec.~\ref{sec: Numerical study}, and conclusions are given in Sec.~\ref{sec: Conclusions}.
The Feynman rules, analytical formulae for effective Higgs vertices and 1PI functions, and the application of the pinch technique to the RHTM are presented in Appendices.

\section{The real Higgs triplet model}
\label{sec: RHTM}

The RHTM contains an $\mathrm{SU(2)_{L}}$ doublet field $\Phi$ with the hypercharge $Y=1/2$ and a real triplet field $\Delta$ with $Y=0$.
Around the EW vacuum, these fields are expanded as
\begin{align}
\Phi = \mqty(\phi^{+} \\ \frac{1}{\sqrt{2}}(v_{\phi}+\phi+iG^{0}))\qc
\Delta =
\frac{1}{\sqrt{2}}\mqty(\Delta^{0} & \Delta^{+}\\
\Delta^{-} & -\Delta^{0})\qq{with}
\Delta^{0}=\frac{1}{\sqrt{2}}\qty(v_{\Delta}+\delta),
\end{align}
where $v=\sqrt{v_{\phi}^{2}+4v_{\Delta}^{2}}\simeq 246\GeV$.

\subsection{The Higgs potential}

The Higgs potential is given by
\begin{align}
V(\Phi, \Delta)
&=
m_{\Phi}^{2}\Phi^{\dagger}\Phi+m_{\Delta}^{2}\Tr(\Delta^{\dagger}\Delta)
+\mu\Phi^{\dagger}\Delta\Phi
\notag \\ &\quad
+\frac{\lambda_{1}}{2}\qty(\Phi^{\dagger}\Phi)^{2}
+\frac{\lambda_{2}}{2}\qty[\Tr(\Delta^{\dagger}\Delta)]^{2}
+\lambda_{3}\qty(\Phi^{\dagger}\Phi)\Tr(\Delta^{\dagger}\Delta),
\label{eq: Higgs_potential}
\end{align}
where all parameters are real, and hence the Higgs potential is CP conserving~\cite{Ashanujjaman:2024lnr}.
The tadpole terms for $\phi$ and $\delta$ are obtained as
\begin{align}
T_{\phi} &= -\eval{\pdv{V}{\phi}}_{0}
= -\qty(m_{\Phi}^{2}+\frac{\lambda_{1}}{2}v_{\phi}^{2}
+\frac{\lambda_{3}}{2}v_{\Delta}^{2}-\frac{\mu}{2} v_{\Delta})v_{\phi}, \\
T_{\delta} &= -\eval{\pdv{V}{\delta}}_{0}
= -\qty(m_{\Delta}^{2}+\frac{\lambda_{2}}{2}v_{\Delta}^{2}
+\frac{\lambda_{3}}{2}v_{\phi}^{2}-\frac{\mu v_{\phi}^{2}}{4v_{\Delta}})v_{\Delta},
\end{align}
where $\cdots |_{0}$ indicates that all fields are set to zero.
For the non-zero VEVs, the stationary conditions, $T_{\phi} = T_{\delta} = 0$, give
\begin{align}
m_{\Phi}^{2}
=
-\frac{\lambda_{1}}{2}v_{\phi}^{2}-\frac{\lambda_{3}}{2}v_{\Delta}^{2}+\frac{\mu}{2} v_{\Delta}\qc
m_{\Delta}^{2}
=
-\frac{\lambda_{2}}{2}v_{\Delta}^{2}-\frac{\lambda_{3}}{2}v_{\phi}^{2}+\frac{\mu v_{\phi}^{2}}{4v_{\Delta}}.
\label{eq: stationary conditions}
\end{align}
In the following expressions, we retain the tadpole-dependent terms in the mass matrices, which are needed for the one-loop calculation in the standard tadpole scheme (see Sec.~\ref{subsec: tadpole renormalization}).

The mass matrices of the charged and CP-even scalar bosons are given by
\begin{align}
\mathcal{M}_{\pm}^{2}
=
\mqty(
\mu v_{\Delta} &
\cfrac{\mu v_{\phi}}{2}
\\
\cfrac{\mu v_{\phi}}{2} &
\cfrac{\mu v_{\phi}^{2}}{4v_{\Delta}}
)
-\mqty(
\cfrac{T_{\phi}}{v_{\phi}} & 0\\
0 & \cfrac{T_{\delta}}{v_{\Delta}}
)\qc
\mathcal{M}_{\mathrm{even}}^{2}
=
\mqty(
\mathcal{M}_{11}^{2} &
\mathcal{M}_{12}^{2} \\
\mathcal{M}_{21}^{2} &
\mathcal{M}_{22}^{2}
)
-\mqty(
\cfrac{T_{\phi}}{v_{\phi}} & 0 \\
0 & \cfrac{T_{\delta}}{v_{\Delta}}
),
\label{eq: mass_mat}
\end{align}
where
\begin{align}
\mathcal{M}_{11}^{2}
=
\lambda_{1}v_{\phi}^{2}\qc
\mathcal{M}_{22}^{2}
=
\lambda_{2}v_{\Delta}^{2}+\frac{\mu v_{\phi}^{2}}{4v_{\Delta}}\qc
\mathcal{M}_{12}^{2}
=
\mathcal{M}_{21}^{2}
=
\lambda_{3}v_{\phi}v_{\Delta}-\frac{\mu v_{\phi}}{2}.
\label{eq: mass_elem_CPeven}
\end{align}

We define the mass eigenstates as
\begin{align}
\mqty(\phi^{\pm} \\ \Delta^{\pm})
=
R(\beta)\mqty(G^{\pm} \\ H^{\pm})\qc
\mqty(\phi \\ \delta)
=
R(\alpha)\mqty(h \\ H),\qq{with}
R(\theta) = \mqty(c_{\theta} & -s_{\theta} \\ s_{\theta} & c_{\theta}),
\end{align}
where $c_{\theta}=\cos{\theta}$ and $s_{\theta}=\sin{\theta}$.
The mixing angles are defined as
\begin{align}
\tan{\beta} = -\frac{2v_{\Delta}}{v_{\phi}}\qc
\tan{2\alpha}
=
\frac{2\mathcal{M}_{12}^{2}}{\mathcal{M}_{11}^{2}-\mathcal{M}_{22}^{2}}
=
\frac{v_{\phi}}{v_{\Delta}}\frac{2\lambda_{3}v_{\Delta}^{2}-\mu v_{\Delta}}{\lambda_{1}v_{\phi}^{2}-\lambda_{2}v_{\Delta}^{2}-\mu v_{\phi}^{2}/(4v_{\Delta})}.
\label{eq: definition of mixing angles}
\end{align}
The tadpoles in the mass eigenbasis are given by
\begin{align}
\mqty(T_{h} \\ T_{H})
=
R(-\alpha)\mqty(T_{\phi} \\ T_{\delta})
=
\mqty(c_{\alpha}T_{\phi}+s_{\alpha}T_{\delta} \\ -s_{\alpha}T_{\phi}+c_{\alpha}T_{\delta}).
\label{eq: T_h and T_H}
\end{align}
The mass matrix of the charged scalar bosons becomes
\begin{align}
R(-\beta)\mathcal{M}_{\pm}^{2}R(\beta)
=
\mqty(0 & 0 \\ 0 & m_{H^{\pm}}^{2})
-
\mqty(T_{G^{+}G^{-}} & T_{G^{+}H^{-}} \\ T_{H^{+}G^{-}} & T_{H^{+}H^{-}}),
\end{align}
where
\begin{align}
m_{H^{\pm}}^{2} &= \frac{\mu}{4v_{\Delta}} v^2,
\end{align}
and
\begin{align}
T_{G^{+}G^{-}} &= \frac{T_{\phi}}{v_{\phi}}c_{\beta}^{2}+\frac{T_{\delta}}{v_{\Delta}}s_{\beta}^{2}\qc
T_{H^{+}H^{-}} = \frac{T_{\phi}}{v_{\phi}}s_{\beta}^{2}+\frac{T_{\delta}}{v_{\Delta}}c_{\beta}^{2}, \\
T_{G^{+}H^{-}} &= T_{H^{+}G^{-}} = -\qty(\frac{T_{\phi}}{v_{\phi}}-\frac{T_{\delta}}{v_{\Delta}})s_{\beta}c_{\beta}.
\end{align}
The mass matrix of the CP-even scalar bosons becomes
\begin{align}
R(-\alpha)\mathcal{M}_{\mathrm{even}}^{2}R(\alpha)
=
\mqty(m_{h}^{2} & 0 \\ 0 & m_{H}^{2})
-
\mqty(T_{hh} & T_{hH} \\ T_{Hh} & T_{HH}),
\end{align}
where
\begin{align}
m_{h}^{2} &= \mathcal{M}_{11}^{2}c_{\alpha}^{2}+\mathcal{M}_{22}^{2}s_{\alpha}^{2}+2\mathcal{M}_{12}^{2}s_{\alpha}c_{\alpha},\\
m_{H}^{2} &= \mathcal{M}_{11}^{2}s_{\alpha}^{2}+\mathcal{M}_{22}^{2}c_{\alpha}^{2}-2\mathcal{M}_{12}^{2}s_{\alpha}c_{\alpha},
\end{align}
and
\begin{align}
T_{hh} = \frac{T_{\phi}}{v_{\phi}}c_{\alpha}^{2}+\frac{T_{\delta}}{v_{\Delta}}s_{\alpha}^{2}\qc
T_{HH} = \frac{T_{\phi}}{v_{\phi}}s_{\alpha}^{2}+\frac{T_{\delta}}{v_{\Delta}}c_{\alpha}^{2}\qc
T_{hH} = T_{Hh} = -\qty(\frac{T_{\phi}}{v_{\phi}}-\frac{T_{\delta}}{v_{\Delta}})s_{\alpha}c_{\alpha}.
\end{align}

The potential parameters can be written as
\begin{align}
\mu &= \frac{4m_{H^{\pm}}^{2}}{v^{2}}v_{\Delta}, \\
\lambda_{1} &= \frac{1}{v_{\phi}^{2}}(m_{h}^{2}c_{\alpha}^{2}+m_{H}^{2}s_{\alpha}^{2}), \\
\lambda_{2} &= \frac{1}{v_{\Delta}^{2}}\qty(m_{h}^{2}s_{\alpha}^{2}+m_{H}^{2}c_{\alpha}^{2}-\frac{v_{\phi}^{2}}{v^{2}}m_{H^{\pm}}^{2}), \\
\lambda_{3} &= \frac{2m_{H^{\pm}}^{2}}{v^{2}}+\frac{m_{h}^{2}-m_{H}^{2}}{v_{\phi}v_{\Delta}}s_{\alpha}c_{\alpha}.
\end{align}
The scalar couplings, such as $g_{\phi_{i}\phi_{j}\phi_{k}}^{}$ and $g_{\phi_{i}\phi_{j}\phi_{k}\phi_{l}}^{}$, are listed in Appendix~\ref{app: Feynman rules}.

\subsection{Kinetic term and Yukawa interaction}

The kinetic term is given by
\begin{align}
\mathcal{L}_{\mathrm{Kin}}
=
(D_{\mu}\Phi)^{\dagger}(D^{\mu}\Phi)+\Tr[(D_{\mu}\Delta)^{\dagger}D^{\mu}\Delta],
\end{align}
where the covariant derivatives are defined as
\begin{align}
D_{\mu}\Phi
=
\qty(\partial_{\mu}-ig\tau^{a}W_{\mu}^{a}-ig'YB_{\mu})\Phi\qc
D_{\mu}\Delta
=
\partial_{\mu}\Delta-ig\qty[\tau^{a}W_{\mu}^{a},\Delta],
\end{align}
where $[A,B]=AB-BA$ denotes the commutator.
The $\mathrm{SU(2)_{L}}$ generators are defined as $\tau^{a}=\sigma^{a}/2$, where $\sigma^{a}$ are the Pauli matrices.
The masses of weak gauge bosons are obtained as
\begin{align}
m_{W}^{2} = \frac{g^{2}}{4}v^2\qc
m_{Z}^{2} = \frac{g_{Z}^{2}}{4}v_{\phi}^{2},
\end{align}
with $g_{Z}=g/c_{W}$ and $\tan{\theta_{W}}=g'/g$.
At tree level, the EW rho parameter is obtained as
\begin{align}
\rho_{0} = \frac{m_{W}^{2}}{m_{Z}^{2}c_{W}^{2}} = 1+\tan^{2}{\beta}.
\label{eq: rho_tree}
\end{align}
The scalar-gauge couplings, such as $g_{\phi V_{1}V_{2}}^{}, g_{\phi_{1} \phi_{2} V}^{}$ and $g_{\phi_{1} \phi_{2} V_{1} V_{2}}^{}$, are listed in Appendix~\ref{app: Feynman rules}.

Since the triplet scalar field $\Delta$ does not couple with fermions, the Yukawa interaction term $\mathcal{L}_{Y}$ is the same as that in the SM.
In the mass eigenbasis, we obtain
\begin{align}
\mathcal{L}_{Y}
&\supset
-\sum_{f=u, d, e}\frac{m_{f}}{v}\qty[\overline{f}f(\zeta_{h}h+\zeta_{H}H)-2iI_{f}\overline{f}\gamma_{5}f \zeta_{G^{0}}G^{0}]
\notag \\ &\quad
+\qty[\qty(\frac{\sqrt{2}m_{u}}{v}V_{ud}\overline{u}P_{L}d-\frac{\sqrt{2}m_{d}}{v}V_{ud}\overline{u}P_{R}d)(\zeta_{G^{\pm}}G^{+}+\zeta_{H^{\pm}}H^{+})+\mathrm{h.c.}]
\notag \\ &\quad
-\qty[\frac{\sqrt{2}m_{e}}{v}\overline{\nu}P_{R}e(\zeta_{G^{\pm}}G^{+}+\zeta_{H^{\pm}}H^{+})+\mathrm{h.c.}],
\end{align}
where $I_f$ denotes the weak isospin of a fermion $f$, and $V_{ud}$ is the CKM matrix.
Since the quark mixing effects are small, we set the CKM matrix to unity in the following.
The coupling modifiers $\zeta_{X}$ are given by
\begin{alignat}{3}
\zeta_{h} &= c_{\alpha}/c_{\beta}\qc&
\zeta_{G^{\pm}} &= 1\qc&
\zeta_{G^{0}} &= 1/c_{\beta}, \\
\zeta_{H} &= -s_{\alpha}/c_{\beta}\qc \quad&
\zeta_{H^{\pm}} &= -\tan\beta. \qquad&
&
\label{eq: zeta_phi}
\end{alignat}

\subsection{Theoretical constraints}
\label{subsec: Th_consts}

The Higgs potential has to be bounded from below at large field values to ensure the stability of the theory.
The necessary and sufficient conditions are given by~\cite{Chabab:2018ert, Ashanujjaman:2024lnr}
\begin{align}
\lambda_{1}>0\qc
\lambda_{2}>0\qc
\lambda_{3}&+\sqrt{\lambda_{1}\lambda_{2}}>0.
\end{align}

Tree-level unitarity requires that the absolute values of the eigenvalues of the two-to-two scalar scattering matrix do not exceed $\zeta_{\mathrm{LQT}}=1$~\cite{Lee:1977yc, Lee:1977eg} or $1/2$~\cite{Luscher:1988gc, Marciano:1989ns}, depending on the choice of unitarity criterion.
In the RHTM, the independent eigenvalues are given by~\cite{Chabab:2018ert, Ashanujjaman:2024lnr}
\begin{align}
16\pi a_{1} &= \lambda_{1}\qc
16\pi a_{2} = \lambda_{2}\qc
16\pi a_{3} = \lambda_{3},
\notag \\
16\pi a_{4}^{\pm} &= \frac{1}{4}\qty[6\lambda_{1}+5\lambda_{2}
\pm\sqrt{\qty(6\lambda_{1}-5\lambda_{2})^{2}+48\lambda_{3}^{2}}].
\end{align}
The tree-level unitarity bound is given by
\begin{align}
\abs{a_{i}} < \zeta_{\mathrm{LQT}}\qc (i = 1, 2, 3, 4^{\pm}).
\end{align}
In this paper, we take $\zeta_{\mathrm{LQT}}=1/2$.

\section{Renormalization of the RHTM}
\label{sec: Renormalization}

In this section, we formulate the renormalization scheme with the scalar mixing in the RHTM at the one-loop level.
We first discuss the tadpole renormalization in the standard tadpole (ST) scheme~\cite{Denner:1991kt} and the Fleischer--Jegerlehner (FJ) scheme~\cite{Fleischer:1980ub} (see also Refs.~\cite{Krause:2016oke, Denner:2016etu}).
We then describe the renormalization of the Higgs sector, including the scalar masses, field renormalization constants, and mixing angles.
Finally, we discuss the renormalization of the gauge sector, which is different from the SM due to $\rho_0\neq1$ at tree level.

\subsection{Tadpole}
\label{subsec: tadpole renormalization}

\subsubsection{Standard tadpole scheme}

In the ST scheme, we shift the bare tadpole as
\begin{align}
T_{S, \mathrm{B}} = T_{S}+\delta T_{S},
\end{align}
where $S$ denotes either $h$ or $H$.
The renormalized one-point functions are given by
\begin{align}
\widehat{\Gamma}_{S} = T_{S}+\Gamma_{S}^{\mathrm{1PI}}+\delta T_{S},
\end{align}
where the 1PI one-point functions $\Gamma_{S}^{\mathrm{1PI}}$ are given in Appendix~\ref{app: 1PFs}.
We impose the following renormalization conditions.
\begin{align}
\widehat{\Gamma}_{S} = 0.
\end{align}
Then, we obtain
\begin{align}
T_{S} = 0\qc
\delta T_{S} = -\Gamma_{S}^{\mathrm{1PI}}.
\end{align}
From Eq.~\eqref{eq: T_h and T_H}, $\delta T_{\phi}$ and $\delta T_{\delta}$ are obtained as
\begin{align}
\delta T_{\phi}
&=
c_{\alpha}\delta T_{h}-s_{\alpha}\delta T_{H}\qc
\delta T_{\delta}
=
s_{\alpha}\delta T_{h}+c_{\alpha}\delta T_{H}.
\end{align}

In the ST scheme, tadpole-inserted diagrams are canceled by the corresponding tadpole counterterm diagrams.
Therefore, we do not need to take into account these diagrams in perturbative calculations.
On the other hand, the tadpole counterterms appear in the mass and mixing counterterms.
We note that the bare tadpoles are not zero in the ST scheme: $T_{S, \mathrm{B}} = \delta T_{S} = -\Gamma_{S}^{\mathrm{1PI}}$.
Since $\Gamma_{S}^{\mathrm{1PI}}$ is gauge-dependent, $T_{S, \mathrm{B}}$ also depends on the gauge choice.

\subsubsection{Fleischer-Jegerlehner scheme}

In the FJ scheme, we shift the bare VEVs as $v_{i, \mathrm{B}}\to v_{i, \mathrm{B}}+\Delta v_{i}\ (i=\phi,\, \Delta)$.
Then, the bare tadpole terms are shifted as
\begin{align}
T_{\phi, \mathrm{B}} &\to T_{\phi, \mathrm{B}}+\Delta T_{\phi}
=
T_{\phi, \mathrm{B}}
+\qty(\frac{T_{\phi, \mathrm{B}}}{v_{\phi, \mathrm{B}}}-\mathcal{M}_{11, \mathrm{B}}^{2})\Delta v_{\phi}
-\mathcal{M}_{12, \mathrm{B}}^{2}\Delta v_{\Delta}, \\
T_{\delta, \mathrm{B}} &\to T_{\delta, \mathrm{B}}+\Delta T_{\delta}
=
T_{\delta, \mathrm{B}}
-\mathcal{M}_{21, \mathrm{B}}^{2}\Delta v_{\phi}
+\qty(\frac{T_{\delta, \mathrm{B}}}{v_{\Delta, \mathrm{B}}}-\mathcal{M}_{22, \mathrm{B}}^{2})\Delta v_{\Delta},
\end{align}
where we have retained terms up to $\mathcal{O}(\Delta v_{i})$.
We choose $v_{\phi, \mathrm{B}}$ and $v_{\Delta, \mathrm{B}}$ so that the bare tadpole terms vanish, $T_{\phi, \mathrm{B}}=T_{\delta, \mathrm{B}}=0$, and we obtain
\begin{align}
\mqty(\Delta T_{\phi} \\ \Delta T_{\delta})
&=
-\left. \mathcal{M}_{\mathrm{even}}^{2}\right|_{T_{\phi}=T_{\delta}=0}\mqty(\Delta v_{\phi} \\ \Delta v_{\Delta}),
\end{align}
where $\mathcal{M}_{\mathrm{even}}^{2}$ is given by Eq.~\eqref{eq: mass_mat}.
Since $\left. \mathcal{M}_{\mathrm{even}}^{2}\right|_{T_{\phi}=T_{\delta}=0}$ is diagonalized by $R(\alpha)$, we obtain
\begin{align}
\mqty(\Delta v_{h} \\ \Delta v_{H})
=
\mqty(-\cfrac{\Delta T_{h}}{m_{h}^{2}} \\ -\cfrac{\Delta T_{H}}{m_{H}^{2}}),
\label{eq: Del_vh and Del_vH}
\end{align}
with
\begin{align}
\mqty(\Delta v_{h} \\ \Delta v_{H}) = R(-\alpha)\mqty(\Delta v_{\phi} \\ \Delta v_{\Delta})\qc
\mqty(\Delta T_{h} \\ \Delta T_{H}) = R(-\alpha)\mqty(\Delta T_{\phi} \\ \Delta T_{\delta}).
\end{align}
The renormalized one-point functions are given by
\begin{align}
\widehat{\Gamma}_{S} = \Gamma_{S}^{\mathrm{1PI}}+\Delta T_{S},
\end{align}
where $S$ denotes either $h$ or $H$.
We impose the following renormalization conditions.
\begin{align}
\widehat{\Gamma}_{S} = 0.
\end{align}
Then, we obtain $\Delta T_{S} = -\Gamma_{S}^{\mathrm{1PI}}$.
From Eq.~\eqref{eq: Del_vh and Del_vH}, we obtain
\begin{align}
\Delta v_{S} = \frac{\Gamma_{S}^{\mathrm{1PI}}}{m_{S}^{2}}.
\end{align}
From this expression, we can interpret terms with $\Delta v_{S}$ in the Lagrangian as contributions of tadpole-inserted diagrams, which contain the tadpole connected by the Higgs propagator with zero momentum transfer.

\subsection{Scalar sector}

In the mass eigenbasis, we shift the bare masses as
\begin{align}
m_{S, \mathrm{B}}^{2} &= m_{S}^{2}+\delta m_{S}^{2},
\end{align}
where $S = h, H, H^{\pm}$.
The bare scalar fields are renormalized as
\begin{align}
\mqty(h_{\mathrm{B}}\\ H_{\mathrm{B}})
&=
Z_{\mathrm{even}}^{1/2}\mqty(h\\ H)
=
\mqty(
1+\cfrac{\delta Z_{hh}}{2} & \cfrac{\delta Z_{hH}}{2} \\
\cfrac{\delta Z_{Hh}}{2} & 1+\cfrac{\delta Z_{HH}}{2}
)\mqty(h\\ H), \label{eq: wf_even} \\
\mqty(G^{\pm}_{\mathrm{B}}\\ H^{\pm}_{\mathrm{B}})
&=
Z_{\pm}^{1/2}\mqty(G^{\pm}\\ H^{\pm})
=
\mqty(
1+\cfrac{\delta Z_{G^{\pm}G^{\pm}}}{2} & \cfrac{\delta Z_{G^{\pm}H^{\pm}}}{2} \\
\cfrac{\delta Z_{H^{\pm}G^{\pm}}}{2} & 1+\cfrac{\delta Z_{H^{\pm}H^{\pm}}}{2}
)\mqty(G^{\pm}\\ H^{\pm}),
\label{eq: wf_singly_charged}
\end{align}
For the inverse propagators, we impose
\begin{align}
\Re\Gamma_{ii}(m_i^2) &= 0\qc
\Re\eval{\dv{\Gamma_{ii}(p^2)}{p^2}}_{p^2=m_i^2} = 1, \\
\Re\Gamma_{ij}(m_i^2) &= 0\qc
\Re\Gamma_{ij}(m_j^2) = 0
\quad (i\neq j),
\end{align}
where $(i,j)$ is either $(h,H)$ or $(G^\pm,H^\pm)$.
For the renormalization of the charged scalar two-point functions, we take $p^{2}=0$ for $G^{\pm}$.

The inverse propagators are given by
\begin{align}
\Gamma_{ij}(p^{2}) =
i\qty[(p^{2}-m_{i}^{2})\delta_{ij}+\widehat{\Pi}_{ij}(p^{2})],
\end{align}
where the renormalized self-energies are given by
\begin{align}
\widehat{\Pi}_{ij}(p^{2})
&=
\Pi_{ij}^{\mathrm{1PI}}(p^{2})
+\Pi_{ij}^{\mathrm{Tad}}
-\delta m_{i}^{2}\delta_{ij}
+\frac{\delta Z_{ij}}{2}\qty(p^{2}-m_{i}^{2})
+\frac{\delta Z_{ji}}{2}\qty(p^{2}-m_{j}^{2}).
\end{align}
The tadpole contributions depend on the renormalization scheme.
For the CP-even Higgs bosons, we obtain
\begin{align}
\Pi_{hh}^{\mathrm{Tad}}|_{\mathrm{ST}}
&=
\delta T_{hh}
=
\frac{\delta T_{\phi}}{v_{\phi}}c_{\alpha}^{2}+\frac{\delta T_{\delta}}{v_{\Delta}}s_{\alpha}^{2}, \\
\Pi_{HH}^{\mathrm{Tad}}|_{\mathrm{ST}}
&=
\delta T_{HH}
=
\frac{\delta T_{\phi}}{v_{\phi}}s_{\alpha}^{2}+\frac{\delta T_{\delta}}{v_{\Delta}}c_{\alpha}^{2}, \\
\Pi_{hH}^{\mathrm{Tad}}|_{\mathrm{ST}}
&=
\delta T_{hH}
=
-\qty(\frac{\delta T_{\phi}}{v_{\phi}}-\frac{\delta T_{\delta}}{v_{\Delta}})s_{\alpha}c_{\alpha}, \\
\Pi_{hh}^{\mathrm{Tad}}|_{\mathrm{FJ}}
&=
6g_{hhh}^{}\Delta v_{h}+2g_{Hhh}^{}\Delta v_{H}, \\
\Pi_{HH}^{\mathrm{Tad}}|_{\mathrm{FJ}}
&=
2g_{hHH}^{}\Delta v_{h}+6g_{HHH}^{}\Delta v_{H}, \\
\Pi_{hH}^{\mathrm{Tad}}|_{\mathrm{FJ}}
&=
2g_{hhH}^{}\Delta v_{h}+2g_{HHh}^{}\Delta v_{H}.
\end{align}
For the charged scalar bosons, we obtain
\begin{align}
\Pi_{H^{+}H^{-}}^{\mathrm{Tad}}|_{\mathrm{ST}}
&=
\delta T_{H^{+}H^{-}}
=
\frac{\delta T_{\phi}}{v_{\phi}}s_{\beta}^{2}+\frac{\delta T_{\delta}}{v_{\Delta}}c_{\beta}^{2}, \\
\Pi_{H^{+}G^{-}}^{\mathrm{Tad}}|_{\mathrm{ST}}
&=
\delta T_{H^{+}G^{-}}
=
-\qty(\frac{\delta T_{\phi}}{v_{\phi}}-\frac{\delta T_{\delta}}{v_{\Delta}})s_{\beta}c_{\beta},
\end{align}
in the ST scheme. Those in the FJ scheme are given by
\begin{align}
\Pi_{H^{+}H^{-}}^{\mathrm{Tad}}|_{\mathrm{FJ}}
&=
g_{hH^{+}H^{-}}^{}\Delta v_{h}+g_{HH^{+}H^{-}}^{}\Delta v_{H}, \\
\Pi_{H^{+}G^{-}}^{\mathrm{Tad}}|_{\mathrm{FJ}}
&=
g_{hH^{+}G^{-}}^{}\Delta v_{h}+g_{HH^{+}G^{-}}^{}\Delta v_{H}.
\end{align}
We note that $\Pi_{G^{+}G^{-}}^{\mathrm{Tad}}$ is irrelevant for the renormalization of the physical sector.
The counterterms are obtained as
\begin{align}
\delta m_{i}^{2}
&= \Re\qty[\Pi_{ii}^{\mathrm{1PI}}(m_{i}^{2})+\Pi_{ii}^{\mathrm{Tad}}]\qc
\delta Z_{ii} = -\eval{\dv{p^{2}}\Re\Pi_{ii}^{\mathrm{1PI}}(p^{2})}_{p^{2}=m_{i}^{2}}, \\
\delta Z_{ij} &= \frac{2}{m_{i}^{2}-m_{j}^{2}}\Re\qty[\Pi_{ij}^{\mathrm{1PI}}(m_{j}^{2})+\Pi_{ij}^{\mathrm{Tad}}].
\label{eq: scalar WF CTs}
\end{align}

\subsubsection{Mixing angles}

We shift the bare mixing angles as
\begin{align}
\alpha_{\rm B} = \alpha + \delta\alpha\qc
\beta_{\rm B} = \beta + \delta\beta.
\end{align}
We determine the mixing-angle counterterms by the off-diagonal components of the wave-function renormalization matrices~\cite{Kanemura:2004mg} (see also Ref.~\cite{Denner:2018opp}).
For the CP-even sector, we shift the bare fields in the weak basis as
\begin{align}
\mqty(\phi_{\rm B}\\ \delta_{\rm B})
=
Z_\mathrm{even, weak}^{1/2}\mqty(\phi\\ \delta).
\end{align}
Then, the bare fields in the mass eigenbasis are renormalized as
\begin{align}
\mqty(h_{\rm B}\\ H_{\rm B})
=
Z_{\rm even, mass}^{1/2}\mqty(h\\ H)
=
R(-\delta\alpha)\tilde{Z}_{\rm even}^{1/2}\mqty(h\\ H),
\label{eq: Z_even_mass relation}
\end{align}
where we have defined $\tilde{Z}_{\rm even}^{1/2} = R(-\alpha)Z_{\rm even, weak}^{1/2}R(\alpha)$.

Let us focus on the structure of the UV-divergent parts.
Since $\phi$ and $\delta$ belong to different $\mathrm{SU(2)_{L}}$ multiplets, the UV-divergent part of $Z_\mathrm{even, weak}^{1/2}$ is diagonal.
Therefore, the UV-divergent part of $\tilde{Z}^{1/2}_{\rm even}$ is symmetric.
\begin{align}
\tilde{Z}_{\rm even}^{1/2}
=
\mqty(
1+\cfrac{\delta Z_{hh}^{\mathrm{UV}}}{2} & \delta C_{hH}^{\mathrm{UV}}\\
\delta C_{hH}^{\mathrm{UV}} & 1+\cfrac{\delta Z_{HH}^{\mathrm{UV}}}{2}
)+\text{UV-finite parts}.
\end{align}
Then, from Eq.~\eqref{eq: Z_even_mass relation}, we obtain
\begin{align}
Z_{\rm even, mass}^{1/2}
=
\mqty(
1+\cfrac{\delta Z_{hh}^{\mathrm{UV}}}{2} & \delta C_{hH}^{\mathrm{UV}}+\delta\alpha^{\mathrm{UV}} \\
\delta C_{hH}^{\mathrm{UV}}-\delta\alpha^{\mathrm{UV}} & 1+\cfrac{\delta Z_{HH}^{\mathrm{UV}}}{2}
)+\text{UV-finite parts}.
\end{align}
Comparing with Eq.~\eqref{eq: wf_even}, we obtain
\begin{align}
\delta\alpha^{\mathrm{UV}}
=
\frac{1}{4}\qty(\delta Z_{hH}^{\mathrm{UV}}-\delta Z_{Hh}^{\mathrm{UV}}).
\label{eq: del_alpha UV}
\end{align}
Similarly, for the charged sector, we obtain
\begin{align}
\delta\beta^{\mathrm{UV}}
=
\frac{1}{4}\qty(\delta Z_{G^{+}H^{+}}^{\mathrm{UV}}-\delta Z_{H^{+}G^{+}}^{\mathrm{UV}}).
\label{eq: del_beta UV}
\end{align}
While the UV-divergent parts of the mixing-angle counterterms are determined, the finite parts are not uniquely fixed and can depend on the choice of renormalization scheme.
We impose the relations in Eqs.~\eqref{eq: del_alpha UV} and \eqref{eq: del_beta UV} not only on the UV-divergent parts but also on the finite parts of the mixing-angle counterterms.
\begin{align}
\delta\alpha
=
\frac{1}{4}\qty(\delta Z_{hH}-\delta Z_{Hh})\qc
\delta\beta
=
\frac{1}{4}\qty(\delta Z_{G^{+}H^{+}}-\delta Z_{H^{+}G^{+}}).
\label{eq: mixing CTs}
\end{align}

The mixing angle counterterms defined by Eqs.~\eqref{eq: mixing CTs} are gauge-dependent.
It is known that the gauge dependence of the mixing angles does not cancel in higher-order amplitudes,
leading to gauge dependence in physical quantities such as partial decay widths~\cite{Krause:2016oke}.
To eliminate the gauge dependence, we define the mixing angle counterterms using the pinch technique~\cite{Papavassiliou:1994pr,Papavassiliou:1989zd,Cornwall:1981zr,Binosi:2009qm,Degrassi:1992ue,Cornwall:1989gv}.
We introduce gauge-independent pinched self-energies, in which the gauge dependence is canceled by adding pinch terms:
\begin{align}
\Pi_{ij}(p^{2})
=
\Pi_{ij}^{\mathrm{1PI}}(p^{2})+\Pi_{ij}^{\mathrm{Tad}}+\Pi_{ij}^{\mathrm{PT}}(p^{2}).
\end{align}
The pinch terms of the $h\text{--}H$ and $H^{+}\text{--}G^{-}$ self-energies are given by
\begin{align}
\Pi_{hH}^{\mathrm{PT}}(p^{2})
&=
-\frac{g^{2}}{32\pi^{2}} \left(2p^{2}-m_{h}^{2}-m_{H}^{2}\right) 
\big[
c_{G^{\pm}hW^{\mp}}^{} c_{G^{\mp}HW^{\pm}}^{}\, B_{0}(p^{2}; W, W)
\notag \\ &\hspace{4.8cm}
+ c_{H^{\pm}hW^{\mp}}^{} c_{H^{\mp}HW^{\pm}}^{}\, B_{0}(p^{2}; W, H^{\pm})
\big]
\notag \\ &\quad
-\frac{g_{Z}^{2}}{64\pi^{2}} \left(2p^{2}-m_{h}^{2}-m_{H}^{2}\right)
c_{G^{0}hZ}^{} c_{G^{0}HZ}^{} \, B_{0}(p^{2}; Z, Z),
\label{eq:PT_CPeven}\\
\Pi_{H^{+}G^{-}}^{\mathrm{PT}}(p^{2})
&=
-\frac{g^{2}}{64\pi^{2}} \left(2p^{2}-m_{H^{\pm}}^{2}\right) 
\big[
c_{H^{\pm}hW^{\mp}}^{} c_{G^{\pm}hW^{\mp}}^{}\, B_{0}(p^{2}; W, h)
\notag \\ &\hspace{4.0cm}
+ c_{H^{\pm}HW^{\mp}}^{} c_{G^{\pm}HW^{\mp}}^{}\, B_{0}(p^{2}; W, H)
\notag \\ &\hspace{4.0cm}
+ c_{H^{\pm}G^0W^{\mp}}^{} c_{G^{\pm}G^0W^{\mp}}^{}\, B_{0}(p^{2}; W, Z)
\big]
\notag \\ &\quad
-\frac{g_{Z}^{2}}{64\pi^{2}} \left(2p^{2}-m_{H^{\pm}}^{2}\right)
\big[
c_{H^+H^-Z}c_{H^\pm G^\mp Z} B_{0}(p^{2}; Z, H^{\pm})
\notag \\ &\hspace{4.0cm}
+c_{G^+G^-Z}c_{H^\pm G^\mp Z}\, B_{0}(p^{2}; W, Z)
\big],\label{eq:PT_Charged}
\end{align}
where
\begin{align}
&c_{G^\pm G^0 W^\mp} = c_\beta, \quad
c_{H^\pm G^0 W^\mp} = -s_\beta, \quad
c_{G^0 hZ} = c_\alpha/c_\beta, \quad
c_{G^0 HZ} = -s_\alpha/c_\beta, \notag\\
&c_{G^\pm h W^\mp} = c_\alpha c_\beta - 2s_\alpha s_\beta, \quad
c_{H^\pm h W^\mp} = -2c_\beta s_\alpha - c_\alpha s_\beta, \notag\\
&c_{G^\pm H W^\mp} = -c_\beta s_\alpha - 2c_\alpha s_\beta, \quad
c_{H^\pm H W^\mp} = -2c_\alpha c_\beta + s_\alpha s_\beta, \notag\\\
&c_{H^+H^-Z} = c_{2W}^{}+c_{\beta}^2,\quad
c_{H^\pm G^\mp Z} = s_\beta c_\beta, \quad 
c_{G^+G^-Z} = c_{2W}^{}+s_{\beta}^2. 
\end{align}
From Eqs.~\eqref{eq: mixing CTs} and \eqref{eq: scalar WF CTs}, the gauge-independent mixing-angle counterterms are obtained as
\begin{align}
\delta\alpha
&=
-\frac{1}{2(m_{H}^{2}-m_{h}^{2})}\Re\qty[\Pi_{hH}(m_{h}^{2})+\Pi_{hH}(m_{H}^{2})], \\
\delta\beta
&=
-\frac{1}{2m_{H^{\pm}}^{2}}\Re\qty[\Pi_{H^{+}G^{-}}(0)+\Pi_{H^{+}G^{-}}(m_{H^{\pm}}^{2})].
\end{align}

\subsection{Gauge sector}

The renormalization of the gauge sector is largely the same as in the SM~\cite{Bohm:1986rj, Hollik:1988ii}.
However, because $\rho_{0}\neq 1$, the gauge-sector counterterms depend on the mixing-angle counterterm $\delta\beta$.
In the mass-eigenbasis, we shift the bare parameters as
\begin{align}
m_{V, \mathrm{B}}^{2} = m_{V}^{2}+\delta m_{V}^{2}\qc
e_{\mathrm{B}} = e+\delta e\qc
v_{\mathrm{B}} = v+\delta v. 
\end{align}
The bare gauge fields are shifted as
\begin{align}
W_{\mu \mathrm{B}}^{\pm} &= \qty(1 + \frac{1}{2}\delta Z_{WW})W_{\mu}^{\pm}, \\
\left(\begin{array}{c}
Z_{\mu \mathrm{B}} \\
A_{\mu \mathrm{B}}
\end{array}\right)
&=
\left(\begin{array}{cc}
1+\cfrac{1}{2}\delta Z_{ZZ} & \cfrac{1}{2}\left(\delta C_{Z\gamma} - \cfrac{\delta s_{W}^{2}}{c_{W}s_{W}}\right)\\
\cfrac{1}{2}\left(\delta C_{Z\gamma}+\cfrac{\delta s_{W}^{2}}{c_{W}s_{W}}\right) & 1+\cfrac{1}{2}\delta Z_{\gamma\gamma}
\end{array}\right)
\left(\begin{array}{c}
Z_{\mu} \\
A_{\mu}
\end{array}\right).
\end{align}
These counterterms are not independent.
When we take $\delta Z_{\gamma\gamma}$ and $\delta C_{Z\gamma}$ as independent, $\delta Z_{WW}$ and $\delta Z_{ZZ}$  are determined by
\begin{align}
\delta Z_{WW} =\delta Z_{\gamma\gamma}+\frac{c_{W}}{s_{W}}\delta C_{Z\gamma}\qc
\delta Z_{ZZ} = \delta Z_{\gamma\gamma} +\frac{c_{W}^{2}-s_{W}^{2}}{c_{W}s_{W}}\delta C_{Z\gamma}. \label{eq: del_ZVV}
\end{align}

From the bare relations $s_{W, \mathrm{B}}^{2} = 1-m_{W, \mathrm{B}}^{2}/(m_{Z, \mathrm{B}}^{2}\rho_{0, \mathrm{B}})$ and $v_{\mathrm{B}}^{2} = m_{W, \mathrm{B}}^{2}s_{W, \mathrm{B}}^{2}/(\pi \alpha_{\mathrm{em}, \mathrm{B}})$, we obtain
\begin{align}
\delta s_{W}^{2}
=
c_{W}^{2}\left(\frac{\delta m_{Z}^{2}}{m_{Z}^{2}}
-\frac{\delta m_{W}^{2}}{m_{W}^{2}}
+\frac{\delta \rho_{0}}{\rho_{0}}
\right)\qc
\delta v
=
\frac{v}{2}\qty(
\frac{\delta m_{W}^{2}}{m_{W}^{2}}
+\frac{\delta s_{W}^{2}}{s_{W}^{2}}
-\frac{\delta \alpha_{\rm em}}{\alpha_{\rm em}}),
\end{align}
where $\delta \rho_{0}/\rho_{0} = 2\tan{\beta}\delta\beta$.
As in the SM, the five counterterms, $\delta m_{W}^{2}$, $\delta m_{Z}^{2}$, $\delta Z_{\gamma\gamma}$, $\delta C_{Z\gamma}$ and $\delta e$ are independent. In the RHTM, an additional counterterm, $\delta\beta$, is required.

The inverse propagators are given by
\begin{align}
\widehat{\Gamma}^{\mu\nu}_{VV'}
=
-ig^{\mu\nu}(p^{2}-m_{V}^{2})\delta_{VV'}
+i\widehat{\Pi}_{VV'}^{\mu\nu}(p^{2}),
\end{align}
where $V, V'= W, Z, \gamma$.
The self-energies are expressed by the sum of the longitudinal polarization part and transverse polarization part,
\begin{align}
\widehat{\Pi}_{VV'}^{\mu\nu}(p^{2}) =
\left(-g^{\mu\nu}+\frac{p^{\mu}p^{\nu}}{p^{2}}\right)\widehat{\Pi}_{VV'}^{T}(p^{2})
+\frac{p^{\mu}p^{\nu}}{p^{2}}\widehat{\Pi}_{VV'}^{L}(p^{2}).
\end{align}
The longitudinal parts are irrelevant to the renormalization of physical particles.
Thus, we omit the $T$ and $L$ superscript in the following.
The renormalized self-energies are written as
\begin{align}
\widehat{\Pi}_{WW}(p^{2})
&=
\Pi_{WW}^{{\rm 1PI}}(p^{2})+\Pi_{WW}^{{\rm Tad}}-\delta m_{W}^{2}+\delta Z_{WW}(p^{2} -m _{W}^{2}), \\
\widehat{\Pi}_{ZZ}(p^{2})
&=
\Pi_{ZZ}^{{\rm 1PI}}(p^{2})+\Pi_{ZZ}^{{\rm Tad}}-\delta m_{Z}^{2}+\delta Z_{ZZ}(p^{2}-m_{Z}^{2}), \\
\widehat{\Pi}_{\gamma\gamma}(p^{2})
&=
\Pi_{\gamma\gamma}^{{\rm 1PI}}(p^{2})+\delta Z_{\gamma\gamma}p^{2}, \\
\widehat{\Pi}_{Z\gamma}(p^{2})
&=
\Pi_{Z\gamma}^{{\rm 1PI}}(p^{2})
+\frac{1}{2}\qty(\delta C_{Z\gamma}-\frac{\delta s_{W}^{2}}{s_{W}c_{W}})(p^{2}-m_{Z}^{2})
+\frac{1}{2}\qty(\delta C_{Z\gamma}+\frac{\delta s_{W}^{2}}{s_{W}c_{W}})p^{2}.
\end{align}
The tadpole inserted self-energies of the gauge bosons $\Pi_{VV}^{\mathrm{Tad}}$ are obtained as
\begin{alignat}{2}
\Pi_{WW}^{{\rm Tad}}|_{\mathrm{ST}} &= 0\qc&
\Pi_{ZZ}^{{\rm Tad}}|_{\mathrm{ST}} &= 0, \\
\Pi_{WW}^{{\rm Tad}}|_{\mathrm{FJ}} &= -g_{hWW}\Delta v_{h}-g_{HWW}\Delta v_{H}\qc&
\Pi_{ZZ}^{{\rm Tad}}|_{\mathrm{FJ}} &= -2(g_{hZZ}\Delta v_{h}+g_{HZZ}\Delta v_{H}).
\end{alignat}

For the inverse propagators, we impose
\begin{align}
\Re \widehat{\Gamma}_{WW}(m_{W}^{2}) = 0&\qc
\Re \widehat{\Gamma}_{ZZ}(m_{Z}^{2}) = 0, \\
\eval{\dv{p^{2}}\Re \widehat{\Gamma}_{\gamma\gamma}(p^{2})}_{p^{2}=0} = 1&\qc
\Re \widehat{\Gamma}_{Z\gamma}(0) = 0.
\end{align}
Then, we obtain
\begin{alignat}{2}
\delta m_{W}^{2} &= \Re \qty[\Pi^{{\rm 1PI}}_{WW}(m_{W}^{2})+\Pi^{{\rm Tad}}_{WW}], &\quad
\delta m_{Z}^{2} &= \Re \qty[\Pi^{{\rm 1PI}}_{ZZ}(m_{Z}^{2})+\Pi^{{\rm Tad}}_{ZZ}],
\label{eq: del_mWsq and del_mZsq} \\
\delta Z_{\gamma\gamma} &= -\Re\left. \dv{\Pi_{\gamma\gamma}^{{\rm 1PI}}(p^{2})}{p^{2}} \right|_{p^{2}=0}, &\quad
\delta C_{Z\gamma} &= \frac{2\Re \Pi^{{\rm 1PI}}_{Z\gamma}(0)}{m_{Z}^{2}}
+\frac{\delta s_{W}^{2}}{s_{W}c_{W}}.
\label{eq: del_ZGG and del_ZZG}
\end{alignat}

We determine $\delta \alpha_{\rm em}$ so that the renormalized photon-electron-positron vertex $\widehat{\Gamma}^{\mu}_{ee\gamma}$ takes tree-level form at the Thomson limit,
\begin{align}
\widehat{\Gamma}^{\mu}_{ee\gamma}(q^{2}=0, \slashed{p}_{1}=\slashed{p}_{2}=m_{e}) = ie\gamma^{\mu}.
\end{align}
We obtain,
\begin{align}
\frac{\delta \alpha_{\rm em}}{\alpha_{\rm em}}
=
2\frac{\delta e}{e}
=
\left. \dv{\Pi^{\mathrm{1PI}}_{\gamma\gamma}(p^{2})}{p^{2}}\right|_{p^{2}=0}
+\frac{2s_{W}}{c_{W}}\frac{\Pi^{\mathrm{1PI}}_{Z\gamma}(0)}{m_{Z}^{2}}.
\label{eq: del_alpha_em_wo_PT}
\end{align}

In this renormalization scheme, the residue of the gauge boson two-point function is not unity.
Therefore, we need to include the derivative of the renormalized self-energy, $-\widehat{\Pi}^{\prime}_{VV}(m_{V}^{2})/2$, for each external $W$ and $Z$ bosons.
Similarly, we need to add $-\widehat{\Pi}_{Z\gamma}(m_{Z}^{2})/m_{Z}^{2}$ if there are $\gamma\text{--}Z$ mixing contributions for external $Z$ bosons.

\subsection{Fermion sector}

In the following, we neglect fermion flavor mixing in the one-loop calculations.
In the mass-eigenbasis, we shift the bare masses and fields as
\begin{align}
m_{f, \mathrm{B}} = m_{f} + \delta m_{f}\qc
\psi_{\mathrm{B}} = \psi+\frac{1}{2}\qty(\delta Z_{V}^{f}-\delta Z_{A}^{f}\gamma_{5})\psi.
\end{align}
The inverse propagators are given by
\begin{align}
\widehat{\Gamma}_{ff}(p)
=
i(\slashed{p}-m_{f})+i\widehat{\Pi}_{ff}(p),
\end{align}
where the fermion self-energies are decomposed into vector, axial, and scalar parts
\begin{align}
\widehat{\Pi}_{ff}(p) &= 
\slashed{p}\widehat{\Pi}_{ff, V}
-\slashed{p}\gamma_{5}\widehat{\Pi}_{ff, A}
+m_{f}\widehat{\Pi}_{ff, S},
\end{align}
and they are given by
\begin{align}
\widehat{\Pi}_{ff, V}(p^{2}) &= \Pi^{{\rm 1PI}}_{ff, V}(p^{2})+\delta Z_{V}^{f}, \\
\widehat{\Pi}_{ff, A}(p^{2}) &= \Pi^{{\rm 1PI}}_{ff, A}(p^{2})+\delta Z_{A}^{f}, \\
\widehat{\Pi}_{ff, S}(p^{2}) &= \Pi^{\rm 1PI}_{ff, S}(p^{2})+\Pi^{\rm Tad}_{ff, S}-\delta Z_{V}^{f}-\frac{\delta m_{f}}{m_{f}}.
\end{align}
The tadpole inserted self-energy of the fermions $\Pi^{\rm Tad}_{ff, S}$ is obtained as
\begin{align}
\Pi^{\mathrm{Tad}}_{ff, S}|_{\mathrm{ST}} = 0\qc
\Pi^{\mathrm{Tad}}_{ff, S}|_{\mathrm{FJ}}
=
-\zeta_{h}\frac{\Delta v_{h}}{v}-\zeta_{H}\frac{\Delta v_{H}}{v}.
\end{align}

For the inverse propagators, we impose
\begin{align}
\left.\Re\widehat{\Gamma}_{ff}(p)u(p)\right|_{p^{2}=m_{f}^{2}} = 0\qc
\left.\frac{\slashed{p}+m_{f}}{p^{2}-m_{f}^{2}}\Re\widehat{\Gamma}_{ff}(p)u(p)\right|_{p^{2}=m_{f}^{2}} = i u(p).
\label{eq: renormalization_conditions_fermion}
\end{align}
Then, we obtain
\begin{align}
\delta m_{f} &= m_{f}\Re\qty[\Pi^{{\rm 1PI}}_{ff, V}(m_{f}^{2}) + \Pi^{{\rm 1PI}}_{ff, S}(m_{f}^{2})+\Pi^{{\rm Tad}}_{ff, S}], \\
\delta Z_{V}^{f}
&=
\left.-\Re\Pi_{ff, V}^{\mathrm{1PI}}(m_{f}^{2})
-2m_{f}^{2}\dv{p^{2}}\Re\qty[\Pi^{\mathrm{1PI}}_{ff, V}(p^{2})+\Pi^{\mathrm{1PI}}_{ff, S}(p^{2})]\right|_{p^{2}=m_{f}^{2}}, \\
\delta Z_{A}^{f} &= -\Re\qty[\Pi_{ff, A}^{{\rm 1PI}}(m_{f}^{2})].
\end{align}

\section{Numerical study}
\label{sec: Numerical study}

In this section, we perform a numerical study of the decays of the SM-like Higgs boson in the RHTM.
The decay rates are expressed in terms of the renormalized vertices given in Appendix~\ref{sec: renormalized vertex}.
The expressions derived for the CHTM in Ref.~\cite{Aiko:2026qzi} can be applied to the RHTM with the following replacements:
\begin{align}
    c_{hZZ} = c_{\alpha}\qc c_{hWW} = c_{\beta}c_{\alpha}-2s_{\beta}s_{\alpha},
\end{align}
where $c_{hVV}\, (V=Z, W^{\pm})$ are the mixing factors in $g_{hVV}^{}$.
In addition, the LO decay rate of $h\to Z f\bar{f}$ is changed as
\begin{align}
\Gamma_{\mathrm{LO}}(h\to Z f\bar{f})
&=
N_{c}^{f}\qty(\frac{c_{hZZ}}{c_{\beta}^{2}})^{2}
\frac{G_{F}^{2}m_{h}m_{Z}^{4}}{24\pi^{3}}(v_{f}^{2}+a_{f}^{2})F\qty(\epsilon_{Z}),
\end{align}
where $v_{f} = I_{f}/2-Q_{f}s_{W}^{2}$ and $a_{f} = I_{f}/2$. 
The kinematical function $F\qty(\epsilon_{Z})$ is given in Ref.~\cite{Keung:1984hn, Aiko:2026qzi}.
The coefficients of the loop functions of the box diagrams, $C^{BZi}$ and $C^{BWi}$, used for the $h\to ZZ^{*}, WW^{*}$ decays, are also changed. We list them in Appendix~\ref{app: 4PFs}.

The resulting analytic formulae are implemented in the \texttt{H-COUP} program~\cite{Kanemura:2017gbi, Kanemura:2019slf, Aiko:2023xui}.
The loop functions are numerically evaluated using \texttt{LoopTools}~\cite{Hahn:1998yk}.
We have numerically checked that the cancellation of the UV divergence of the renormalized quantities.
In addition, we have checked that there is no scheme difference between the ST and FJ scheme in the renormalized quantities\footnote{This is non-trivial when $\overline{\mathrm{MS}}$ renormalized counterterms are involved, as in the 2HDM and the HSM~\cite{Kanemura:2017wtm}.}.
The infrared (IR) divergences are regularized by introducing a finite photon mass.
The photon-mass dependence is canceled by including the decay rate for real-photon emission following Refs.~\cite{Denner:1991kt, Goodsell:2017pdq, Kanemura:2019kjg}.
We have numerically checked that the inclusive decay rates are independent of the photon mass.
The SM input parameters used in the one-loop calculations are listed in Ref.~\cite{Aiko:2023xui}, while their numerical values are taken from the latest results~\cite{ParticleDataGroup:2026mpi}.

The remainder of this section is organized as follows.
We first discuss the input scheme adopted for the one-loop calculations in the RHTM.
We then examine the constraints from electroweak precision observables, collider searches, and Higgs signal measurements.
Finally, we investigate deviations in the decay of the SM-like Higgs boson and compare the predictions of the RHTM with those of other extended Higgs models, such as the HSM, IDM, 2HDMs, and CHTM.

\subsection{Input scheme}
\label{subsec: Input scheme}

We here describe the input scheme and the procedure for generating the model parameters.
In the on-shell renormalization scheme, $m_{W}$ is treated as an independent renormalized parameter.
For the numerical analysis, however, we replace $m_{W}$ by $G_{F}$ as an input parameter because $G_{F}$ is measured with high precision.
For this replacement, we use
\begin{align}
G_F
= \frac{1}{\sqrt{2}v^{2}(1-\Delta r)}
\qq{with}
v^{2} = \frac{m_{W}^{2}}{\pi\alpha_{\mathrm{em}}}\left(1-\frac{m_{W}^{2}}{\rho_{0}m_{Z}^{2}}\right),
\label{eq:G_F}
\end{align}
where $\Delta r$ is determined by the muon decay process and is given by~\cite{Sirlin:1980nh, Hollik:1988ii}
\begin{align}
\Delta r
&=
\frac{\Re\widehat{\Pi}_{WW}(0)}{m_{W}^{2}}+\delta_{VB},
\label{eq: delta_r}
\end{align}
with~\cite{Blank:1997qa}
\begin{align}
\delta_{VB}
&=
\frac{\alpha_{\rm em}}{4\pi s_{W}^{2}}\qty[
6+\frac{10(1-s_{W}^{2})-3(R/c_{W}^{2})(1-2s_{W}^{2})}{2(1-R)}\ln{R}]
\qq{with} R=\frac{m_{W}^{2}}{m_{Z}^{2}}.
\end{align}
The weak mixing angle and the $W$ boson mass are written as
\begin{align}
    s_{W}^{2}
    =
    \frac{1}{2}\qty[
    1-\sqrt{1-\frac{4A_{0}}{\rho_{0}m_{Z}^{2}(1-\Delta r)}}]\qc
    m_{W}^{2}
    =
    \rho_{0}m_{Z}^{2}c_{W}^{2}
    \qq{with}
    A_{0}=\frac{\pi\alpha_{\mathrm{em}}}{\sqrt{2}G_{F}}.
    \label{eq: swsq and mwsq}
\end{align}
We note that Eqs.~\eqref{eq: swsq and mwsq} are self-consistent equations because $\Delta r$ depends on $m_{W}$.

In order to evaluate $\Delta r$, we introduce the tree-level $W$ boson mass, $\overline{m}_{W}$, as~\cite{Kanemura:2019kjg}
\begin{align}
    \overline{m}_{W}^{2}
    =
    \rho_{0}m_{Z}^{2}\overline{c}_{W}^{2}
    \qq{with}
    \overline{c}_{W}^{2} = 1-\overline{s}_{W}^{2},
\end{align}
where $\overline{s}_{W}$ is given by
\begin{align}
    \overline{s}_{W}^{2}
    &=
    \frac{1}{2}\left(1-\sqrt{1-\frac{4A_{0}}{\rho_{0}m_{Z}^{2}}}\right).
\end{align}
We evaluate $\Delta r$ using $\overline{m}_{W}$ instead of $m_{W}$. Since $\Delta r$ is a one-loop quantity, the difference would only appear at the two-loop level.
Similarly, we use $\overline{m}_{W}$ and $\overline{s}_{W}$ for one-loop calculations in subsequent analysis.

Let us discuss the RHTM parameter generation.
In addition to $v$ and $m_{h}$, we choose the following parameters as the independent renormalized parameters of the Higgs sector:
\begin{align}
    m_{H}\qc m_{H^{\pm}}\qc \alpha\qc \beta.
\end{align}
Since these parameters are subject to nontrivial correlations imposed by the theoretical constraints, it is not numerically efficient to generate parameter points directly in terms of them.
Instead, we introduce auxiliary tree-level quantities and generate the parameters according to the corresponding tree-level relations.

First, we introduce $\overline{v}_\Delta$ through
\begin{align}
    \overline{v}_{\phi}^{2}+4\overline{v}_{\Delta}^2
    = \overline{v}^{2}
    \qq{with}
    \overline{v}^{2} = \frac{1}{\sqrt{2}G_F},
\end{align}
and obtain $\beta$ using the tree-level relation in Eq.~\eqref{eq: definition of mixing angles},
\begin{align}
    \tan\beta
    =
    -\frac{2\overline{v}_\Delta}{\overline{v}_\phi}.
\end{align}
Next, we introduce $\overline{\lambda}_{2}$ and $\overline{\lambda}_{3}$.
For given $m_{H^{\pm}}$, we have
\begin{align}
    m_{H}^{2} = \overline{\mathcal{M}}_{22}^{2}+\frac{(\overline{\mathcal{M}}_{12}^{2})^{2}}{\overline{\mathcal{M}}_{22}^{2}-m_{h}^{2}},
\end{align}
where $\overline{\mathcal{M}}_{22}^{2}$ and $\overline{\mathcal{M}}_{12}^{2}$ are obtained as
\begin{align}
    \overline{\mathcal{M}}_{22}^{2} = \overline{\lambda}_{2}\overline{v}_{\Delta}^{2}+\frac{\overline{v}_{\phi}^{2}}{\overline{v}^{2}}m_{H^{\pm}}^{2}\qc
    \overline{\mathcal{M}}_{12}^{2} = \overline{v}_{\phi}\overline{v}_{\Delta}\qty(\overline{\lambda}_{3}-\frac{2m_{H^{\pm}}^{2}}{\overline{v}^{2}}).
\end{align}
Finally, $\alpha$ is obtained as
\begin{align}
    \sin(2\alpha) = \frac{2\overline{v}_{\phi}\overline{v}_{\Delta}}{m_{H}^{2}-m_{h}^{2}}\qty(\frac{2m_{H^{\pm}}^{2}}{\overline{v}^{2}}-\overline{\lambda}_{3}).
\end{align}
Thus, for given $\{m_{H^{\pm}}, \overline{\lambda}_{2}, \overline{\lambda}_{3}, \overline{v}_{\Delta}\}$, one can generate $\{m_{H}, \alpha, \beta\}$.
We note that the following conditions must be satisfied for a given set of inputs:
\begin{align}
m_{H}^{2}>0\qc
(m_{H}^{2}-m_{h}^{2})^{2} \geq 4\left(\overline{\mathcal{M}}_{12}^{2}\right)^{2},
\end{align}
where the second condition ensures $\abs{\sin(2\alpha)}\leq 1$.
We also introduce $\overline{\lambda}_{1}$ by using the tree-level relation.
\begin{align}
\overline{\lambda}_{1}
&=
\frac{1}{2\overline{v}_{\phi}^{2}}\bigg[
   m_{H}^{2}+m_{h}^{2}
   -\qty(m_{H}^{2}-m_{h}^{2})\cos{2\alpha}
   \bigg].
\end{align}

The barred quantities denote tree-level quantities and receive loop corrections when expressed in the $G_{F}$ input scheme.
For example, $v_{\Delta}$ is shifted as
\begin{align}
    v_{\Delta}^{2}
    =
    \frac{s_{\beta}^{2}}{2(1+c_{\beta}^{2})}
    \frac{1}{\sqrt{2}G_{F}(1-\Delta r)}.
\end{align}
Similarly, $\lambda_{1}, \lambda_{2}$ and $\lambda_{3}$ also receive a $\Delta r$ shift.
For the theoretical constraints, we use the tree-level couplings $\overline{\lambda}_{i}$, since the difference between $\overline{\lambda}_{i}$ and $\lambda_{i}$ is of one-loop order.

\subsection{Experimental constraints}
\label{subsec: Constraints}

\begin{figure}[t]
 \centering
 \includegraphics[scale=0.475]{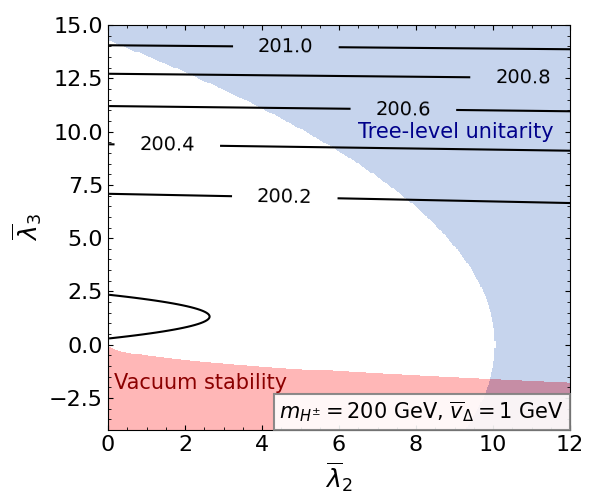}
 \includegraphics[scale=0.475]{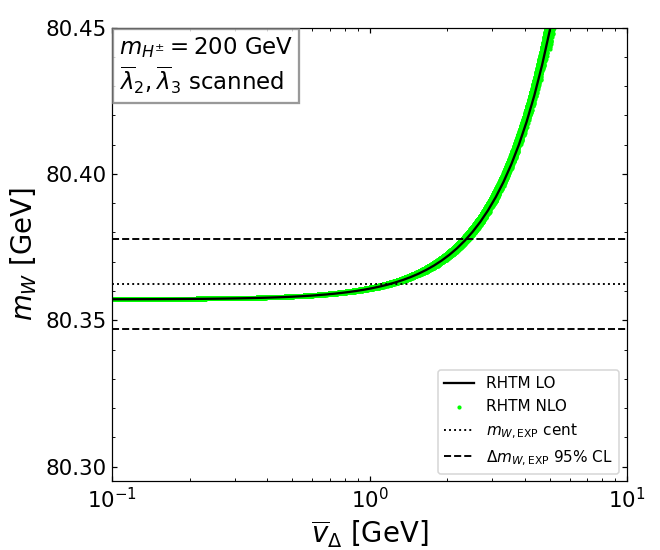}
 \caption{
 (Left) Contours of $m_{H}$ in the $\overline{\lambda}_{2}$--$\overline{\lambda}_{3}$ plane with $m_{H^{\pm}}=200$~GeV and $\overline{v}_{\Delta}=1~\GeV$. The red (blue) region is excluded by vacuum stability (tree-level unitarity).
 (Right) Prediction for the $W$-boson mass with $m_{H^{\pm}}=200~\GeV$. The black solid line shows the LO prediction, while the green points are the NLO predictions.
 The dotted and dashed lines indicate the central value and $2\sigma$ uncertainty of the observed $W$-boson mass, respectively. For the NLO prediction, $\overline{\lambda}_{2}$ and $\overline{\lambda}_{3}$ are scanned under the theoretical bounds.
 }
 \label{fig: th_bounds_and_mWGF_mch200}
\end{figure}

First, we discuss the constraint from the measurement of EWPOs.
From Eq.~\eqref{eq: rho_tree}, we can see that a nonzero $v_{\Delta}$ increases $\rho_{0}$ from unity.
From the global fit result\footnote{We have used $\rho_{\mathrm{exp}}=1+\alpha_{\mathrm{em}}(m_{Z})T$, where the $T$ parameter result is obtained by $STU$ global fit: $T=0.04\pm0.12$.}, $\rho_{\mathrm{exp}} = 1.0003\pm 0.0009$~\cite{ParticleDataGroup:2026mpi}, we obtain $v_{\Delta}\lesssim 5.8~\mathrm{GeV}$ at tree level.
At one-loop level, the $\rho$ parameter receives loop contributions.
For $v_{\Delta}\ll v$, the leading one-loop effect is obtained as~\cite{Lavoura:1993nq, Forshaw:2001xq, Cheng:2022hbo}
\begin{align}
\Delta \rho \approx \frac{G_F}{3\sqrt{2}\pi^{2}}(m_{H}-m_{H^\pm})^{2},
\end{align}
which is positive and therefore further increases the $\rho$ parameter.
In addition, the size of the mass difference is constrained by the theoretical bounds.
In the left panel of Fig.~\ref{fig: th_bounds_and_mWGF_mch200}, we show the contour of $m_{H}$ in $\overline{\lambda}_{2}$--$\overline{\lambda}_{3}$ plane with $m_{H^{\pm}}=200~\GeV$ and $\overline{v}_{\Delta} = 1~\GeV$.
The red (blue) colored region is excluded by vacuum stability (tree-level unitarity).
The mass-squared difference is suppressed by $\overline{v}_{\Delta}^{2}$.
\begin{align}
m_{H}^{2}-m_{H^{\pm}}^{2} = \overline{v}_{\Delta}^{2}\qty[\qty(\overline{\lambda}_{3}-\frac{2m_{H^{\pm}}^{2}}{\overline{v}^{2}})^{2}\frac{\overline{v}^{2}}{m_{H^{\pm}}^{2}-m_{h}^{2}}
+\overline{\lambda}_{2}-\frac{4m_{H^{\pm}}^{2}}{\overline{v}^{2}}]+\mathcal{O}(\overline{v}_{\Delta}^{4}).
\end{align}
Therefore, the mass difference is small and remains below $1$~GeV for $\overline{v}_{\Delta}=1$~GeV.
It is mainly determined by $\overline{\lambda}_{3}$ and weakly depends on $\overline{\lambda}_{2}$.
We note that the allowed region in the $\overline{\lambda}_{2}$--$\overline{\lambda}_{3}$ plane is almost independent of the values of $m_{H^{\pm}}$ and $\overline{v}_{\Delta}$.

For the EWPO constraints, we require that $m_{W},\, \rho,\, s_{e}^{2}$ and $\Gamma(Z\to \ell\overline{\ell})$ are consistent with the experimental data.
The $1\sigma$ errors of these quantities are given by~\cite{ParticleDataGroup:2026mpi}
\begin{align}
    |\Delta m_{W}| \leq 7.7~\mathrm{MeV}\qc
    |\rho-1| &\leq 9.3\times 10^{-4}\qc
    |\Delta s_{e}^{2}| \leq 1.3\times 10^{-4},
    \notag \\
    |\Delta \Gamma(Z\to \ell\overline{\ell})| &\leq 8.5\times 10^{-2}~\mathrm{MeV}.
\end{align}
Here, we have taken the collider average for $|\Delta s_{e}^{2}|$ and the result obtained under the assumption of lepton universality for $|\Delta \Gamma(Z\to \ell\overline{\ell})|$.
For the $W$-boson mass, we have adopted the averaged value excluding the CDF Run-II measurement~\cite{CDF:2022hxs}\footnote{For the interpretation of the CDF Run-II result in the RHTM, see Refs.~\cite{FileviezPerez:2022lxp, Cheng:2022hbo}.}.
We calculate the differences from the SM predictions at the one-loop level and impose the bounds at $2\sigma$ level.
In the right panel of Fig.~\ref{fig: th_bounds_and_mWGF_mch200}, we have evaluated $\Delta m_{W} = m_{W}^{\mathrm{RHTM}}-m_{W}^{\mathrm{SM}}$ at LO and NLO using \texttt{H-COUP} and plotted $m_{W}=80.357+\Delta m_{W}$, where the first term is the SM prediction taken from PDG~\cite{ParticleDataGroup:2026mpi}\footnote{The NLO value of $m_{W}^{\mathrm{SM}}$ obtained by \texttt{H-COUP} is $80.526$~GeV and relatively larger than the PDG value due to the lack of higher order corrections~\cite{Awramik:2003rn}.}.
The deviation in the $W$-boson mass is mainly induced by $\rho_{0}$, while the NLO contribution is almost negligible.
The upper bound on $\overline{v}_{\Delta}$ is about 2~GeV.
Since it is mainly determined by the LO prediction, the upper bound is almost independent of the other parameters such as $m_{H^{\pm}}, \overline{\lambda}
_{2}$ and $\overline{\lambda}_{3}$.

Next, we review the constraint from the collider searches.
When $v_{\Delta}<\mathcal{O}(1)~\GeV$, the additional Higgs bosons are mainly produced via the Drell--Yan process.
The constraints from multi-lepton final states are studied in Refs.~\cite{Butterworth:2023rnw, Ashanujjaman:2024lnr}.
$H^{\pm}$ with mass below $110~\GeV$ is excluded via the $H^{\pm}\to\tau\nu$ decay.
On the other hand, $H$ with mass in the $180\text{--}200~\GeV$ is excluded for small $\lambda_{3}$ via the $H\to ZZ$ decay.
For $m_{H^{\pm}}<m_{t}$, the top-quark decay into $H^{\pm}\bar{b}$ followed by $H^{\pm}\to WZ$ is studied in Ref.~\cite{Ashanujjaman:2025una}, and $v_{\Delta}$ above $1\text{-}2~\GeV$ is excluded.

Finally, we discuss the constraint from Higgs measurements.
We define the coupling modifier for the Higgs decays as
\begin{align}
    \kappa_{X}^{2} = \frac{\Gamma(h\to XX)}{\Gamma(h\to XX)_{\mathrm{SM}}}.
\end{align}
We use a symmetric bound around the SM value, \ie, $\kappa_{X}=1$, using the largest uncertainty among the ATLAS and CMS measurements~\cite{ATLAS:2025qxq, CMS:2026zxu, ParticleDataGroup:2026mpi}.
The $1\sigma$ errors of $\kappa_{X}$ are given by
\begin{align}
    \abs{\Delta \kappa_{W}}      \leq 0.06\qc
    \abs{\Delta \kappa_{Z}}      \leq 0.06\qc&
    \abs{\Delta \kappa_{\gamma}} \leq 0.07\qc
    \abs{\Delta \kappa_{g}}      \leq 0.07,
    \notag \\
    \abs{\Delta \kappa_{b}}       \leq 0.13\qc
    \abs{\Delta \kappa_{\tau}}    \leq& 0.07\qc
    \abs{\Delta \kappa_{Z\gamma}} \leq 0.36,
\end{align}
where $\Delta\kappa_{X} = \kappa_{X} - 1$.
We impose these bounds at $2\sigma$ level.

The trilinear Higgs coupling is sensitive to loop effects of new particles.
The experimental constraint on the coupling modifier is obtained from Higgs-pair production processes by scaling the SM trilinear Higgs coupling as $g_{hhh}^{\mathrm{SM}}\to \kappa_{\lambda}^{\mathrm{exp}}g_{hhh}^{\mathrm{SM}}$.
The current bound is $-0.71<\kappa_{\lambda}^{\mathrm{exp}}<6.1$~\cite{CMS:2026nuu}.
In this paper, we define the coupling modifier with the zero-momentum effective trilinear Higgs vertex.
\begin{align}
    \kappa_{\lambda} = 1+\frac{\widehat{\Gamma}_{hhh}(0, 0, 0)-\widehat{\Gamma}_{hhh}^{\mathrm{SM}}(0, 0, 0)}{\Gamma_{hhh}^{\mathrm{tree, SM}}}.
\end{align}
This definition isolates the effects of new physics from the SM prediction and approaches unity in the decoupling limit.

\subsection{Deviation in the SM-like Higgs boson decays}
\label{subsec: Deviations}

We study the deviations in the SM-like Higgs boson decays at NLO.
Following the input scheme described in Sec.~\ref{subsec: Input scheme}, we determine $\{\beta, m_{H}, \alpha\}$ from the randomly generated parameters $\{m_{H^{\pm}}, \overline{v}_{\Delta}, \overline{\lambda}_{2}, \overline{\lambda}_{3}\}$, and perform the one-loop calculations using the resulting parameter set.
We scan the following ranges of parameters and apply the theoretical and experimental constraints as discussed in the previous subsection.
\begin{align}
0.1~\GeV \leq \overline{v}_{\Delta} &< 3~\GeV\qc
200~\GeV \leq m_{H^{\pm}} < 1000~\GeV,
\notag \\
0 & \leq \overline{\lambda}_{2} < 4\pi\qc
-2.5 \leq \overline{\lambda}_{3} < 15.
\label{eq: scan_RHTM}
\end{align}

\subsubsection{Results in the RHTM}

\begin{figure}[t]
 \centering
 \includegraphics[scale=0.525]{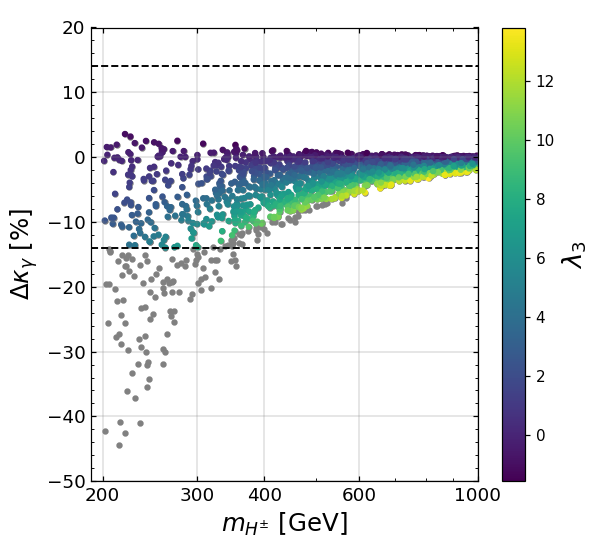}
 \includegraphics[scale=0.525]{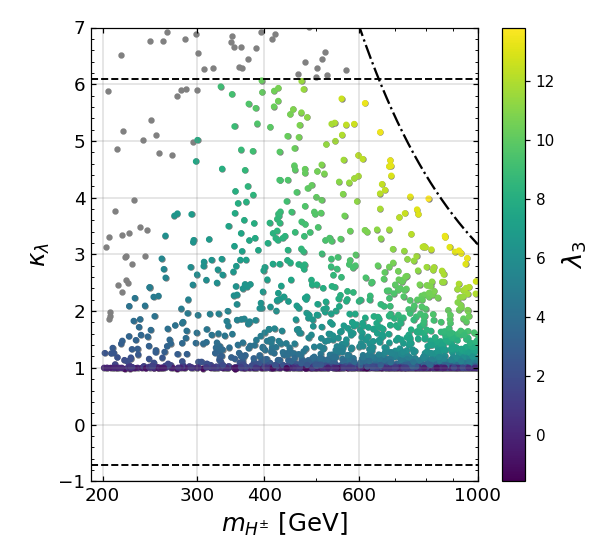}
 \caption{
 (Left) Prediction for $\Delta \kappa_{\gamma}$ and (Right) $\kappa_{\lambda}$ with theoretical and experimental constraints.
 The dashed lines indicate the constraints on $\kappa_{\gamma}$ and $\kappa_{\lambda}$.
 The colored (gray) points are allowed (excluded) by the Higgs signal measurements.
 In the right panel, the dot-dashed line indicates the maximal deviation allowed by the theoretical constraints.
 }
 \label{fig: dkappa_gam_mch_NLO and kappa_lam_mch_NLO}
\end{figure}

In the left panel of Fig.~\ref{fig: dkappa_gam_mch_NLO and kappa_lam_mch_NLO}, we show the prediction for $\Delta \kappa_{\gamma}$ as a function of $m_{H^{\pm}}$ and $\lambda_{3}$.
The colored (gray) points are allowed (excluded) by the Higgs signal measurements.
For $m_{H^{\pm}}\lesssim 400$~GeV, the magnitude of the $\lambda_{3}$ coupling is constrained by the $\kappa_{\gamma}$ measurement.
This is because $g_{H^{+}H^{-}h}\approx -\lambda_{3}v$, and the loop contribution of the charged Higgs boson is controlled by $\lambda_{3}$.
$\Delta \kappa_{\gamma}$ becomes negative (positive) for positive (negative) $\lambda_{3}$.
The possible value of negative $\lambda_{3}$ is constrained by the vacuum stability condition, $\lambda_{3}+\sqrt{\lambda_{1}\lambda_{2}}>0$, and $\Delta \kappa_{\gamma}$ can be as large as $+4\%$.

In the right panel of Fig.~\ref{fig: dkappa_gam_mch_NLO and kappa_lam_mch_NLO}, we show the prediction for $\kappa_{\lambda}$.
The dominant contribution arises from the additional Higgs boson loop effects in the 1PI diagrams~\cite{Kanemura:2002vm, Kanemura:2004mg, Kanemura:2015mxa}.
In the limit $v_{\Delta}\ll v$ and $m_{h}\ll m_{H^{\pm}},m_{H}$, the leading contribution can be approximated as
\begin{align}
\kappa_{\lambda}
\approx
1+\frac{\lambda_{3}^{3}v^{4}}{32\pi^{2} m_{h}^{2}m_{H^{\pm}}^{2}}.
\label{eq: approximate kappa_lambda}
\end{align}
Thus, the loop contribution is positive (negative) for positive (negative) $\lambda_{3}$.
Since negative $\lambda_{3}$ is constrained by vacuum stability, $\kappa_{\lambda}$ is typically larger than unity in the RHTM.
The magnitude of $\kappa_{\lambda}$ is constrained by tree-level unitarity, and the maximal deviation obtained from Eq.~\eqref{eq: approximate kappa_lambda} is shown by the dot-dashed line in Fig.~\ref{fig: dkappa_gam_mch_NLO and kappa_lam_mch_NLO}.
For $m_{H^{\pm}}\lesssim 600~\GeV$, the magnitude of the $\lambda_{3}$ coupling is constrained by the $\kappa_{\lambda}$ measurement, which provides a stronger bound on $\lambda_{3}$ than $\Delta\kappa_{\gamma}$ for $m_{H^{\pm}}\gtrsim 300~\GeV$.
On the other hand, $\Delta\kappa_{\gamma}$ provides a stronger bound in the region $m_{H^{\pm}}\lesssim 300~\GeV$.

\begin{figure}[t]
 \centering
 \includegraphics[scale=0.525]{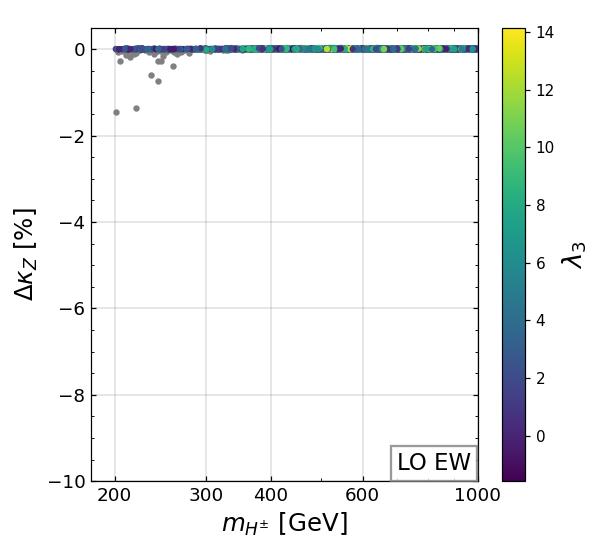}
 \includegraphics[scale=0.525]{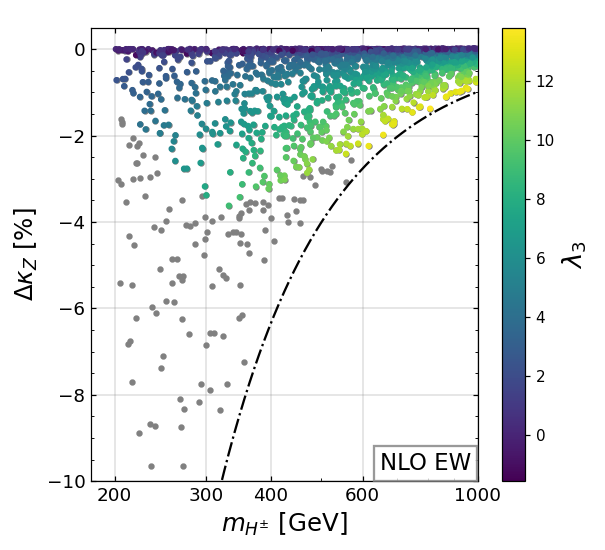}
 \caption{
 Prediction for $\Delta \kappa_Z$ at LO (left) and NLO (right) subject to theoretical and experimental constraints.
 The colored (gray) points are allowed (excluded) by the Higgs signal measurements.
 In the right panel, the dot-dashed line indicates the maximal deviation allowed by the theoretical constraints.
 }
 \label{fig: dkappa_Z_mch_NLO}
\end{figure}

In Fig.~\ref{fig: dkappa_Z_mch_NLO}, we show the prediction for $\Delta \kappa_Z$ at LO and NLO.
At LO, $\Delta \kappa_Z$ is induced by the scalar mixing effects, but these are suppressed due to the small $v_{\Delta}$.
At NLO, for $h\to f\bar{f}$ and $h\to VV^{*}$, the dominant contribution arises from the wave-function renormalization of the SM-like Higgs boson, $\delta Z_{hh}$~\cite{Kanemura:2004mg, Kanemura:2015mxa}.
In the limit $v_{\Delta}\ll v$ and $m_h\ll m_{H^\pm},m_H$, the leading contribution to $h\to ZZ^{*}$ can be approximated as
\begin{align}
\kappa_Z^2-1
\approx
-\frac{\lambda_3^2v^2}{64\pi^2m_{H^\pm}^2}.
\label{eq: approximate dkappa_Z}
\end{align}
Thus, $\Delta\kappa_Z$ becomes more negative for larger $|\lambda_3|$.
The magnitude of $\Delta\kappa_Z$ is constrained by tree-level unitarity, and this upper bound obtained by Eq.~\eqref{eq: approximate dkappa_Z} is shown by the dot-dashed line in Fig.~\ref{fig: dkappa_Z_mch_NLO}.
As shown in Fig.~\ref{fig: dkappa_gam_mch_NLO and kappa_lam_mch_NLO}, $\lambda_3$ is also constrained by $\Delta\kappa_\gamma$ and $\kappa_\lambda$ for $m_{H^\pm}\lesssim 600$~GeV.

\subsubsection{Model discrimination}

Let us study the distinguishability of the RHTM from other extended Higgs models, such as the HSM, the IDM, the type-I and type-II 2HDMs (2HDM-I and 2HDM-II), and the CHTM in the heaviest $H^{\pm\pm}$ scenario (HS) and the lightest $H^{\pm\pm}$ scenario (LS)
We parameterize the deviations in the decay rates and branching ratios as
\begin{align}
\Delta R(h\to XY) = \frac{\Gamma(h\to XY)}{\Gamma(h\to XY)_{\mathrm{SM}}}-1\qc
\Delta \mu(h\to XY) = \frac{\BR(h\to XY)}{\BR(h\to XY)_{\mathrm{SM}}}-1.
\end{align}
The decay rates of $h$ are calculated including higher-order corrections for all of these models using \texttt{H-COUP}~\cite{Kanemura:2017gbi, Kanemura:2019slf, Aiko:2023xui} (see Ref.~\cite{Aiko:2026qzi} for the CHTM).

We scan the input parameters defined in Refs.~\cite{Aiko:2023xui, Aiko:2026qzi} as follows\footnote{We adopt the $b\to s\gamma$ constraint for the Type-II 2HDM from Ref.~\cite{Misiak:2026sqy}, which is slightly weaker than that obtained in Ref.~\cite{Misiak:2020vlo}.}:
\begin{align}
\text{HSM:}\ \ &
    400~\GeV \leq m_{H} < 1000~\GeV\qc
    \lambda_{S} = 0.1\qc
    \mu_{s} = 0~\GeV,
\notag \\ &
    0.95 \leq c_{\alpha} < 1\qc
    0 \leq m_{s} < m_{H}+500~\GeV, \\
\text{IDM:}\ \ &
    m_{H} = 60~\GeV\qc
    m_{A}=m_{H^{\pm}}\qc
    \mu_{2}^{2} = 3581~\GeV^{2},
\notag \\ &
    100~\GeV \leq m_{H^{\pm}} < 1000~\GeV\qc
    0 \leq \lambda_{2} < 4\pi, \\
\text{2HDM Type-I (II):}\ \ &
    400\ (670)~\GeV \leq m_{H} < 1000~\GeV\qc
    m_{A}=m_{H^{\pm}}=m_{H},
\notag \\ &
    0.99\ (0.995) \leq s_{\beta-\alpha} < 1\qc
    1.5\ (1) \leq \tan{\beta} < 10,
\notag \\ &
    0 \leq M < m_H+500~\GeV, \\
\text{CHTM HS (LS):}\ \ &
    0.1~\GeV \le \overline{v}_{\Delta} < 18\ (12)~\GeV,
\notag \\ &
    300\ (400)~\GeV \le m_{\mathrm{Lightest}} < 1000~\GeV,
\notag \\ &
    m_{H} = m_{A}\qc m_{H^{\pm}}^{2}-m_{A}^{2} = m_{H^{\pm\pm}}^{2}-m_{H^{\pm}}^{2},
\notag \\ &
    0 \le m_{H^{\pm}} - m_{\mathrm{Lightest}} < 150~\GeV\qc
    0\ (-5) \le \overline{\lambda}_{4} < 4\pi,
\end{align}
where $m_{\mathrm{Lightest}} = m_{A}\, (m_{H^{\pm\pm}})$ in the HS (LS) of the CHTM.
The scan regions are chosen such that the direct-search and flavor constraints are satisfied.
For the HSM, IDM, and 2HDMs, we impose the constraints from the EWPOs using the $ST$ global fit result with $U=0$~\cite{ParticleDataGroup:2026mpi}.
For the CHTM, we impose the EWPO constraints following the discussion in Sec.~\ref{subsec: Constraints} (see also Ref.~\cite{Aiko:2026qzi}).
The constraints from the Higgs signal measurements are imposed using $\Delta \kappa_{X}$ and $\kappa_{\lambda}$, as discussed in Sec.~\ref{subsec: Constraints}.

\begin{figure}[t]
 \centering
 \includegraphics[scale=0.535]{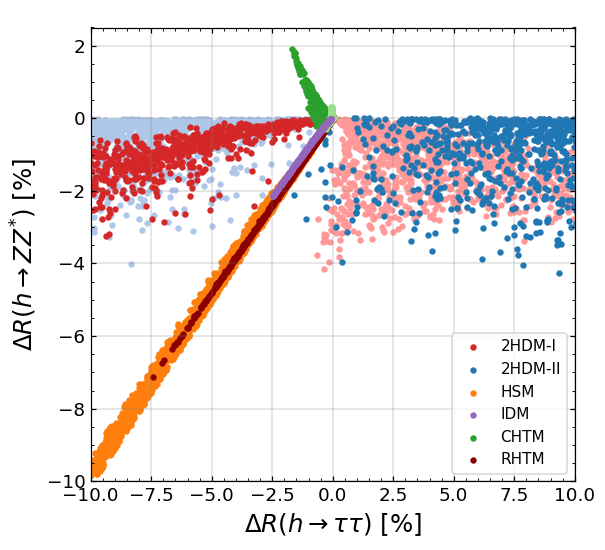}
 \includegraphics[scale=0.535]{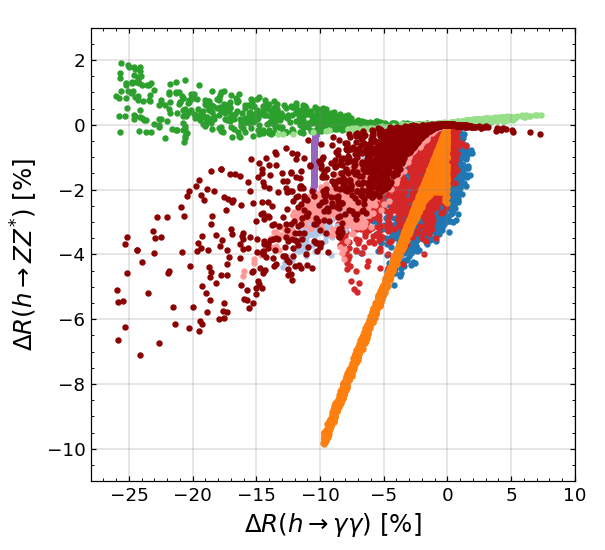}
 \\
 \includegraphics[scale=0.535]{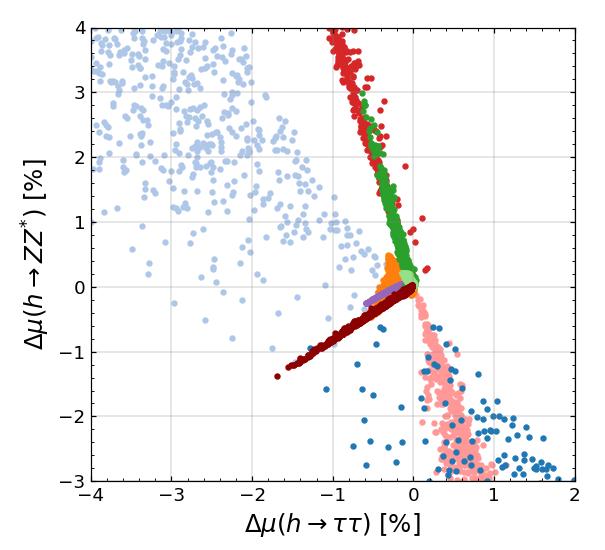}
 \includegraphics[scale=0.535]{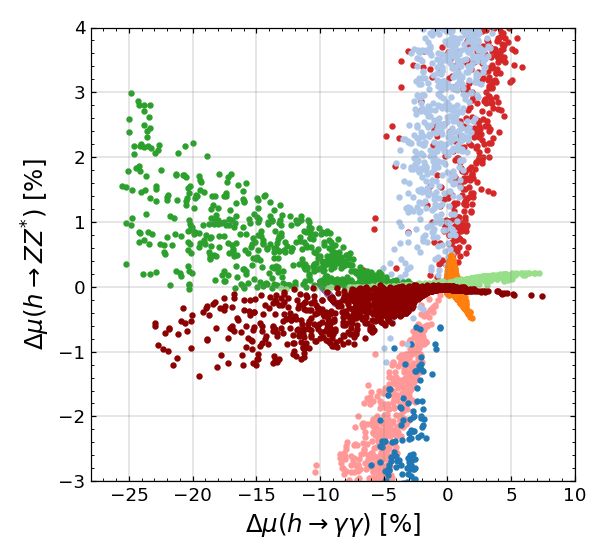} 
 \caption{Correlations between the deviations in the decay rates (upper panels) and branching ratios (lower panels) of the SM-like Higgs boson at NLO EW.
 The left and right panels show the predictions on the $h\to\tau\tau$ and $h\to ZZ^{*}$ and on the $h\to\gamma\gamma$ and $h\to ZZ^{*}$ channels, respectively.
 Different colors indicate the different extended Higgs models shown in the upper-left panel.
 In the 2HDM, predictions with $c_{\beta-\alpha}<0$ ($c_{\beta-\alpha}>0$) are shown in darker (lighter) colors. Similarly, in the CHTM, predictions with HS (LS) are shown in darker (lighter) colors.}
 \label{fig: dR_tautau_ZZstar and dR_gamgam_ZZstar}
\end{figure}

In the upper panels of Fig.~\ref{fig: dR_tautau_ZZstar and dR_gamgam_ZZstar}, we show the correlations of the deviations in the decay rates of the SM-like Higgs boson.
The darkred, orange, red, blue, purple, and green points are the predictions in the RHTM, HSM, 2HDM-I, 2HDM-II, IDM, and CHTM, respectively.
In the 2HDM, predictions with $c_{\beta-\alpha}<0$ ($c_{\beta-\alpha}>0$) are shown in darker (lighter) colors. Similarly, in the CHTM, predictions with HS (LS) are shown in darker (lighter) colors.
In the RHTM, both $\Delta R(h\to \tau\tau)$ and $\Delta R(h\to ZZ^{*})$ decrease because the leading correction commonly arises from $\delta Z_{hh}$.
A similar effect appears in the IDM and the CHTM with small $\overline{v}_{\Delta}$, but the magnitudes are relatively small.
In the HSM, a similar correlation appears mainly due to the LO mixing effects, and the predictions mostly overlap with those in the RHTM.
In the 2HDMs, $\Delta R(h\to \tau\tau)$ is sensitive to the mixing and loop effects, and it tends to be larger than $\Delta R(h\to ZZ^{*})$.
In the CHTM, $\Delta R(h\to ZZ^{*})$ becomes positive for large $\overline{v}_{\Delta}$ due to the LO mixing effects.
In the HSM, $\Delta R(h\to \gamma\gamma)$ arises mainly from the scalar mixing, and the sizes of $\Delta R(h\to \gamma\gamma)$ and $\Delta R(h\to ZZ^{*})$ are comparable.
On the other hand, in the RHTM, the charged Higgs loop leads to a larger deviation in $\Delta R(h\to \gamma\gamma)$ than in $\Delta R(h\to ZZ^{*})$.

In the lower panels of Fig.~\ref{fig: dR_tautau_ZZstar and dR_gamgam_ZZstar}, we show the correlations of $\Delta \mu(h\to XY)$.
The effect of the Higgs total width changes the correlations from those in $\Delta R(h\to XY)$, especially in the HSM and 2HDMs.
In the HSM, the LO mixing effects commonly appear in the Higgs decay processes.
Therefore, the effects in the decay rates and the total width are partially canceled, and the deviations in the branching ratios become smaller than those in the decay rates.
In the RHTM, the size of $\Delta \mu(h\to XY)$ also becomes smaller because both the Higgs decay rates and the total width decrease due to the loop effects, while the cancellation is moderate compared with that in the HSM.
In the 2HDMs, the deviation in the Higgs total width is sizable, and the correlation is qualitatively changed from those in the Higgs decay rates.
In the CHTM, the loop effects decrease the Higgs total width via $h\to b\bar{b}$, and the positive deviation in $\Delta \mu(h\to ZZ^{*})$ is larger than that in $\Delta R(h\to ZZ^{*})$.
In the $h\to \tau\tau$ and $h\to ZZ^{*}$ plane, the predictions in the RHTM overlap with those in the 2HDM-II.
However, in the $h\to \gamma\gamma$ and $h\to ZZ^{*}$ plane, the RHTM can be distinguished from the 2HDM-II due to the effect of the Higgs total width.

\section{Conclusions}
\label{sec: Conclusions}

In this paper, we investigated the electroweak one-loop corrections to the decay rates of the SM-like Higgs boson in the RHTM.
The EW gauge and Higgs sectors are renormalized in the on-shell scheme, where the scalar mixing angles are renormalized in a gauge-independent way using the pinch technique.
The renormalized form factors of the effective three-point vertices of the SM-like Higgs boson are constructed from the 1PI vertex functions and the corresponding counterterms.
The two-body decays ($h\to f\bar{f}$) and three-body decays ($h\to VV^\ast$) are calculated including the NLO EW and QCD corrections.
The loop-induced processes $h\to\gamma\gamma$, $Z\gamma$, and $gg$ are calculated  including higher-order QCD corrections, while the EW corrections are at LO.
The resulting analytic formulae are implemented in the \texttt{H-COUP} program~\cite{Kanemura:2017gbi, Kanemura:2019slf, Aiko:2023xui}, and the correlations of the deviations in the decay rates and branching ratios of the SM-like Higgs boson are studied in comparison with various extended Higgs models.

The LO deviations in the RHTM are rather small due to the constraint on $v_{\Delta}$ from the EWPO measurements.
However, sizable NLO effects can arise in the decay rates through the Higgs wave-function (external-leg) corrections.
In addition, the Higgs di-photon decay rate and the Higgs trilinear coupling are also sensitive to loop contributions.
We have shown that the RHTM can be distinguished from other extended Higgs models, such as the HSM, IDM, 2HDMs, and CHTM, by studying the correlations among the $h\to\tau\tau$, $h\to ZZ^{*}$, and $h\to\gamma\gamma$ decay modes.
Thus, small LO deviations do not imply negligible effects in Higgs phenomenology in the RHTM, and precision measurements of the SM-like Higgs boson at future collider experiments will play an important role in probing extended Higgs sectors through higher-order effects.

\begin{acknowledgments}
SK was supported, in part, by Japan Society for Promotion of Science (JSPS) under the Bilateral Program between Japan and China (JPJSBP120267416).
KS was supported by the JSPS KAKENHI Grant Numbers 25K17379 and Tsuruoka National College of Technology Promotion Association.
ST was supported by JST SPRING, Grant Number JPMJSP2138.
\end{acknowledgments}

\appendix

\section{Feynman rules}
\label{app: Feynman rules}

\begin{table}[t]
	\begin{tabular}{p{6em}p{12em}p{6em}p{12em}}
		\hline
		Vertices & $g_{\phi V_{1}V_{2}}^{}$ & Vertices & $g_{\phi V_{1}V_{2}}^{}$\\
		\hline
		$hW^{+}W^{-}$ & $gm_{W}(c_{\beta}c_{\alpha}-2s_{\beta}s_{\alpha})$ &
		$hZZ$ & $\frac{g_{Z}m_{Z}}{2}c_{\alpha}$ \\
		$HW^{+}W^{-}$ & $-gm_{W}(c_{\beta}s_{\alpha}+2s_{\beta}c_{\alpha})$ &
		$HZZ$ & $-\frac{g_{Z}m_{Z}}{2}s_{\alpha}$ \\
		$G^{\pm}ZW^{\mp}$ & $-g_{Z}m_{W}(s_{W}^{2}-s_{\beta}^{2})$ &
		$G^{\pm}\gamma W^{\mp}$ & $em_{W}$ \\
		$H^{\pm}ZW^{\mp}$ & $g_{Z}m_{W}s_{\beta}c_{\beta}$ &
		\\
		\hline
		\end{tabular}
		\caption{The scalar-gauge-gauge vertices.}
    \label{table: SVV}
\end{table}

\begin{table}[t]
	\begin{tabular}{p{6em}p{12em}p{6em}p{12em}}
		\hline
		Vertices & $g_{\phi_{1}\phi_{2}V}^{}$ & Vertices & $g_{\phi_{1}\phi_{2}V}^{}$\\
		\hline
		$G^{\pm}hW^{\mp}$ & $\pm i\frac{g}{2}(c_{\alpha}c_{\beta}-2s_{\alpha}s_{\beta})$ &
		$H^{\pm}hW^{\mp}$ & $\mp i\frac{g}{2}(c_{\alpha}s_{\beta}+2s_{\alpha}c_{\beta})$ \\
		$G^{\pm}HW^{\mp}$ & $\mp i\frac{g}{2}(s_{\alpha}c_{\beta}+2c_{\alpha}s_{\beta})$ &
		$H^{\pm}HW^{\mp}$ & $\pm i\frac{g}{2}(s_{\alpha}s_{\beta}-2c_{\alpha}c_{\beta})$ \\
		$G^{\pm}G^{0}W^{\mp}$ & $\frac{g}{2}c_{\beta}$ &
		$H^{\pm}G^{0}W^{\mp}$ & $-\frac{g}{2}s_{\beta}$ \\
		$G^{+}G^{-}Z$ & $i\frac{g_{Z}}{2}(c_{2W}+s_{\beta}^{2})$ &
		$G^{+}G^{-}\gamma$ & $ie$ \\
		$H^{+}H^{-}Z$ & $i\frac{g_{Z}}{2}(c_{2W}+c_{\beta}^{2})$ &
		$H^{+}H^{-}\gamma$ & $ie$ \\
		$H^{\pm}G^{\mp}Z$ & $\pm i\frac{g_{Z}}{2}c_{\beta}s_{\beta}$ &
		$G^{0}hZ$ & $\frac{g_{Z}}{2}c_{\alpha}$ \\
		 & &
		$G^{0}HZ$ & $-\frac{g_{Z}}{2}s_{\alpha}$ \\
		\hline
		\end{tabular}
		\caption{The scalar-scalar-gauge vertices.}
    \label{table: SSV}
\end{table}

\begin{table}[t]
	\begin{tabular}{p{6em}p{12em}p{6em}p{12em}}
		\hline
		Vertices & $g_{\phi_{1}\phi_{2}V_{1}V_{2}}^{}$ & Vertices & $g_{\phi_{1}\phi_{2}V_{1}V_{2}}^{}$\\
		\hline
		$G^{+}G^{-}W^{+}W^{-}$ & $\frac{g^{2}}{2}(1+s_{\beta}^{2})$ &
		$G^{+}G^{-}ZZ$ & $\frac{g_{Z}^{2}}{4}(4c_{W}^{4}s_{\beta}^{2}+c_{2W}^{2}c_{\beta}^{2})$ \\
		$H^{+}H^{-}W^{+}W^{-}$ & $\frac{g^{2}}{2}(1+c_{\beta}^{2})$ &
		$H^{+}H^{-}ZZ$ & $\frac{g_{Z}^{2}}{4}(4c_{W}^{4}c_{\beta}^{2}+c_{2W}^{2}s_{\beta}^{2})$ \\
		$H^{\pm}G^{\mp}W^{+}W^{-}$ & $\frac{g^{2}}{2}s_{\beta}c_{\beta}$ &
		$H^{\pm}G^{\mp}ZZ$ & $\frac{g_{Z}^{2}}{4}(4c_{W}^{4}-c_{2W}^{2})s_{\beta}c_{\beta}$ \\
		$hhW^{+}W^{-}$ & $\frac{g^{2}}{4}(1+3s_{\alpha}^{2})$ &
		$hhZZ$ & $\frac{g_{Z}^{2}}{8}c_{\alpha}^{2}$ \\
		$HHW^{+}W^{-}$ & $\frac{g^{2}}{4}(1+3c_{\alpha}^{2})$ &
		$HHZZ$ & $\frac{g_{Z}^{2}}{8}s_{\alpha}^{2}$ \\
		$HhW^{+}W^{-}$ & $\frac{3g^{2}}{2}s_{\alpha}c_{\alpha}$ &
		$HhZZ$ & $-\frac{g_{Z}^{2}}{4}s_{\alpha}c_{\alpha}$ \\
		$G^{0}G^{0}W^{+}W^{-}$ & $\frac{g^{2}}{4}$ &
		$G^{0}G^{0}ZZ$ & $\frac{g_{Z}^{2}}{8}$ \\
		$G^{\pm}hW^{\mp}Z$ & $-\frac{gg_{Z}}{2}(s_{W}^{2}c_{\beta}c_{\alpha}+2c_{W}^{2}s_{\beta}s_{\alpha})$ &
		$H^{\pm}hW^{\mp}Z$ & $\frac{gg_{Z}}{2}(s_{W}^{2}s_{\beta}c_{\alpha}-2c_{W}^{2}c_{\beta}s_{\alpha})$ \\
		$G^{\pm}HW^{\mp}Z$ & $\frac{gg_{Z}}{2}(s_{W}^{2}c_{\beta}s_{\alpha}-2c_{W}^{2}s_{\beta}c_{\alpha})$ &
		$H^{\pm}HW^{\mp}Z$ & $-\frac{gg_{Z}}{2}(s_{W}^{2}s_{\beta}s_{\alpha}+2c_{W}^{2}c_{\beta}c_{\alpha})$ \\
		$G^{\pm}G^{0}W^{\mp}Z$ & $\pm i\frac{gg_{Z}}{2}s_{W}^{2}c_{\beta}$ &
		$H^{\pm}G^{0}W^{\mp}Z$ & $\mp i\frac{gg_{Z}}{2}s_{W}^{2}s_{\beta}$ \\
		$G^{\pm}hW^{\mp}\gamma$ & $\frac{eg}{2}(c_{\beta}c_{\alpha}-2s_{\beta}s_{\alpha})$ &
		$H^{\pm}hW^{\mp}\gamma$ & $-\frac{eg}{2}(s_{\beta}c_{\alpha}+2c_{\beta}s_{\alpha})$ \\
		$G^{\pm}HW^{\mp}\gamma$ & $-\frac{eg}{2}(c_{\beta}s_{\alpha}+2s_{\beta}c_{\alpha})$ &
		$H^{\pm}HW^{\mp}\gamma$ & $\frac{eg}{2}(s_{\beta}s_{\alpha}-2c_{\beta}c_{\alpha})$ \\
		$G^{\pm}G^{0}W^{\mp}\gamma$ & $\mp i\frac{eg}{2}c_{\beta}$ &
		$H^{\pm}G^{0}W^{\mp}\gamma$ & $\pm i\frac{eg}{2}s_{\beta}$ \\
		$G^{\pm}G^{\pm}W^{\mp}W^{\mp}$ & $-\frac{g^{2}}{2}s_{\beta}^{2}$ &			$G^{+}G^{-}\gamma Z$ & $eg_{Z}(c_{2W}+s_{\beta}^{2})$ \\
		$H^{\pm}H^{\pm}W^{\mp}W^{\mp}$ & $-\frac{g^{2}}{2}c_{\beta}^{2}$ &
		$H^{+}H^{-}\gamma Z$ & $eg_{Z}(c_{2W}+c_{\beta}^{2})$ \\
		$H^{\pm}G^{\pm}W^{\mp}W^{\mp}$ & $-g^{2}s_{\beta}c_{\beta}$ &
		$H^{\pm}G^{\mp}\gamma Z$ & $eg_{Z}s_{\beta}c_{\beta}$ \\
		$G^{+}G^{-}\gamma\gamma$ & $e^{2}$ &
		$H^{+}H^{-}\gamma\gamma$ & $e^{2}$ \\
		\hline
		\end{tabular}
		\caption{The scalar-scalar-gauge-gauge vertices.}
    \label{table: SSVV}
\end{table}

\begin{table}[t]
	\begin{tabular}{p{6em}p{12em}p{6em}p{12em}}
		\hline
		Vertices & $g_{\phi \bar{c}_{1}c_{2}}^{}$ & Vertices & $g_{\phi \bar{c}_{1}c_{2}}^{}$\\
		\hline
		$h\bar{c}^{\pm}c^{\mp}$ & $-\frac{g\xi_{W}m_{W}}{2}(c_{\beta}c_{\alpha}-2s_{\beta}s_{\alpha})$ &
		$h\bar{c}_{Z}c_{Z}$ & $-\frac{g_{Z}\xi_{Z}m_{Z}}{2}c_{\alpha}$ \\
		$H\bar{c}^{\pm}c^{\mp}$ & $\frac{g\xi_{W}m_{W}}{2}(c_{\beta}s_{\alpha}+2s_{\beta}c_{\alpha})$ &
		$H\bar{c}_{Z}c_{Z}$ & $\frac{g_{Z}\xi_{Z}m_{Z}}{2}s_{\alpha}$ \\
		$G^{0}\bar{c}^{\pm}c^{\mp}$ & $\pm i\frac{g\xi_{W}m_{W}}{2}c_{\beta}$
		& \\
		$G^{\pm}\bar{c}^{\mp}c_{Z}$ & $-\frac{g_{Z}\xi_{W}m_{W}}{2}(c_{2W}c_{\beta}^{2}+2c_{W}^{2}s_{\beta}^{2})$ &
		$G^{\pm}\bar{c}_{Z}c^{\mp}$ & $\frac{g\xi_{Z}m_{Z}}{2}c_{\beta}$ \\
		$H^{\pm}\bar{c}^{\mp}c_{Z}$ & $-\frac{g_{Z}\xi_{W}m_{W}}{2}c_{\beta}s_{\beta}$ &
		$H^{\pm}\bar{c}_{Z}c^{\mp}$ & $-\frac{g\xi_{Z}m_{Z}}{2}s_{\beta}$ \\
		$G^{\pm}\bar{c}^{\mp}c_{A}$ & $-e\xi_{W}m_{W}$ &
		& \\
		\hline
		\end{tabular}
		\caption{The scalar-ghost-ghost vertices.}
    \label{table: Scc}
\end{table}

The interaction terms among the scalar and the gauge bosons are given by
\begin{align}
\mathcal{L} \supset 
g_{\phi V_{1}V_{2}}g_{\mu\nu}\phi V_{1}^{\mu}V_{2}^{\nu}
+g_{\phi_{1}\phi_{2}V}(\partial^{\mu}\phi_{1}\phi_{2}-\phi_{1}\partial^{\mu}\phi_{2})V_{\mu}
+g_{\phi_{1}\phi_{2}V_{1}V_{2}}g_{\mu\nu}\phi_{1}\phi_{2}V_{1}^{\mu}V_{2}^{\nu}.
\end{align}
We define the charge flow such that particles (anti-particles) are incoming to (outgoing from) the vertex.
The $g_{\phi V_{1}V_{2}}^{}, g_{\phi_{1}\phi_{2}V}$, and $g_{\phi_{1}\phi_{2}V_{1}V_{2}}$ couplings are listed in Tables~\ref{table: SVV}, \ref{table: SSV}, and ~\ref{table: SSVV}, respectively.
The interaction terms among the Higgs bosons and the ghost fields are given by
\begin{align}
\mathcal{L} \supset 
g_{\phi \bar{c}_{1}c_{2}}\phi\bar{c}_{1}c_{2}.
\end{align}
The coefficients of the scalar-ghost-ghost vertex $g_{\phi \bar{c}_{1}c_{2}}$ are listed in Table~\ref{table: Scc}.

The scalar trilinear and quartic couplings are given by
\begin{align}
\mathcal{L} \supset
g_{\phi_{i}\phi_{j}\phi_{k}}^{} \phi_{i}\phi_{j}\phi_{k}
+g_{\phi_{i}\phi_{j}\phi_{k}\phi_{l}}^{} \phi_{i}\phi_{j}\phi_{k}\phi_{l}.
\end{align}
The coefficients of the scalar trilinear coupling are given by
\begin{align}
g_{H^{+}H^{-}h}^{}
&=
-\frac{1}{2}\mu s_{\beta}\left(s_{\alpha} s_{\beta}-2 c_{\alpha} c_{\beta}\right)
-\lambda_{1} c_{\alpha} s_{\beta}^{2} v_{\phi}
-\lambda_{2} s_{\alpha} c_{\beta}^{2} v_{\Delta}
\notag \\ &\quad
+\lambda_{3} \left(-c_{\alpha} c_{\beta}^{2} v_{\phi}-s_{\alpha} s_{\beta}^{2} v_{\Delta}\right), \\
g_{H^{+}H^{-}H}^{}
&=
-\frac{1}{2} \mu s_{\beta} \left(c_{\alpha} s_{\beta}+2 s_{\alpha} c_{\beta}\right)
+\lambda_{1} s_{\alpha} s_{\beta}^{2} v_{\phi}
-\lambda_{2} c_{\alpha} c_{\beta}^{2} v_{\Delta}
\notag \\ &\quad
+\lambda_{3} \left(s_{\alpha} c_{\beta}^{2} v_{\phi}-c_{\alpha} s_{\beta}^{2} v_{\Delta}\right), \\
g_{G^{+}G^{-}h}^{}
&=
-\frac{1}{2} \mu  c_{\beta} \left(s_{\alpha} c_{\beta}+2 c_{\alpha} s_{\beta}\right)
-\lambda_{1} c_{\alpha} c_{\beta}^{2} v_{\phi}
-\lambda_{2} s_{\alpha} s_{\beta}^{2} v_{\Delta}
\notag \\ &\quad
+\lambda_{3} \left(-c_{\alpha} s_{\beta}^{2} v_{\phi}-s_{\alpha} c_{\beta}^{2} v_{\Delta}\right), \\
g_{G^{+}G^{-}H}^{}
&=
-\frac{1}{2}\mu c_{\beta} \left(c_{\alpha} c_{\beta} -2 s_{\alpha} s_{\beta}\right)
+\lambda_{1} s_{\alpha} c_{\beta}^{2} v_{\phi}
-\lambda_{2} c_{\alpha} s_{\beta}^{2} v_{\Delta}
\notag \\ &\quad
+\lambda_{3} \left(s_{\alpha} s_{\beta}^{2} v_{\phi}-c_{\alpha} c_{\beta}^{2} v_{\Delta}\right), \\
g_{G^{\pm}H^{\mp}h}^{}
&=
\frac{1}{2} \mu  \left(s_{\alpha} c_{\beta} s_{\beta}-c_{\alpha} \left(c_{\beta}^{2}-s_{\beta}^{2}\right)\right)
+\lambda_{1} c_{\alpha} c_{\beta} s_{\beta} v_{\phi}
-\lambda_{2} s_{\alpha} c_{\beta} s_{\beta} v_{\Delta}
\notag \\ &\quad
+\lambda_{3} c_{\beta} s_{\beta} \left(s_{\alpha} v_{\Delta}-c_{\alpha} v_{\phi}\right), \\
g_{G^{\pm}H^{\mp}H}^{}
&=
\frac{1}{2} \mu \left(c_{\alpha} c_{\beta} s_{\beta} +s_{\alpha} \left(c_{\beta}^{2}  -s_{\beta}^{2}\right)\right)
-\lambda_{1} s_{\alpha} c_{\beta} s_{\beta} v_{\phi}
-\lambda_{2} c_{\alpha} c_{\beta} s_{\beta} v_{\Delta}
\notag \\ &\quad
+\lambda_{3} c_{\beta} s_{\beta} \left(c_{\alpha} v_{\Delta}+s_{\alpha} v_{\phi}\right), \\
g_{G^{\pm}H^{\mp}G^{0}}^{}
&=
\pm \frac{1}{2} i \mu, \\
g_{G^{0}G^{0}h}^{}
&=
\frac{1}{4} \mu s_{\alpha}-\frac{1}{2} \lambda_{1} c_{\alpha} v_{\phi}-\frac{1}{2} \lambda_{3} s_{\alpha} v_{\Delta}, \\
g_{G^{0}G^{0}H}^{}
&=
\frac{1}{4} \mu c_{\alpha}
+\frac{1}{2} \lambda_{1} s_{\alpha} v_{\phi}
-\frac{1}{2} \lambda_{3} c_{\alpha} v_{\Delta}, \\
g_{hhh}^{}
&=
\frac{1}{4} \mu  c_{\alpha}^{2} s_{\alpha}
-\frac{1}{2} \lambda_{1} c_{\alpha}^{3} v_{\phi}
-\frac{1}{2} \lambda_{2} s_{\alpha}^{3} v_{\Delta}
-\frac{1}{2} \lambda_{3} c_{\alpha} s_{\alpha} \left(c_{\alpha} v_{\Delta}+s_{\alpha} v_{\phi}\right), \\
g_{Hhh}^{}
&=
\frac{1}{4} \mu c_{\alpha}\left(c_{\alpha}^{2}-2 s_{\alpha}^{2}\right)
+\frac{3}{2} \lambda_{1} c_{\alpha}^{2} s_{\alpha} v_{\phi}
-\frac{3}{2} \lambda_{2} c_{\alpha} s_{\alpha}^{2} v_{\Delta}
\notag \\ &\quad
+\frac{1}{2} \lambda_{3} \left(c_{\alpha} \left(2 s_{\alpha}^{2}-c_{\alpha}^{2} \right) v_{\Delta}+s_{\alpha} \left(s_{\alpha}^{2}-2 c_{\alpha}^{2} \right) v_{\phi}\right), \\
g_{HHh}^{}
&=
\frac{1}{4} \mu s_{\alpha} \left(s_{\alpha}^{2}-2 c_{\alpha}^{2} \right)
-\frac{3}{2} \lambda_{1} c_{\alpha} s_{\alpha}^{2} v_{\phi}
-\frac{3}{2} \lambda_{2} c_{\alpha}^{2} s_{\alpha} v_{\Delta}
\notag \\ &\quad
-\frac{1}{2} \lambda_{3} \left(s_{\alpha}\left(s_{\alpha}^{2}-2 c_{\alpha}^{2} \right) v_{\Delta}+c_{\alpha} \left(c_{\alpha}^{2}-2 s_{\alpha}^{2} \right) v_{\phi}\right), \\
g_{HHH}^{}
&=
\frac{1}{4} \mu  c_{\alpha} s_{\alpha}^{2}+\frac{1}{2} \lambda_{1} s_{\alpha}^{3} v_{\phi}-\frac{1}{2} \lambda_{2} c_{\alpha}^{3} v_{\Delta}+\frac{1}{2} \lambda_{3} c_{\alpha} s_{\alpha} \left(c_{\alpha} v_{\phi}-s_{\alpha} v_{\Delta}\right).
\end{align}
The coefficients of the scalar quartic coupling are given by
\begin{align}
g_{G^{+}G^{-}G^{+}G^{-}}^{}
&=
-\frac{1}{2} \lambda_{1} c_{\beta}^{4}-\frac{1}{2} \lambda_{2} s_{\beta}^{4}-\lambda_{3} c_{\beta}^{2} s_{\beta}^{2}, \\
g_{H^{+}H^{-}H^{+}H^{-}}^{}
&=
-\frac{1}{2} \lambda_{1} s_{\beta}^{4}-\frac{1}{2} \lambda_{2} c_{\beta}^{4}-\lambda_{3} c_{\beta}^{2} s_{\beta}^{2}, \\
g_{H^{+}H^{-}G^{+}G^{-}}^{}
&=
-2 \lambda_{1} c_{\beta}^{2} s_{\beta}^{2}-2 \lambda_{2} c_{\beta}^{2} s_{\beta}^{2}-\lambda_{3} \left(c_{\beta}^{2}-s_{\beta}^{2}\right)^{2}, \\
g_{G^{+}G^{-}G^{\pm}H^{\mp}}^{}
&=
\lambda_{1} c_{\beta}^{3} s_{\beta}-\lambda_{2} c_{\beta} s_{\beta}^{3}-\lambda_{3} c_{\beta} s_{\beta} \left(c_{\beta}^{2}-s_{\beta}^{2}\right), \\
g_{H^{+}H^{-}H^{\pm}G^{\mp}}^{}
&=
\lambda_{1} c_{\beta} s_{\beta}^{3}-\lambda_{2} c_{\beta}^{3} s_{\beta}+\lambda_{3} c_{\beta} s_{\beta} \left(c_{\beta}^{2}-s_{\beta}^{2}\right), \\
g_{H^{\pm}G^{\mp}H^{\pm}G^{\mp}}^{}
&=
-\frac{1}{2} \lambda_{1} c_{\beta}^{2} s_{\beta}^{2}-\frac{1}{2} \lambda_{2} c_{\beta}^{2} s_{\beta}^{2}+\lambda_{3} c_{\beta}^{2} s_{\beta}^{2}, \\
g_{H^{+}H^{-}hh}^{}
&=
-\frac{1}{2} \lambda_{1} c_{\alpha}^{2} s_{\beta}^{2}
-\frac{1}{2} \lambda_{2} s_{\alpha}^{2} c_{\beta}^{2}
-\frac{1}{2} \lambda_{3} \left(c_{\alpha}^{2} c_{\beta}^{2}+s_{\alpha}^{2} s_{\beta}^{2}\right), \\
g_{H^{+}H^{-}HH}^{}
&=
-\frac{1}{2} \lambda_{1} s_{\alpha}^{2} s_{\beta}^{2}
-\frac{1}{2} \lambda_{2} c_{\alpha}^{2} c_{\beta}^{2}
-\frac{1}{2} \lambda_{3} \left(s_{\alpha}^{2} c_{\beta}^{2}+c_{\alpha}^{2} s_{\beta}^{2}\right), \\
g_{H^{+}H^{-}Hh}^{}
&=
\lambda_{1} c_{\alpha} s_{\alpha} s_{\beta}^{2}-\lambda_{2} c_{\alpha} s_{\alpha} c_{\beta}^{2}+\lambda_{3} c_{\alpha} s_{\alpha} \left(c_{\beta}^{2}-s_{\beta}^{2}\right), \\
g_{H^{\pm}H^{\mp}G^{0}G^{0}}^{}
&=
-\frac{1}{2} \lambda_{1} s_{\beta}^{2}-\frac{1}{2} \lambda_{3} c_{\beta}^{2}, \\
g_{G^{+}G^{-}hh}^{}
&=
-\frac{1}{2} \lambda_{1} c_{\alpha}^{2} c_{\beta}^{2}-\frac{1}{2} \lambda_{2} s_{\alpha}^{2} s_{\beta}^{2}-\frac{1}{2} \lambda_{3} \left(s_{\alpha}^{2} c_{\beta}^{2}+c_{\alpha}^{2} s_{\beta}^{2}\right), \\
g_{G^{+}G^{-}Hh}^{}
&=
\lambda_{1} c_{\alpha} s_{\alpha} c_{\beta}^{2}-\lambda_{2} c_{\alpha} s_{\alpha} s_{\beta}^{2}-\lambda_{3} c_{\alpha} s_{\alpha} \left(c_{\beta}^{2}-s_{\beta}^{2}\right), \\
g_{G^{+}G^{-}HH}^{}
&=
-\frac{1}{2} \lambda_{1} s_{\alpha}^{2} c_{\beta}^{2}-\frac{1}{2} \lambda_{2} c_{\alpha}^{2} s_{\beta}^{2}-\frac{1}{2} \lambda_{3} \left(c_{\alpha}^{2} c_{\beta}^{2}+s_{\alpha}^{2} s_{\beta}^{2}\right), \\
g_{G^{+}G^{-}G^{0}G^{0}}^{}
&=
-\frac{1}{2} \lambda_{1} c_{\beta}^{2}-\frac{1}{2} \lambda_{3} s_{\beta}^{2}, \\
g_{H^{\pm}G^{\mp}hh}^{}
&=
\frac{1}{2} \lambda_{1} c_{\alpha}^{2} c_{\beta} s_{\beta}
-\frac{1}{2} \lambda_{2} s_{\alpha}^{2} c_{\beta} s_{\beta}
-\frac{1}{2} \lambda_{3} c_{\beta} s_{\beta} \left(c_{\alpha}^{2}-s_{\alpha}^{2}\right), \\
g_{H^{\pm}G^{\mp}HH}^{}
&=
\frac{1}{2} \lambda_{1} s_{\alpha}^{2} c_{\beta} s_{\beta}
-\frac{1}{2} \lambda_{2} c_{\alpha}^{2} c_{\beta} s_{\beta}
+\frac{1}{2} \lambda_{3} c_{\beta} s_{\beta} \left(c_{\alpha}^{2}-s_{\alpha}^{2}\right), \\
g_{H^{\pm}G^{\mp}Hh}^{}
&=
-\lambda_{1} c_{\alpha} s_{\alpha} c_{\beta} s_{\beta}-\lambda_{2} c_{\alpha} s_{\alpha} c_{\beta} s_{\beta}+2 \lambda_{3} c_{\alpha} s_{\alpha} c_{\beta} s_{\beta}, \\
g_{H^{\pm}G^{\mp}G^{0}G^{0}}^{}
&=
\frac{1}{2} \lambda_{1} c_{\beta} s_{\beta}-\frac{1}{2} \lambda_{3} c_{\beta} s_{\beta}, \\
g_{G^{0}G^{0}hh}^{}
&=
-\frac{1}{4} \lambda_{1} c_{\alpha}^{2}-\frac{1}{4} \lambda_{3} s_{\alpha}^{2}, \\
g_{G^{0}G^{0}Hh}^{}
&=
\frac{1}{2} \lambda_{1} c_{\alpha} s_{\alpha}-\frac{1}{2} \lambda_{3} c_{\alpha} s_{\alpha}, \\
g_{G^{0}G^{0}HH}^{}
&=
-\frac{1}{4} \lambda_{1} s_{\alpha}^{2}-\frac{1}{4} \lambda_{3} c_{\alpha}^{2}, \\
g_{G^{0}G^{0}G^{0}G^{0}}^{}
&=
-\frac{\lambda_{1}}{8}, \\
g_{hhhh}^{}
&=
-\frac{1}{8} \lambda_{1} c_{\alpha}^{4}-\frac{1}{8} \lambda_{2} s_{\alpha}^{4}-\frac{1}{4} \lambda_{3} c_{\alpha}^{2} s_{\alpha}^{2}, \\
g_{Hhhh}^{}
&=
\frac{1}{2} \lambda_{1} c_{\alpha}^{3} s_{\alpha}-\frac{1}{2} \lambda_{2} c_{\alpha} s_{\alpha}^{3}-\frac{1}{2} \lambda_{3} c_{\alpha} s_{\alpha} \left(c_{\alpha}^{2}-s_{\alpha}^{2}\right), \\
g_{HHhh}^{}
&=
-\frac{3}{4} \lambda_{1} c_{\alpha}^{2} s_{\alpha}^{2}-\frac{3}{4} \lambda_{2} c_{\alpha}^{2} s_{\alpha}^{2}-\frac{1}{4} \lambda_{3} \left(c_{\alpha}^{4}-4c_{\alpha}^{2}s_{\alpha}^{2}+s_{\alpha}^{4}\right), \\
g_{HHHh}^{}
&=
\frac{1}{2} \lambda_{1} c_{\alpha} s_{\alpha}^{3}-\frac{1}{2} \lambda_{2} c_{\alpha}^{3} s_{\alpha}+\frac{1}{2} \lambda_{3} c_{\alpha} s_{\alpha} \left(c_{\alpha}^{2}-s_{\alpha}^{2}\right), \\
g_{HHHH}^{}
&=
-\frac{1}{8} \lambda_{1} s_{\alpha}^{4}-\frac{1}{8} \lambda_{2} c_{\alpha}^{4}-\frac{1}{4} \lambda_{3} c_{\alpha}^{2} s_{\alpha}^{2}.
\end{align}

\section{Renormalized vertex}
\label{sec: renormalized vertex}
In this section, we give the analytic expressions for the renormalized three-point vertices of the SM-like Higgs boson.

\subsection{\texorpdfstring{$hf\bar{f}$}{} vertex}
\label{sec: hff_vetex}

The renormalized $hf\bar{f}$ vertex can be decomposed as \cite{Kanemura:2015mxa}
\begin{align}
\widehat{\Gamma}_{hf\bar{f}}(p_{1}, p_{2}, q)&=
\widehat{\Gamma}_{hf\bar{f}}^{S}+\gamma_{5}\widehat{\Gamma}_{hf\bar{f}}^{P}
+\slashed{p}_{1}\widehat{\Gamma}_{hf\bar{f}}^{V_{1}}
+\slashed{p}_{2}\widehat{\Gamma}_{hf\bar{f}}^{V_{2}} \notag\\
&\quad
+\slashed{p}_{1}\gamma_{5}\widehat{\Gamma}_{hf\bar{f}}^{A_{1}}
+\slashed{p}_{2}\gamma_{5}\widehat{\Gamma}_{hf\bar{f}}^{A_{2}}
+\slashed{p}_{1}\slashed{p}_{2}\widehat{\Gamma}_{hf\bar{f}}^{T}
+\slashed{p}_{1}\slashed{p}_{2}\gamma_{5}\widehat{\Gamma}_{hf\bar{f}}^{PT}, \label{eq:hff-form}
\end{align}
where $p_{1}^{\mu} (p_{2}^{\mu})$ denotes the incoming four-momentum of the fermion (anti-fermion), and $q^{\mu}$ is the outgoing four-momentum of the SM-like Higgs boson.
The renormalized form factors are composed of the tree-level and the one-loop part as
\begin{align}
\widehat{\Gamma}^X_{hf\bar{f}}={\Gamma}^{X,\mathrm{tree}}_{hf\bar{f}}+{\Gamma}^{X,\mathrm{loop}}_{hf\bar{f}}\qc (X = S, P, V_{1}, V_{2}, A_{1}, A_{2}, T, PT).
\end{align}
The tree-level form factors are given by
\begin{align}
\Gamma^{S,{\rm tree}}_{hf\bar{f}}=-\frac{m_{f}}{v}\zeta_{h}\qc
\Gamma^{X,{\rm tree}}_{hf\bar{f}}=0\ \ \ (X\neq S). 
\end{align}
The one-loop-level parts are decomposed into the contributions from 1PI diagrams and the counterterms,
\begin{align}
\Gamma^{X,{\rm loop}}_{hf\bar{f}}=\Gamma^{X,{\rm 1PI}}_{hf\bar{f}}+\delta \Gamma^{X}_{hf\bar{f}}.
\end{align}
The analytic expressions for the 1PI contributions are given in Appendix~\ref{app: hff_1PI}.
The counterterms are given by
\begin{align}
\delta \Gamma^{S}_{hf\bar{f}}
&=
\Gamma^{S, \mathrm{tree}}_{hf\bar{f}}\qty(
\frac{\delta m_{f}}{m_{f}}-\frac{\delta v}{v}
+\frac{\delta \zeta_{h}}{\zeta_{h}}
+\delta Z_{V}^{f}+\frac{\delta Z_{hh}}{2})
+\Gamma^{S,{\rm tree}}_{Hf\bar{f}}\frac{\delta Z_{Hh}}{2}
\notag \\
&=
\Gamma^{S, \mathrm{tree}}_{hf\bar{f}}\qty(
\frac{\delta m_{f}}{m_{f}}-\frac{\delta v}{v}+\tan{\beta}\delta\beta
+\delta Z_{V}^{f}+\frac{\delta Z_{hh}}{2})
+\Gamma^{S, \mathrm{tree}}_{Hf\bar{f}}\qty(\delta\alpha+\frac{\delta Z_{Hh}}{2}), \\
\delta \Gamma^{X}_{hf\bar{f}} &= 0\qquad (X\neq S),
\end{align}
where $\Gamma^{S, \mathrm{tree}}_{Hf\bar{f}}=-m_{f}\zeta_{H}/v$, and we have used 
\begin{align}
\frac{\delta\zeta_{h}}{\zeta_{h}}
=
\frac{\zeta_{H}}{\zeta_{h}}\delta\alpha
+\tan{\beta}\delta\beta.
\end{align}

\subsection{\texorpdfstring{$hVV$}{} vertex}
\label{sec: hVV_vetex}

The renormalized $hVV$ vertex ($V=Z, W^{\pm}$) is defined by
\begin{align}
\widehat{\Gamma}_{hVV}^{\mu\nu}(p_{1}, p_{2}, q) &= \qty[
g^{\mu\nu}\widehat{\Gamma}_{hVV}^1
+\frac{p_1^\nu p_2^\mu}{m_V^2}\widehat{\Gamma}_{hVV}^2
+i\epsilon^{\mu\nu\rho\sigma}\frac{p_{1\rho} p_{2\sigma}}{m_V^2}\widehat{\Gamma}_{hVV}^3
](p_{1}^{2}, p_{2}^{2}, q^{2}),
\end{align}
where $p_{1}^{\mu}$ and $p_{2}^{\mu}$ are the incoming four-momenta of the weak bosons, and $q^{\mu}$ is the outgoing four-momentum of the Higgs boson.
The renormalized form factors are composed of the tree-level and one-loop-level parts as
\begin{align}
\widehat{\Gamma}^{i}_{hVV}={\Gamma}^{i,{\rm tree}}_{hVV}+{\Gamma}^{i,{\rm loop}}_{hVV}\qc (i=1,2,3).
\end{align}
The tree-level form factors are given by
\begin{align}
\Gamma^{1, \mathrm{tree}}_{hVV}= c_{V}g_{hVV}\qc
\Gamma^{2, \mathrm{tree}}_{hVV}= \Gamma^{3, \mathrm{tree}}_{hVV}=0,
\end{align}
where $c_{Z}=2$ and $c_{W}=1$.
The one-loop parts are decomposed into the contributions from 1PI diagrams, tadpole diagrams, and counterterms,
\begin{align}
{\Gamma}^{i, \mathrm{loop}}_{hVV} = \Gamma^{i, \mathrm{1PI}}_{hVV}+\Gamma^{i, \mathrm{Tad}}_{hVV}+\delta \Gamma^{i}_{hVV}
\end{align}
The analytic expressions for the 1PI diagrams are given in Appendix~\ref{app: hVV_1PI}.
The tadpole contributions are given by,
\begin{align}
\Gamma^{1, \mathrm{Tad}}_{hVV}\big|_{\mathrm{ST}} = 0\qc
\Gamma^{1, \mathrm{Tad}}_{hVV}\big|_{\mathrm{FJ}}
=
c_{V}\qty(2g_{hhVV}\Delta v_{h}+g_{hHVV}\Delta v_{H}),
\end{align}
and $\Gamma^{2, \mathrm{Tad}}_{hVV} = \Gamma^{3, \mathrm{Tad}}_{hVV} = 0$ in both schemes.
The counterterms are given by
\begin{align}
\delta \Gamma^{1}_{hZZ}
&=
\Gamma^{1, \mathrm{tree}}_{hZZ}\qty(
\frac{\delta m_{Z}^{2}}{m_{Z}^{2}}-\frac{\delta v}{v}
+\tan{\beta}\delta\beta
+\frac{\delta Z_{hh}}{2}+\delta Z_{ZZ})
+\Gamma^{1, \mathrm{tree}}_{HZZ}\qty(\delta\alpha+\frac{\delta Z_{Hh}}{2}), \\
\delta \Gamma^{1}_{hWW}
&=
\Gamma^{1, \mathrm{tree}}_{hWW}\qty(
\frac{\delta m_{W}^{2}}{m_{W}^{2}}-\frac{\delta v}{v}+\frac{\delta Z_{hh}}{2}+\delta Z_{WW})
+\Gamma^{1, \mathrm{tree}}_{HWW}\qty(\delta\alpha+\frac{\delta Z_{Hh}}{2})
\notag \\ &\quad
-gm_{W}\qty(
s_{\beta}c_{\alpha}+2c_{\beta}s_{\alpha}
)\delta\beta,
\end{align}
and $\delta \Gamma^{2}_{hVV} = \delta \Gamma^{3}_{hVV} = 0$.

\subsection{\texorpdfstring{$hhh$}{} vertex}
\label{sec: hhh_vetex}

The renormalized $hhh$ vertex is defined by
\begin{align}
\widehat{\Gamma}_{hhh}(p_{1}^{2}, p_{2}^{2}, q^{2})
&=
\Gamma^{\mathrm{tree}}_{hhh}+\Gamma^{\mathrm{loop}}_{hhh}(p_{1}^{2}, p_{2}^{2}, q^{2}),
\end{align}
where $p_{1}^{\mu}$ and $p_{2}^{\mu}$ are the incoming four-momenta and $q^{\mu}$ is the outgoing four-momentum of the Higgs boson.
The tree-level vertex is given by,
\begin{align}
\Gamma^{\mathrm{tree}}_{hhh} = 6g_{hhh}^{}.
\end{align}
The one-loop parts are decomposed into the contributions from 1PI diagrams, tadpole diagrams, and the counterterms,
\begin{align}
\Gamma^{\mathrm{loop}}_{hhh}
=
\Gamma^{\mathrm{1PI}}_{hhh}+\Gamma^{\mathrm{Tad}}_{hhh}+\delta \Gamma_{hhh}.
\end{align}
The analytic expression for the 1PI diagram is given in Appendix~\ref{app: hhh_1PI}.
The tadpole contribution is given by,
\begin{align}
\Gamma^{\mathrm{Tad}}_{hhh}\big|_{\mathrm{ST}} = 0\qc
\Gamma^{\mathrm{Tad}}_{hhh}\big|_{\mathrm{FJ}}
=
24g_{hhhh}^{} \Delta v_{h} + 6g_{Hhhh}^{} \Delta v_{H}.
\end{align}
The counterterm is given by
\begin{align}
\delta \Gamma_{hhh}
&=
6\qty(
\delta g_{hhh}^{}
+\frac{3}{2} g_{hhh}^{} \delta Z_{hh}
+\frac{1}{2} g_{Hhh}^{} \delta Z_{Hh}
),
\end{align}
where
\begin{align}
\delta g_{hhh}^{}
&=
-\left(
\frac{c_{\alpha}^{3}}{c_{\beta}}-\frac{2s_{\alpha}^{3}}{s_{\beta}}
\right)
\frac{\delta m_{h}^{2}}{2v}
-\frac{c_{\beta}s_{\alpha}^{2}}{s_{\beta}}\left(c_{\beta}s_{\alpha}+s_{\beta}c_{\alpha}\right)
\frac{\delta m_{H^{\pm}}^{2}}{v}
-g_{hhh}^{} \frac{\delta v}{v}
\notag \\ &\quad
+\left\{
-\left(\frac{c_{\alpha}^{3}s_{\beta}}{c_{\beta}^{2}}+\frac{2s_{\alpha}^{3}c_{\beta}}{s_{\beta}^{2}}\right)
\frac{m_{h}^{2}}{2v}
+s_{\alpha}^{2}\left[
s_{\beta}c_{\alpha}+\left(2+\frac{c_{\beta}^{2}}{s_{\beta}^{2}}\right)c_{\beta}s_{\alpha}
\right]
\frac{m_{H^{\pm}}^{2}}{v}
\right\} \delta\beta
\notag \\ &\quad
+\left[
g_{Hhh}^{}+\left(\frac{c_{\alpha}^{2}s_{\alpha}}{c_{\beta}}+\frac{2s_{\alpha}^{2}c_{\alpha}}{s_{\beta}}\right)\frac{m_{h}^{2}-m_{H}^{2}}{2v}
\right]
\delta\alpha.
\end{align}

\subsection{Loop-induced \texorpdfstring{$h\mathcal{V}\mathcal{V'}$}{} vertex}
\label{sec: hVV'_vertex}

The loop-induced $h\mathcal{V}\mathcal{V'}$ vertex ($(\mathcal{V}, \mathcal{V'}) = (g,g), (\gamma, \gamma), (Z, \gamma)$) is defined by
\begin{align}
\widehat{\Gamma}_{h\mathcal{V}\mathcal{V'}}^{\mu\nu}(p_{1}, p_{2}, q)
&=
\qty[
g^{\mu\nu}\widehat{\Gamma}_{h\mathcal{V}\mathcal{V'}}^{1}
+p_{1}^{\nu}p_{2}^{\mu}\widehat{\Gamma}_{h\mathcal{V}\mathcal{V'}}^{2}
+i\epsilon^{\mu\nu\rho\sigma} p_{1\rho}p_{2\sigma}\widehat{\Gamma}_{h\mathcal{V}\mathcal{V'}}^{3}
](p_{1}^{2}, p_{2}^{2}, q^{2}),
\end{align}
where $p_{1}^{\mu}$ and $p_{2}^{\mu}$ are the incoming four-momenta of the gauge bosons, and $q^{\mu}$ is the outgoing four-momentum of the Higgs boson.
For an on-shell photon or gluon with four-momentum $p_2^\mu$, the Ward--Takahashi identity leads to
\begin{align}
\widehat{\Gamma}_{h\mathcal{V}\mathcal{V'}}^{1}(m_{\mathcal{V}}^{2}, 0, m_{h}^{2})
=
-p_{1}\cdot p_{2}\widehat{\Gamma}_{h\mathcal{V}\mathcal{V'}}^{2}(m_{\mathcal{V}}^{2}, 0, m_{h}^{2}).
\end{align}
In the absence of CP violation, we have $\widehat{\Gamma}_{h\mathcal{V}\mathcal{V'}}^{3}(m_{\mathcal{V}}^{2}, 0, m_{h}^{2}) = 0$.
The renormalized form factors are decomposed into the contributions from 1PI diagrams and the counterterms,
\begin{align}
\widehat{\Gamma}^{i}_{h\mathcal{V}\mathcal{V'}}
= \Gamma^{i, \mathrm{1PI}}_{h\mathcal{V}\mathcal{V'}}+\delta \Gamma^{i}_{h\mathcal{V}\mathcal{V'}}\quad (i=1,2,3).
\end{align}
The analytic expressions for the 1PI diagrams are listed in Appendix~\ref{app: hVV'_1PI}.
For $(\mathcal{V},\mathcal{V}') = (g,g), (\gamma,\gamma)$, the vertex counterterms vanish at the one-loop level, \ie, $\delta \Gamma^{i}_{hgg} = \delta \Gamma^{i}_{h\gamma\gamma} = 0$.
On the other hand, for $(\mathcal{V},\mathcal{V}') = (Z,\gamma)$, we obtain
\begin{align}
\delta \Gamma^{1}_{hZ\gamma}
=
g_{hZZ}\qty(\delta C_{Z\gamma}-\frac{\delta s_{W}^{2}}{c_{W}s_{W}})
=
\frac{2g_{hZZ}}{m_{Z}^{2}}\Re\Pi_{Z\gamma}^{\mathrm{1PI}}(0).
\end{align}

\section{Loop functions}
\label{app: 1PIs}

In this section, we give the analytic expressions for the 1PI diagram contributions to one-, two-, three- and four-point functions in terms of the Passarino--Veltman functions~\cite{Passarino:1978jh}.
The 1PI diagrams are calculated in the 't Hooft--Feynman gauge, \ie, $\xi_{W}=\xi_{Z}=1$.
The fermion-loop and boson-loop contributions to the 1PI diagrams with bosonic external lines are separately presented with the subscript $F$ and $B$, respectively. 

\subsection{One-point functions} \label{app: 1PFs}
The fermionic contributions to the 1PI tadpole diagrams for $h$ and $H$ are given by
\begin{align}
\Gamma_{h, F}^{\mathrm{1PI}} &= \zeta_{h}\Gamma_{h, F,\, \SM}^{\mathrm{1PI}}\qc
\Gamma_{H, F}^{\mathrm{1PI}} = \zeta_{H}\Gamma_{h, F,\, \SM}^{\mathrm{1PI}},
\end{align}
where
\begin{align}
16\pi^{2}\Gamma_{h, F,\, \SM}^{\mathrm{1PI}}
=
-\sum_{f} N_{c}^{f} \frac{4m_{f}^{2}}{v}A(f).    
\end{align}
The bosonic contributions are given by
\begin{align}
16\pi^{2}\Gamma_{h,B}^{\text{1PI}}
&=
gm_{W}(c_{\beta}c_{\alpha}-2s_{\beta}s_{\alpha})\qty[4A(W)-2m_{W}^{2}]
+\frac{g_{Z}m_{Z}}{2}c_{\alpha}\qty[4A(Z)-2m_{Z}^{2}]
\notag\\ &\quad
-gm_{W}(c_{\beta}c_{\alpha}-2s_{\beta}s_{\alpha})A(c^{\pm})
-\frac{g_{Z}m_{Z}}{2}c_{\alpha}A(c_{Z})
\notag\\ &\quad
-g_{G^{+}G^{-}h}^{} A(G^{\pm})
-g_{G^{0}G^{0}h}^{} A(G^{0})
-3g_{hhh}^{} A(h)
\notag\\ &\quad
-g_{H^{+}H^{-}h}^{} A(H^{\pm})
-g_{HHh}^{} A(H),\\
16\pi^{2}\Gamma_{H,B}^{\text{1PI}}
&=
-gm_{W}(c_{\beta}s_{\alpha}+2s_{\beta}c_{\alpha})\qty[4A(W)-2m_{W}^{2}]
-\frac{g_{Z}m_{Z}}{2}s_{\alpha}\qty[4A(Z)-2m_{Z}^{2}]
\notag\\ &\quad
+gm_{W}(c_{\beta}s_{\alpha}+2s_{\beta}c_{\alpha})A(c^{\pm})
+\frac{g_{Z}m_{Z}}{2}s_{\alpha}A(c_{Z})
\notag\\ &\quad
-g_{G^{+}G^{-}H}^{} A(G^{\pm})
-g_{G^{0}G^{0}H}^{} A(G^{0})
-g_{hhH}^{} A(h)
\notag\\ &\quad
-g_{H^{+}H^{-}H}^{} A(H^{\pm})
-3g_{HHH}^{} A(H).
\end{align}

\subsection{Two-point functions} \label{app: 2PFs}

\subsubsection{Scalar boson}

The fermionic contributions to the scalar boson two-point functions are given by
\begin{alignat}{2}
\Pi_{hh}^{\mathrm{1PI}}(p^{2})_{F}
&=
\zeta_{h}^{2}\Pi_{hh}^{\mathrm{1PI}}(p^{2})_{F,\, \SM}\qc&
\Pi_{HH}^{\mathrm{1PI}}(p^{2})_{F}
&=
\zeta_{H}^{2}\Pi_{hh}^{\mathrm{1PI}}(p^{2})_{F,\, \SM}, \\
\Pi_{Hh}^{\mathrm{1PI}}(p^{2})_{F}
&=
\zeta_{h}\zeta_{H}\Pi_{hh}^{\mathrm{1PI}}(p^{2})_{F,\, \SM},& &\\
\Pi_{H^{+}H^{-}}^{\mathrm{1PI}}(p^{2})_{F}
&=
\zeta_{H^{\pm}}^{2}\Pi_{G^{+}G^{-}}^{\mathrm{1PI}}(p^{2})_{F,\, \SM}\qc&
\Pi_{H^{+}G^{-}}^{\mathrm{1PI}}(p^{2})_{F}
&=
\zeta_{H^{\pm}}\Pi_{G^{+}G^{-}}^{\mathrm{1PI}}(p^{2})_{F,\, \SM},
\end{alignat}
where
\begin{align}
&16\pi^{2}\Pi_{hh}^{\rm 1PI}(p^{2})_{F,\, \SM}
=
-\sum_{f} N_{c}^{f}\frac{4m_{f}^{2}}{v^{2}}
\qty[A(f)+\qty(2m_{f}^{2}-\frac{p^{2}}{2})B_{0}(p^{2}; f, f)],\\
&16\pi^{2}{\Pi_{G^{+}G^{-}}^{{\rm 1PI}}(p^{2})}_{F,\, \SM}
\notag \\
&=
-2N_{c}
\bigg\{\frac{m_{u}^{2}+m_{d}^{2}}{v^{2}}\qty[A(u)+A(d)+(m_{u}^{2}+m_{d}^{2}-p^{2})B_{0}(p^{2}; u, d)]
-\frac{4m_{u}^{2}m_{d}^{2}}{v^{2}}B_{0}(p^{2}; u, d)\bigg\}
\notag \\ &\quad
-2\frac{m_{\ell}^{2}}{v^{2}}[A(\ell)+(m_{\ell}^{2}-p^{2})B_{0}(p^{2}; 0, \ell)].
\end{align}
The bosonic contributions to the scalar boson two-point functions are given by
\begin{align}
&16\pi^{2}\Pi_{hh}^{\text{1PI}}(p^{2})_{B}
\notag\\
&=
g^{2}m_{W}^{2}(c_{\beta}c_{\alpha}-2s_{\beta}s_{\alpha})^{2}
\qty[4B_{0}(p^2; W, W)-2]
+\frac{g^{2}}{2}(1+3s_{\alpha}^{2})\qty[4A(W)-2m_{W}^{2}]
\notag\\ &\quad
-\frac{g^{2}}{2}(c_{\beta}c_{\alpha}-2s_{\beta}s_{\alpha})^{2}
\qty[2A(W)-A(G^{\pm})+(2m_{G^{\pm}}^{2}-m_{W}^{2}+2p^{2})B_{0}(p^{2}; W, G^{\pm})]
\notag\\ &\quad
-\frac{g^{2}}{2}(s_{\beta}c_{\alpha}+2c_{\beta}s_{\alpha})^{2}
\qty[2A(W)-A(H^{\pm})+(2m_{H^{\pm}}^{2}-m_{W}^{2}+2p^{2})B_{0}(p^{2}; W, H^{\pm})]
\notag\\ &\quad
-\frac{g^{2}m_{W}^{2}}{2}(c_{\beta}c_{\alpha}-2s_{\beta}s_{\alpha})^{2}
B_{0}(p^{2}; c^{\pm}, c^{\pm})
\notag\\ &\quad
+\frac{g_{Z}^{2}m_{Z}^{2}}{2}c_{\alpha}^{2}
\qty[4B_{0}(p^2; Z, Z)-2]
+\frac{g_{Z}^{2}}{4}c_{\alpha}^{2}\qty[4A(Z)-2m_{Z}^{2}]
\notag\\ &\quad
-\frac{g_{Z}^{2}}{4}c_{\alpha}^{2}
\qty[2A(Z)-A(G^{0})+(2m_{G^{0}}^{2}-m_{Z}^{2}+2p^{2})B_{0}(p^{2}; Z, G^{0})]
\notag\\ &\quad
-\frac{g_{Z}^{2}m_{Z}^{2}}{4}c_{\alpha}^{2}
B_{0}(p^{2}; c_{Z}, c_{Z})
\notag\\ &\quad
-2g_{H^{+}H^{-}hh}^{} A(H^{\pm})
-2g_{G^{+}G^{-}hh}^{} A(G^{\pm})
\notag\\ &\quad
-2g_{G^{0}G^{0}hh}^{} A(G^{0})
-2g_{HHhh}^{} A(H)
-12g_{hhhh}^{} A(h)
\notag\\ &\quad
+g_{H^{+}H^{-}h}^{2}B_{0}(p^{2}; H^{\pm}, H^{\pm})
+g_{G^{+}G^{-}h}^{2}B_{0}(p^{2}; G^{\pm}, G^{\pm})
+2g_{H^{+}G^{-}h}^{2}B_{0}(p^{2}; H^{\pm}, G^{\pm})
\notag\\ &\quad
+2g_{G^{0}G^{0}h}^{2}B_{0}(p^2; G^{0}, G^{0})
\notag\\ &\quad
+2g_{HHh}^{2}B_{0}(p^{2}; H, H)
+18g_{hhh}^{2}B_{0}(p^{2}; h, h)
+4g_{Hhh}^{2}B_{0}(p^{2}; h, H),
\\
&16\pi^{2}\Pi_{Hh}^{\rm 1PI}(p^{2})_{B}
\notag \\
&=
-g^{2}m_{W}^{2}(c_{\beta}c_{\alpha}-2s_{\beta}s_{\alpha})(c_{\beta}s_{\alpha}+2s_{\beta}c_{\alpha})
\qty[4B_{0}(p^2; W, W)-2]
+\frac{3g^{2}}{2}s_{\alpha}c_{\alpha}\qty[4A(W)-2m_{W}^{2}]
\notag\\ &\quad
+\frac{g^{2}}{2}(c_{\beta}c_{\alpha}-2s_{\beta}s_{\alpha})(c_{\beta}s_{\alpha}+2s_{\beta}c_{\alpha})
\qty[2A(W)-A(G^{\pm})+(2m_{G^{\pm}}^{2}-m_{W}^{2}+2p^{2})B_{0}(p^{2}; W, G^{\pm})]
\notag\\ &\quad
+\frac{g^{2}}{2}(s_{\beta}c_{\alpha}+2c_{\beta}s_{\alpha})(s_{\beta}s_{\alpha}-2c_{\beta}c_{\alpha})
\qty[2A(W)-A(H^{\pm})+(2m_{H^{\pm}}^{2}-m_{W}^{2}+2p^{2})B_{0}(p^{2}; W, H^{\pm})]
\notag\\ &\quad
+\frac{g^{2}m_{W}^{2}}{2}(c_{\beta}c_{\alpha}-2s_{\beta}s_{\alpha})(c_{\beta}s_{\alpha}+2s_{\beta}c_{\alpha})
B_{0}(p^{2}; c^{\pm}, c^{\pm})
\notag\\ &\quad
-\frac{g_{Z}^{2}m_{Z}^{2}}{2}c_{\alpha}s_{\alpha}
\qty[4B_{0}(p^2; Z, Z)-2]
-\frac{g_{Z}^{2}}{4}s_{\alpha}c_{\alpha}\qty[4A(Z)-2m_{Z}^{2}]
\notag\\ &\quad
+\frac{g_{Z}^{2}}{4}c_{\alpha}s_{\alpha}
\qty[2A(Z)-A(G^{0})+(2m_{G^{0}}^{2}-m_{Z}^{2}+2p^{2})B_{0}(p^{2}; Z, G^{0})]
\notag\\ &\quad
+\frac{g_{Z}^{2}m_{Z}^{2}}{4}c_{\alpha}s_{\alpha}
B_{0}(p^{2}; c_{Z}, c_{Z})
\notag\\ &\quad
-g_{H^{+}H^{-}Hh}^{} A(H^{\pm})
-g_{G^{+}G^{-}Hh}^{} A(G^{\pm})
\notag\\ &\quad
-g_{G^{0}G^{0}Hh}^{} A(G^{0})
-3g_{HHHh}^{} A(H)
-3g_{Hhhh}^{} A(h)
\notag\\ &\quad
+g_{H^{+}H^{-}h}^{} g_{H^{+}H^{-}H}^{} B_{0}(p^{2}; H^{\pm}, H^{\pm})
+g_{G^{+}G^{-}h}^{} g_{G^{+}G^{-}H}^{} B_{0}(p^{2}; G^{\pm}, G^{\pm})
\notag\\ &\quad
+2g_{H^{+}G^{-}h}^{} g_{H^{+}G^{-}H}^{} B_{0}(p^{2}; H^{\pm}, G^{\pm})
\notag\\ &\quad
+2g_{G^{0}G^{0}h}^{} g_{G^{0}G^{0}H}^{} B_{0}(p^{2}; G^{0}, G^{0})
\notag\\ &\quad
+6g_{HHh}^{} g_{HHH}^{} B_{0}(p^{2}; H, H)
+6g_{hhh}^{} g_{Hhh}^{} B_{0}(p^{2}; h, h)
+4g_{Hhh}^{} g_{HHh}^{} B_{0}(p^{2}; H, h),
\\
&16\pi^{2}{\Pi_{H^{+}H^{-}}^{{\rm 1PI}}(p^{2})}_{B}
\notag \\
&=
-\frac{g^{2}}{4}\bigg\{
(c_{\alpha}s_{\beta}+2s_{\alpha}c_{\beta})^{2}
[2A(W)-A(h)+(2p^{2}+2m_{h}^{2}-m_{W}^{2})B_{0}(p^{2}; W, h)]
\notag \\ &\quad
+(s_{\alpha}s_{\beta}-2c_{\alpha}c_{\beta})^{2}
[2A(W)-A(H)+(2p^{2}+2m_{H}^{2}-m_{W}^{2})B_{0}(p^{2}; W, H)]
\notag \\ &\quad
+s_{\beta}^{2}
\left[2A(W)-A(G^{0})+(2p^{2}+2m_{G^{0}}^{2}-m_{W}^{2})B_{0}\right](p^{2}; W, G^{0})
\bigg\}
\notag \\ &\quad
-\frac{g_{Z}^{2}}{4}\bigg\{
c_{\beta}^{2}s_{\beta}^{2}
[2A(Z)-A(G^{\pm})+(2p^{2}+2m_{G^{\pm}}^{2}-m_{Z}^{2})B_{0}(p^{2}; Z, G^{\pm})]
\notag \\ &\quad
+(c_{2W}+c_{\beta}^{2})^{2}
[2A(Z)-A(H^{\pm})+(2p^{2}+2m_{H^{\pm}}^{2}-m_{Z}^{2})B_{0}(p^{2}; Z, H^{\pm})]
\bigg\}
\notag \\ &\quad
-e^{2}[2A(\gamma)-A(H^{\pm})+(2p^{2}+2m_{H^{\pm}}^{2}-m_{\gamma}^{2})B_{0}(p^{2}; \gamma, H^{\pm})]
\notag \\ &\quad
+\frac{g^{2}}{2}(1+c_{\beta}^{2})
\left[4A(W)-2m_{W}^{2}\right]
+\frac{g_{Z}^{2}}{4}(4c_{W}^{4}c_{\beta}^{2}+c_{2W}^{2}s_{\beta}^{2})
\left[4A(Z)-2m_{Z}^{2}\right]
\notag \\ &\quad
+e^{2}\left[4A(\gamma)-2m_{\gamma}^{2}\right]
\notag \\ &\quad
+g_{Z}^{2}m_{W}^{2}s_{\beta}^{2}c_{\beta}^{2}
\left[4B_{0}(p^{2}; W, Z)-2-\frac{1}{2}B_{0}(p^{2}; c^{\pm}, c_{Z})\right]
\notag \\ &\quad
+g_{H^{+}G^{-}h}^{2}B_{0}(p^{2}; G^{\pm}, h)
+g_{H^{+}H^{-}h}^{2}B_{0}(p^{2}; H^{\pm}, h)
\notag \\ &\quad
+g_{H^{+}G^{-}H}^{2}B_{0}(p^{2}; G^{\pm}, H)
+g_{H^{+}H^{-}H}^{2}B_{0}(p^{2}; H^{\pm}, H)
\notag \\ &\quad
+\abs{g_{G^{+}H^{-}G^{0}}}^{2}B_{0}(p^{2}; G^{\pm}, G^{0})
\notag \\ &\quad
-g_{H^{+}H^{-}hh}^{} A(h)
-g_{H^{+}H^{-}G^{0}G^{0}}{} A(G^{0})
-g_{H^{+}H^{-}G^{+}G^{-}}{} A(G^{\pm})
\notag \\ &\quad
-g_{H^{+}H^{-}HH}{} A(H)
-4g_{H^{+}H^{-}H^{+}H^{-}}{} A(H^{\pm}), \\
&16\pi^{2}{\Pi_{H^{+}G^{-}}^{{\rm 1PI}}(p^{2})}_{B}
\notag \\
&=
\frac{g^{2}}{4}\bigg\{
\left(c_{\alpha}c_{\beta}-2s_{\alpha}s_{\beta}\right)
\left(c_{\alpha}s_{\beta}+2s_{\alpha}c_{\beta}\right)
\left[2A(W)-A(h)+(2p^{2}+2m_{h}^{2}-m_{W}^{2})B_{0}\right](p^{2}; W, h)
\notag \\ &\quad
+\left(s_{\alpha}c_{\beta}+2c_{\alpha}s_{\beta}\right)
\left(s_{\alpha}s_{\beta}-2c_{\alpha}c_{\beta}\right)
\left[2A(W)-A(H)+(2p^{2}+2m_{H}^{2}-m_{W}^{2})B_{0}\right](p^{2}; W, H)
\notag \\ &\quad
+c_{\beta}s_{\beta}
\left[2A(W)-A(G^{0})+(2p^{2}+2m_{G^{0}}^{2}-m_{W}^{2})B_{0}\right](p^{2}; W, G^{0})
\bigg\}
\notag \\ &\quad
-\frac{g_{Z}^{2}}{4}c_{\beta}s_{\beta}\bigg\{
(c_{2W}+s_{\beta}^{2})
\left[2A(Z)-A(G^{\pm})+(2p^{2}+2m_{G^{\pm}}^{2}-m_{Z}^{2})B_{0}\right](p^{2}; Z, G^{\pm})
\notag \\ &\quad
+(c_{2W}+c_{\beta}^{2})
\left[2A(Z)-A(H^{\pm})+(2p^{2}+2m_{H^{\pm}}^{2}-m_{Z}^{2})B_{0}\right](p^{2}; Z, H^{\pm})
\bigg\}
\notag \\ &\quad
+\frac{g^{2}}{2}s_{\beta}c_{\beta}
\left[4A(W)-2m_{W}^{2}\right]
+\frac{g_{Z}^{2}}{4}(4c_{W}^{4}-c_{2W}^{2})s_{\beta}c_{\beta}
\left[4A(Z)-2m_{Z}^{2}\right]
\notag \\ &\quad
-g_{Z}^{2}m_{W}^{2}s_{\beta}c_{\beta}(s_{W}^{2}-s_{\beta}^{2})
\left[4B_{0}(p^{2}; W, Z)-2-\frac{1}{2}B_{0}(p^{2}; c^{\pm}, c_{Z})\right]
\notag \\ &\quad
+g_{G^{+}G^{-}h}{} g_{G^{+}H^{-}h}{} B_{0}(p^{2}; G^{\pm}, h)
+g_{G^{+}H^{-}h}{} g_{H^{+}H^{-}h}{} B_{0}(p^{2}; H^{\pm}, h)
\notag \\ &\quad
+g_{G^{+}G^{-}H}{} g_{G^{+}H^{-}H}{} B_{0}(p^{2}; G^{\pm}, H)
+g_{G^{+}H^{-}H}{} g_{H^{+}H^{-}H}{} B_{0}(p^{2}; H^{\pm}, H)
\notag \\ &\quad
-g_{H^{-}G^{+}hh}{} A(h)
-g_{H^{-}G^{+}G^{0}G^{0}}{} A(G^{0})
-2g_{G^{+}G^{-}G^{+}H^{-}}{} A(G^{\pm})
\notag \\ &\quad
-g_{H^{-}G^{+}HH}{} A(H)
-2g_{H^{+}H^{-}H^{-}G^{+}}{} A(H^{\pm}).
\end{align}

\subsubsection{Gauge boson}
The fermionic contributions to the gauge boson two-point functions are
the same as those in the SM.
These are given by
\begin{align}
16\pi^{2}\Pi_{\gamma\gamma}^{\rm 1PI}(p^{2})_{F}
&=
e^{2}\sum_{f} 8N_{c}^{f}Q_{f}^{2}p^{2}B_{3}(p^{2}; f, f),\\
16\pi^{2}\Pi_{Z\gamma}^{\rm 1PI}(p^{2})_{F}
&=
eg_{Z}\sum_{f} N_{c}^{f} \qty[2p^{2}\qty(2I_{f}Q_{f}-4s_{W}^{2}Q_{f}^{2})B_{3}](p^{2}; f, f), \\
16\pi^{2}\Pi_{ZZ}^{\rm 1PI}(p^{2})_{F}
&=
g_{Z}^{2}\sum_{f} N_{c}^{f} \qty[2p^{2}\qty(4s_{W}^{4}Q_{f}^{2}-4s_{W}^{2}Q_{f}I_{f}+2I_{f}^{2})B_{3}-2I_{f}^{2}m_{f}^{2}B_{0}](p^{2}; f, f), \\
16\pi^{2}\Pi_{WW}^{\rm 1PI}(p^{2})_{F}
&=
g^{2}\sum_{f, f'} N_{c}^{f} \qty[2p^{2}B_{3}-B_{4}](p^{2}; f, f').
\end{align}
The bosonic contributions to the gauge boson two-point functions are given by
\begin{align}
&16\pi^{2}\Pi_{\gamma\gamma}^{\rm 1PI}(p^{2})_{B}
\notag \\
&=
e^{2}\bigg[
B_{5}(p^{2}; H^{\pm}, H^{\pm})
+B_{5}(p^{2}; G^{\pm}, G^{\pm})
+2m_{W}^{2}B_{0}(p^{2}; W, G^{\pm})
\bigg]
\notag\\ &\quad
-e^{2}\bigg[
\qty[16B_{22}+p^{2}(2B_{21}+2B_{1}+5B_{0})](p^{2}; W, W)
-6A(W)
-(2m_{W}^{2}-p^{2})
\bigg],\\
&16\pi^{2}\Pi_{Z\gamma}^{\rm 1PI}(p^{2})_{B}
\notag \\
&=
eg_{Z}\bigg[
\frac{1}{2}(c_{2W}+c_{\beta}^{2})B_{5}(p^{2}; H^{\pm}, H^{\pm})
\notag \\ &\qquad
+\frac{1}{2}(c_{2W}+s_{\beta}^{2})B_{5}(p^{2}; G^{\pm}, G^{\pm})
-2m_{W}^{2}(s_{W}^{2}-s_{\beta}^{2})B_{0}(p^{2}; G^{\pm}, W)
\bigg]
\notag \\ &\quad
-eg_{Z}c_{W}^{2}\bigg[
\qty[16B_{22}+p^{2}(2B_{21}+2B_{1}+5B_{0})](p^{2}; W, W)
-6A(W)
-(2m_{W}^{2}-p^{2})
\bigg],\\
&16\pi^{2}\Pi_{ZZ}^{\rm 1PI}(p^{2})_{B}
\notag \\
&=
\frac{g_{Z}^{2}}{4}\bigg[
(c_{2W}+c_{\beta}^{2})^{2}B_{5}(p^{2}; H^{\pm}, H^{\pm})
\notag \\ &\qquad
+(c_{2W}+s_{\beta}^{2})^{2}B_{5}(p^{2}; G^{\pm}, G^{\pm})
+2s_{\beta}^{2}c_{\beta}^{2}B_{5}(p^{2}; H^{\pm}, G^{\pm})
\notag \\ &\qquad
+c_{\alpha}^{2}B_{5}(p^{2}; h, G^{0})
+s_{\alpha}^{2}B_{5}(p^{2}; H, G^{0})
\bigg]
\notag\\ &\quad
+g_{Z}^{2}\bigg[
m_{Z}^{2}\qty[
c_{\alpha}^{2}B_{0}(p^{2}; h, Z)
+s_{\alpha}^{2}B_{0}(p^{2}; H, Z)]
\notag \\ &\qquad
+m_{W}^{2}\qty[
2c_{\beta}^{2}s_{\beta}^{2}B_{0}(p^{2}; H^{\pm}, W)
+2(s_{W}^{2}-s_{\beta}^{2})^{2}B_{0}(p^{2}; G^{\pm}, W)]
\bigg]
\notag \\ &\quad
-g_{Z}^{2}c_{W}^{4}\bigg[
\qty[16B_{22}+p^{2}(2B_{21}+2B_{1}+5B_{0})](p^{2}; W, W)
-6A(W)
-(2m_{W}^{2}-p^{2})
\bigg], \\
&16\pi^{2}\Pi_{WW}^{\rm 1PI}(p^{2})_{B}
\notag \\
&=
\frac{g^{2}}{4}\bigg[
\left(c_{\alpha}c_{\beta}-2s_{\alpha}s_{\beta}\right)^{2} B_{5}(p^{2}; h, G^{\pm})
+\left(c_{\alpha}s_{\beta}+2s_{\alpha}c_{\beta}\right)^{2} B_{5}(p^{2}; h, H^{\pm})
\notag\\ &\qquad
+\left(s_{\alpha}c_{\beta}+2c_{\alpha}s_{\beta}\right)^{2} B_{5}(p^{2}; H, G^{\pm})
+\left(s_{\alpha}s_{\beta}-2c_{\alpha}c_{\beta}\right)^{2} B_{5}(p^{2}; H, H^{\pm})
\notag\\ &\qquad
+c_{\beta}^{2} B_{5}(p^{2}; G^{\pm}, G^{0})
+s_{\beta}^{2} B_{5}(p^{2}; H^{\pm}, G^{0})
\bigg]
\notag\\ &\quad
+g^{2}m_{W}^{2}\bigg[
(c_{\beta}c_{\alpha}-2s_{\beta}s_{\alpha})^{2} B_{0}(p^{2}; h, W)
+(c_{\beta}s_{\alpha}+2s_{\beta}c_{\alpha})^{2} B_{0}(p^{2}; H, W)\bigg]
\notag\\ &\quad
+g_{Z}^{2}m_{W}^{2}\bigg[
(s_{W}^{2}-s_{\beta}^{2})^{2} B_{0}(p^{2}; G^{\pm}, Z)
+c_{\beta}^{2}s_{\beta}^{2} B_{0}(p^{2}; H^{\pm}, Z)
\bigg]
\notag \\ &\quad
+e^{2}m_{W}^{2} B_{0}(p^{2}; G^{\pm}, \gamma)
\notag\\ &\quad
+g^{2}\bigg[
-s_{W}^{2}\qty[16B_{22}+p^{2}(2B_{21}+2B_{1}+5B_{0})](p^{2}; W, \gamma)
\notag\\ &\qquad
-c_{W}^{2}\qty[16B_{22}+p^{2}(2B_{21}+2B_{1}+5B_{0})](p^{2}; W, Z)
+3\qty[c_{W}^{2}A(Z)+A(W)]
\notag\\ &\qquad
+m_{W}^{2}+c_{W}^{2}m_{Z}^{2}-p^{2}
\bigg].
\end{align}

\subsubsection{Fermion}
The fermion two-point functions are given by
\begin{align}
&16\pi^{2}\Pi_{ff, S}^{\rm 1PI}(p^{2})
\notag \\
&=
-2e^{2}Q_{f}^{2}(2B_{0}-1)(p^{2}; f, \gamma)
-2g_{Z}^{2}(v_{f}^{2}-a_{f}^{2})(2B_{0}-1)(p^{2}; f, Z)
\notag\\ &\quad
+\frac{m_{f}^{2}}{v^{2}}
\qty[\zeta_{h}^{2}B_{0}(p^{2}; f, h)+\zeta_{H}^{2}B_{0}(p^{2}; f, H)-\zeta_{G^{0}}^{2}B_{0}(p^{2}; f, G^{0})]
\notag\\ &\quad
-2\frac{m_{f'}^{2}}{v^{2}}\qty[\zeta_{G^{\pm}}^{2}B_{0}(p^{2}; f', G^{\pm})+\zeta_{H^{\pm}}^{2}B_{0}(p^{2}; f', H^{\pm})], \\
&16\pi^{2}\Pi_{ff, V}^{\rm 1PI}(p^{2})
\notag \\
&=
-e^{2}Q_{f}^{2}(2B_{1}+1)(p^{2}; f, \gamma)
-g_{Z}^{2}(v_{f}^{2}+a_{f}^{2})(2B_{1}+1)(p^{2}; f, Z)
-\frac{g^{2}}{4}(2B_{1}+1)(p^{2}; f', W)
\notag\\ &\quad
-\frac{m_{f}^{2}}{v^{2}}
\qty[\zeta_{h}^{2}B_{1}(p^{2}; f, h)+\zeta_{H}^{2}B_{1}(p^{2}; f, H)
+\zeta_{G^{0}}^{2}B_{1}(p^{2}; f, G^{0})]
\notag\\ &\quad
-\frac{m_{f}^{2}+m_{f'}^{2}}{v^{2}}
\qty[\zeta_{G^{\pm}}^{2}B_{1}(p^{2}; f', G^{\pm})+\zeta_{H^{\pm}}^{2}B_{1}(p^{2}; f', H^{\pm})], \\
&16\pi^{2}\Pi_{ff, A}^{\rm 1PI}(p^{2})
\notag \\
&=
-2g_{Z}^{2}v_{f}a_{f}(2B_{1}+1)(p^{2}; f, Z)
-\frac{g^{2}}{4}(2B_{1}+1)(p^{2}; f',W)
\notag\\ &\quad
+\frac{m_{f}^{2}-m_{f'}^{2}}{v^{2}}
\qty[\zeta_{G^{\pm}}^{2}B_{1}(p^{2}; f', G^{\pm})+\zeta_{H^{\pm}}^{2}B_{1}(p^{2}; f', H^{\pm})].
\end{align}

\subsection{Three-point functions} \label{app: 3PFs}
\subsubsection{\texorpdfstring{$hhh$}{} vertex}
\label{app: hhh_1PI}
The fermionic contributions to the 1PI $hhh$ vertices are obtained as
\begin{align}
\Gamma_{hhh}^{\mathrm{1PI}}(p_{1}^{2}, p_{2}^{2}, q^{2})_{F}
=
\zeta_{h}^{3}
\Gamma_{hhh}^{\mathrm{1PI}}(p_{1}^{2}, p_{2}^{2}, q^{2})_{F,\, \SM},    
\end{align}
where
\begin{align}
16\pi^{2}\Gamma_{hhh}^{\mathrm{1PI}}(p_{1}^{2}, p_{2}^{2}, q^{2})_{F,\, \SM}
&=
-\sum_{f} N_{c}^{f} \frac{8m_{f}^{4}}{v^{3}}
\Bigg\{B_{0}(p_{1}^{2}, f, f)+B_{0}(p_{2}^{2}, f, f)+B_{0}(q^{2}, f, f)
\notag\\ &\quad
+\qty(4m_{f}^{2}-\frac{q^{2}+p_{1}^{2}+p_{2}^{2}}{2})C_{0}(f, f, f)
\Bigg\}.
\end{align}
The bosonic contributions obtained as
\begin{align}
&16\pi^{2}\Gamma_{hhh}^{\rm 1PI}(p_{1}^{2}, p_{2}^{2}, q^{2})_{B}
\notag \\
&=
2g^{3}m_{W}(c_{\beta}c_{\alpha}-2s_{\beta}s_{\alpha})(1+3s_{\alpha}^{2})
\qty[B_{0}(p_{1}^{2}, W, W)+B_{0}(p_{2}^{2}, W, W)+B_{0}(q^{2}, W, W)-\frac{3}{2}]
\notag\\ &\quad
+8g^{3}m_{W}^{3}(c_{\beta}c_{\alpha}-2s_{\beta}s_{\alpha})^{3}C_{0}(W, W, W)
\notag\\ &\quad
-\frac{g^{3}}{2}m_{W}(c_{\beta}c_{\alpha}-2s_{\beta}s_{\alpha})^{3}
\qty[C_{SSS}^{SVV}(G^{\pm},W,W)+C_{SSS}^{VSV}(W, G^{\pm}, W)+C_{SSS}^{VVS}(W, W, G^{\pm})]
\notag\\ &\quad
-\frac{g^{3}}{2}m_{W}(c_{\beta}c_{\alpha}-2s_{\beta}s_{\alpha})
(s_{\beta}c_{\alpha}+2c_{\beta}s_{\alpha})^{2}
\notag\\ &\quad \times
\qty[C_{SSS}^{SVV}(H^{\pm},W,W)+C_{SSS}^{VSV}(W, H^{\pm}, W)+C_{SSS}^{VVS}(W, W, H^{\pm})]
\notag\\ &\quad
+\frac{g^{2}}{2}g_{G^{+}G^{-}h}^{}
(c_{\beta}c_{\alpha}-2s_{\beta}s_{\alpha})^{2}
\qty[C_{SSS}^{VSS}(W, G^{\pm}, G^{\pm})+C_{SSS}^{SVS}(G^{\pm}, W, G^{\pm})+C_{SSS}^{SSV}(G^{\pm}, G^{\pm}, W)]
\notag\\ &\quad
+\frac{g^{2}}{2}g_{H^{+}H^{-}h}^{}
(s_{\beta}c_{\alpha}+2c_{\beta}s_{\alpha})^{2}
\qty[C_{SSS}^{VSS}(W, H^{\pm}, H^{\pm})+C_{SSS}^{SVS}(H^{\pm}, W, H^{\pm})+C_{SSS}^{SSV}(H^{\pm}, H^{\pm}, W)]
\notag\\ &\quad
-\frac{g^{2}}{2}g_{H^{+}G^{-}h}^{}
(c_{\beta}c_{\alpha}-2s_{\beta}s_{\alpha})
(s_{\beta}c_{\alpha}+2c_{\beta}s_{\alpha})
\notag\\ &\quad \times
\big[C_{SSS}^{VSS}(W, G^{\pm}, H^{\pm})+C_{SSS}^{SVS}(G^{\pm}, W, H^{\pm})+C_{SSS}^{SSV}(G^{\pm}, H^{\pm}, W)
\notag\\ &\qquad
+C_{SSS}^{VSS}(W, H^{\pm}, G^{\pm})+C_{SSS}^{SVS}(H^{\pm}, W, G^{\pm})+C_{SSS}^{SSV}(H^{\pm}, G^{\pm}, W)\big]
\notag\\ &\quad
-\frac{g^{3}}{2}m_{W}^{3}(c_{\beta}c_{\alpha}-2s_{\beta}s_{\alpha})^{3}
C_{0}(c^{\pm}, c^{\pm}, c^{\pm})
\notag\\ &\quad
+g_{Z}^{3}m_{Z}c_{\alpha}^{3}
\qty[B_{0}(p_{1}^{2}, Z, Z)+B_{0}(p_{2}^{2}, Z, Z)+B_{0}(q^{2}, Z, Z)-\frac{3}{2}]
\notag\\ &\quad
+4g_{Z}^{3}m_{Z}^{3}c_{\alpha}^{3}
C_{0}(Z, Z, Z)
\notag\\ &\quad
-\frac{g_{Z}^{3}}{4}m_{Z}c_{\alpha}^{3}
\qty[C_{SSS}^{SVV}(G^{0}, Z, Z)+C_{SSS}^{VSV}(Z, G^{0}, Z)+C_{SSS}^{VVS}(Z, Z, G^{0})]
\notag\\ &\quad
+\frac{g_{Z}^{2}}{2}g_{G^{0}G^{0}h}^{}
c_{\alpha}^{2}
\qty[C_{SSS}^{VSS}(Z, G^{0}, G^{0})+C_{SSS}^{SVS}(G^{0}, Z, G^{0})+C_{SSS}^{SSV}(G^{0}, G^{0}, Z)]
\notag\\ &\quad
-\frac{g_{Z}^{3}}{4}m_{Z}^{3}c_{\alpha}^{3}
C_{0}(c_{Z}, c_{Z}, c_{Z})
\notag\\ &\quad
+2g_{H^{+}H^{-}h}^{} g_{H^{+}H^{-}hh}^{}
\qty[B_{0}(p_{1}^{2}, H^{\pm}, H^{\pm})+B_{0}(p_{2}^{2}, H^{\pm}, H^{\pm})+B_{0}(q^{2}, H^{\pm}, H^{\pm})]
\notag\\ &\quad
+2g_{hG^{+}G^{-}}^{} g_{G^{+}G^{-}hh}^{}
\qty[B_{0}(p_{1}^{2}, G^{\pm}, G^{\pm})+B_{0}(p_{2}^{2}, G^{\pm}, G^{\pm})+B_{0}(q^{2}, G^{\pm}, G^{\pm})]
\notag\\ &\quad
+4g_{H^{+}G^{-}h}^{} g_{H^{+}G^{-}hh}^{}
\qty[B_{0}(p_{1}^{2}, H^{\pm}, G^{\pm})+B_{0}(p_{2}^{2}, H^{\pm}, G^{\pm})+B_{0}(q^{2}, H^{\pm}, G^{\pm})]
\notag\\ &\quad
+4g_{G^{0}G^{0}h}^{} g_{G^{0}G^{0}hh}^{}
\qty[B_{0}(p_{1}^{2}, G^{0}, G^{0})+B_{0}(p_{2}^{2}, G^{0}, G^{0})+B_{0}(q^{2}, G^{0}, G^{0})]
\notag\\ &\quad
+4g_{HHh}^{} g_{HHhh}^{}
\qty[B_{0}(p_{1}^{2}, H, H)+B_{0}(p_{2}^{2}, H, H)+B_{0}(q^{2}, H, H)]
\notag\\ &\quad
+12g_{Hhh}^{} g_{Hhhh}^{}
\qty[B_{0}(p_{1}^{2}, h, H)+B_{0}(p_{2}^{2}, h, H)+B_{0}(q^{2}, h, H)]
\notag\\ &\quad
+72g_{hhh}^{} g_{hhhh}^{}
\qty[B_{0}(p_{1}^{2}, h, h)+B_{0}(p_{2}^{2}, h, h)+B_{0}(q^{2}, h, h)]
\notag\\ &\quad
-2g_{H^{+}H^{-}h}^{3} C_{0}(H^{\pm}, H^{\pm}, H^{\pm})
-2g_{G^{+}G^{-}h}^{3} C_0({G^\pm},{G^\pm},{G^\pm})
\notag\\ &\quad
-8g_{G^{0}G^{0}h}^{3} C_{0}(G^{0}, G^{0}, G^{0})
-8g_{HHh}^{3} C_{0}(H, H, H)
-216g_{hhh}^{3} C_{0}(h, h, h)
\notag\\ &\quad
-2g_{H^{+}H^{-}h}^{} g_{H^{+}G^{-}h}^{2}
\qty[C_{0}(G^{\pm}, H^{\pm}, H^{\pm})+C_{0}(H^{\pm}, G^{\pm}, H^{\pm})+C_{0}(H^{\pm}, H^{\pm}, G^{\pm})]
\notag\\ &\quad
-2g_{G^{+}G^{-}h}^{} g_{H^{+}G^{-}h}^{2}
\qty[C_{0}(H^{\pm}, G^{\pm}, G^{\pm})+C_{0}(G^{\pm}, H^{\pm}, G^{\pm})+C_{0}(G^{\pm}, G^{\pm}, H^{\pm})]
\notag\\ &\quad
-8g_{HHh}^{} g_{Hhh}^{2}
\qty[C_{0}(h, H, H)+C_{0}(H, H, h)+C_{0}(H, h, H)]
\notag\\ &\quad
-24g_{hhh}^{} g_{Hhh}^{2}
\qty[C_{0}(h, h, H)+C_{0}(H, h, h)+C_{0}(h, H, h)],
\end{align}
where
\begin{align}
C_{SSS}^{SVV}(X,Y,Z)
&=
-\frac{1}{2}B_{0}(p_{1}^{2};X,Y)+2B_{0}(p_{2}^{2};Y,Z)-\frac{1}{2}B_{0}(q^{2};X,Z)
\notag \\ &\quad
+\qty(p_{1}^{2}-\frac{p_{2}^{2}}{2}+q^{2}+2m_{X}^{2}-\frac{m_{Y}^{2}}{2}-\frac{m_{Z}^{2}}{2})C_{0}(X, Y, Z), \\
C_{SSS}^{VSV}(X,Y,Z)
&=
-\frac{1}{2}B_{0}(p_{1}^{2};X,Y)-\frac{1}{2}B_{0}(p_{2}^{2};Y,Z)+2B_{0}(q^{2};X,Z)
\notag \\ &\quad
+\qty(p_{1}^{2}+p_{2}^{2}-\frac{q^{2}}{2}-\frac{m_{X}^{2}}{2}+2m_{Y}^{2}-\frac{m_{Z}^{2}}{2})C_{0}(X, Y, Z), \\
C_{SSS}^{VVS}(X,Y,Z)
&=
2B_{0}(p_{1}^{2};X,Y)-\frac{1}{2}B_{0}(p_{2}^{2};Y,Z)-\frac{1}{2}B_{0}(q^{2};X,Z)
\notag \\ &\quad
+\qty(-\frac{p_{1}^{2}}{2}+p_{2}^{2}+q^{2}-\frac{m_{X}^{2}}{2}-\frac{m_{Y}^{2}}{2}+2m_{Z}^{2})C_{0}(X, Y, Z), \\
C_{SSS}^{VSS}(X,Y,Z)
&=
B_{0}(p_{1}^{2};X,Y)-B_{0}(p_{2}^{2};Y,Z)+B_{0}(q^{2};X,Z)
\notag \\ &\quad
+\qty(p_{1}^{2}-2p_{2}^{2}+q^{2}-m_{X}^{2}+m_{Y}^{2}+m_{Z}^{2})C_{0}(X, Y, Z), \\
C_{SSS}^{SVS}(X, Y, Z)
&=
B_{0}(p_{1}^{2};X,Y)+B_{0}(p_{2}^{2};Y,Z)-B_{0}(q^{2};X,Z)
\notag \\ &\quad
+\qty(p_{1}^{2}+p_{2}^{2}-2q^{2}+m_{X}^{2}-m_{Y}^{2}+m_{Z}^{2})C_{0}(X, Y, Z), \\
C_{SSS}^{SSV}(X, Y, Z)
&=
-B_{0}(p_{1}^{2};X,Y)+B_{0}(p_{2}^{2};Y,Z)+B_{0}(q^{2};X,Z)
\notag \\ &\quad
+\qty(-2p_{1}^{2}+p_{2}^{2}+q^{2}+m_{X}^{2}+m_{Y}^{2}-m_{Z}^{2})C_{0}(X, Y, Z).
\end{align}

\subsubsection{\texorpdfstring{$hVV$}{} vertex}
\label{app: hVV_1PI}

The fermionic contributions to the 1PI $hZZ$ vertices are obtained as
\begin{align}
\Gamma^{i, \mathrm{1PI}}_{hZZ}(p_{1}^{2}, p_{2}^{2}, q^{2})_{F}
=
\zeta_{h}\Gamma^{i, \mathrm{1PI}}_{hZZ}(p_{1}^{2}, p_{2}^{2}, q^{2})_{F,\, \SM}\qquad (i=1,2,3),
\end{align}
where
\begin{align}
&16\pi^{2}\Gamma^{1,{\rm 1PI}}_{hZZ}(p_{1}^{2}, p_{2}^{2}, q^{2})_{F,\, \SM}
\notag \\
&=
\sum_{f}N_{c}^{f}\frac{4g_{Z}^{2}m_{f}^{2}}{v}\Bigg\{
(v_{f}^{2}+a_{f}^{2})\Big[B_{0}(p_{1}^{2}, f, f)+B_{0}(p_{2}^{2}, f, f)+2B_{0}(q^{2}, f, f)
\notag\\ &\quad
+(4m_{f}^{2}-p_{1}^{2}-p_{2}^{2})C_{0}(f, f, f)-8C_{24}(f, f, f)\Big]
\notag\\ &\quad
-(v_{f}^{2}-a_{f}^{2})\Big[B_{0}(p_{1}^{2}, f, f)+B_{0}(p_{2}^{2}, f, f)+(4m_{f}^{2}-q^{2})C_{0}(f, f, f)\Big]
\Bigg\}, \\
&\qty(m_{Z}^{2})^{-1}16\pi^{2}\Gamma^{2, \mathrm{1PI}}_{hZZ}(p_{1}^{2}, p_{2}^{2}, q^{2})_{F,\, \SM}
\notag \\
&=
-\sum_{f}N_{c}^{f}\frac{8g_{Z}^{2}m_{f}^{2}}{v}
\Big[(v_f^2+a_f^2)
(4C_{23}+3C_{12}+C_{11}+C_0)
\notag \\ &\quad
+(v_{f}^{2}-a_{f}^{2})(C_{12}-C_{11})\Big](f, f, f), \\
&\qty(m_{Z}^{2})^{-1}16\pi^{2}\Gamma^{3, \mathrm{1PI}}_{hZZ}(p_{1}^{2}, p_{2}^{2}, q^{2})_{F,\, \SM}
\notag \\
&=
\sum_{f}N_{c}^{f}\frac{16g_{Z}^{2}m_{f}^{2}}{v}v_{f}a_{f}(C_{11}+C_{12}+C_{0})(f, f, f).
\end{align}
The bosonic contributions are obtained as
\begin{align}
&16\pi^{2}\Gamma^{1, {\rm 1PI}}_{hZZ}(p_{1}^{2}, p_{2}^{2}, q^{2})_{B}
\notag \\
&=
2g^{3}m_{W}c_{W}^{2}(c_{\alpha}c_{\beta}-2s_{\alpha}s_{\beta})
C_{hVV1}^{VVV}(W, W, W)
\notag \\ &\quad
+g^{3}m_{W}(c_{\alpha}c_{\beta}-2s_{\alpha}s_{\beta})(s_{W}^{2}-s_{\beta}^{2})
\qty[C_{hVV1}^{SVV}(G^{\pm}, W, W)+C_{hVV1}^{VVS}(W, W, G^{\pm})]
\notag \\ &\quad
+g^{3}m_{W}(c_{\alpha}s_{\beta}+2s_{\alpha}c_{\beta})s_{\beta}c_{\beta}
\qty[C_{hVV1}^{SVV}(H^{\pm}, W, W)+C_{hVV1}^{VVS}(W, W, H^{\pm})]
\notag \\ &\quad
-gg_{Z}^{2}m_{W}(c_{\alpha}c_{\beta}-2s_{\alpha}s_{\beta})(c_{2W}+s_{\beta}^{2})(s_{W}^{2}-s_{\beta}^{2})
\qty[C_{24}(W, G^{\pm}, G^{\pm})+C_{24}(G^{\pm}, G^{\pm}, W)]
\notag \\ &\quad
-gg_{Z}^{2}m_{W}(c_{\alpha}s_{\beta}+2s_{\alpha}c_{\beta})(c_{2W}+c_{\beta}^{2})s_{\beta}c_{\beta}
\qty[C_{24}(W, H^{\pm}, H^{\pm})+C_{24}(H^{\pm}, H^{\pm}, W)]
\notag \\ &\quad
+gg_{Z}^{2}m_{W}(c_{\alpha}c_{\beta}-2s_{\alpha}s_{\beta})s_{\beta}^{2}c_{\beta}^{2}
\qty[C_{24}(W, H^{\pm}, G^{\pm})+C_{24}(G^{\pm}, H^{\pm}, W)]
\notag \\ &\quad
+gg_{Z}^{2}m_{W}(c_{\alpha}s_{\beta}+2s_{\alpha}c_{\beta})s_{\beta}c_{\beta}(s_{W}^{2}-s_{\beta}^{2})
\qty[C_{24}(W, G^{\pm}, H^{\pm})+C_{24}(H^{\pm}, G^{\pm}, W)]
\notag \\ &\quad
-2gg_{Z}^{2}m_{W}^{3}(c_{\alpha}c_{\beta}-2s_{\alpha}s_{\beta})(s_{W}^{2}-s_{\beta}^{2})^{2}C_{0}(W, G^{\pm}, W)
\notag \\ &\quad
-2gg_{Z}^{2}m_{W}^{3}(c_{\alpha}c_{\beta}-2s_{\alpha}s_{\beta})s_{\beta}^{2}c_{\beta}^{2}C_{0}(W, H^{\pm}, W)
\notag \\ &\quad
+2g_{G^{+}G^{-}h}g_{Z}^{2}m_{W}^{2}(s_{W}^{2}-s_{\beta}^{2})^{2}
C_{0}(G^{\pm}, W, G^{\pm})
+2g_{H^{+}H^{-}h}g_{Z}^{2}m_{W}^{2}s_{\beta}^{2}c_{\beta}^{2}
C_{0}(H^{\pm}, W, H^{\pm})
\notag \\ &\quad
-2g_{H^{+}G^{-}h}g_{Z}^{2}m_{W}^{2}s_{\beta}c_{\beta}(s_{W}^{2}-s_{\beta}^{2})\qty[C_{0}(H^{\pm}, W, G^{\pm})+C_{0}(G^{\pm}, W, H^{\pm})]
\notag \\ &\quad
-2g^{3}m_{W}c_{W}^{2}(c_{\alpha}c_{\beta}-2s_{\alpha}s_{\beta})
C_{24}(c^{\pm}, c^{\pm}, c^{\pm})
\notag \\ &\quad
-g^{3}c_{W}^{2}m_{W}(c_{\alpha}c_{\beta}-2s_{\alpha}s_{\beta})
\qty[6B_{0}(q^{2}; W, W)-4]
\notag \\ &\quad
-gg_{Z}^{2}m_{W}\qty(c_{\alpha}c_{\beta}s_{W}^{2}+2c_{W}^{2}s_{\alpha}s_{\beta})(s_{W}^{2}-s_{\beta}^{2})\qty[B_{0}(p_{1}^{2}; G^{\pm}, W)+B_{0}(p_{2}^{2}; W, G^{\pm})]
\notag \\ &\quad
-gg_{Z}^{2}m_{W}\qty(c_{\alpha}s_{\beta}s_{W}^{2}-2c_{W}^{2}s_{\alpha}c_{\beta})s_{\beta}c_{\beta}\qty[B_{0}(p_{1}^{2}; H^{\pm}, W)+B_{0}(p_{2}^{2}; W, H^{\pm})]
\notag \\ &\quad
+\frac{g_{Z}^{3}}{2}m_{Z}c_{\alpha}^{3}
\qty[C_{24}(Z, h, G^{0})+C_{24}(G^{0}, h, Z)]
+\frac{g_{Z}^{3}}{2}m_{Z}c_{\alpha}s_{\alpha}^{2}
\qty[C_{24}(Z, H, G^{0})+C_{24}(G^{0}, H, Z)]
\notag\\ &\quad
-g_{Z}^{3}m_{Z}^{3}c_{\alpha}^{3}
C_{0}(Z, h, Z)
-g_{Z}^{3}m_{Z}^{3}c_{\alpha}s_{\alpha}^{2}
C_{0}(Z, H, Z)
\notag\\ &\quad
+6g_{hhh}g_{Z}^{2}m_{Z}^{2}c_{\alpha}^{2}
C_{0}(h, Z, h)
+2g_{HHh}g_{Z}^{2}m_{Z}^{2}s_{\alpha}^{2}
C_{0}(H, Z, H)
\notag\\ &\quad
-2g_{Hhh}g_{Z}^{2}m_{Z}^{2}c_{\alpha}s_{\alpha}
\qty[C_{0}(h, Z, H)+C_{0}(H, Z, h)]
\notag\\ &\quad
-\frac{g_{Z}^{3}}{2}m_{Z}c_{\alpha}^{3}
\qty[B_{0}(p_{1}^{2}; h, Z)+B_{0}(p_{2}^{2}; h, Z)]
-\frac{g_{Z}^{3}}{2}m_{Z}c_{\alpha}s_{\alpha}^{2}
\qty[B_{0}(p_{1}^{2}; H, Z)+B_{0}(p_{2}^{2}; H, Z)]
\notag\\ &\quad
-4g_{Z}^{2}\bigg[
\frac{1}{2}g_{H^{+}H^{-}h}(c_{2W}+c_{\beta}^{2})^{2} 
C_{24}(H^{\pm},H^{\pm},H^{\pm})
\notag\\ &\quad
+\frac{1}{2}g_{H^{+}H^{-}h}c_{\beta}^{2}s_{\beta}^{2}
C_{24}(H^{\pm},G^{\pm},H^{\pm})
+\frac{1}{2}g_{G^{+}G^{-}h}c_{\beta}^{2}s_{\beta}^{2}
C_{24}(G^{\pm},H^{\pm},G^{\pm})
\notag\\ &\quad
+\frac{1}{2}g_{H^{+}G^{-}h}c_{\beta}s_{\beta}(c_{2W}+c_{\beta}^{2})
\qty[C_{24}(H^{\pm},H^{\pm},G^{\pm})+C_{24}(G^{\pm},H^{\pm},H^{\pm})]
\notag\\ &\quad
+\frac{1}{2}g_{H^{+}G^{-}h}c_{\beta}s_{\beta}(c_{2W}+s_{\beta}^{2})
\qty[C_{24}(H^{\pm},G^{\pm},G^{\pm})+C_{24}(G^{\pm},G^{\pm},H^{\pm})]
\notag\\ &\quad
+\frac{1}{2}g_{G^{+}G^{-}h}(c_{2W}+s_{\beta}^{2})^{2} 
C_{24}(G^{\pm},G^{\pm},G^{\pm})
\notag\\ &\quad
+\frac{1}{2}g_{G^{0}G^{0}h}s_{\alpha}^{2}
C_{24}(G^{0},H,G^{0})
+\frac{1}{2}g_{G^{0}G^{0}h}c_{\alpha}^{2}
C_{24}(G^{0},h,G^{0})
\notag\\ &\quad
+\frac{1}{2}g_{HHh}s_{\alpha}^{2}
C_{24}(H,G^{0},H)
+\frac{3}{2}g_{hhh}c_{\alpha}^{2}
C_{24}(h,G^{0},h)
\notag\\ &\quad
-\frac{1}{2}g_{Hhh}s_{\alpha}c_{\alpha}
\qty[C_{24}(H,G^{0},h)+C_{24}(h,G^{0},H)]
\bigg]
\notag \\ &\quad
+g_{Z}^{2}\bigg[
\frac{1}{2}g_{G^{+}G^{-}h}
(4c_{W}^{4}s_{\beta}^{2}+c_{2W}^{2}c_{\beta}^{2})
B_{0}(q^{2}, G^{\pm}, G^{\pm})
\notag\\ &\quad
+\frac{1}{2}g_{H^{+}H^{-}h}
(4c_{W}^{4}c_{\beta}^{2}+c_{2W}^{2}s_{\beta}^{2})
B_{0}(q^{2}, H^{\pm}, H^{\pm})
\notag\\ &\quad
+g_{H^{+}G^{-}h}
(4c_{W}^{4}-c_{2W}^{2})s_{\beta}c_{\beta}
B_{0}(q^{2}, H^{\pm}, G^{\pm})
+\frac{1}{2}g_{G^{0}G^{0}h}
B_{0}(q^{2}, G^{0}, G^{0})
\notag\\ &\quad
+\frac{3}{2}g_{hhh}c_{\alpha}^{2}
B_{0}(q^{2},h,h)
+\frac{1}{2}g_{HHh}s_{\alpha}^{2}
B_{0}(q^{2},H,H)
-g_{Hhh}s_{\alpha}c_{\alpha}
B_{0}(q^{2},h,H)
\bigg], \\
&(m_{Z}^{2})^{-1}16\pi^{2}\Gamma_{hZZ}^{2, {\rm 1PI}}(p_{1}^{2},p_{2}^{2},q^{2})_{B}
\notag \\
&=
2g^{3}m_{W}c_{W}^{2}
(c_{\alpha}c_{\beta}-2s_{\alpha}s_{\beta})
C_{hVV2}^{VVV}(W, W, W)
\notag \\ &\quad
+g^{3}m_{W}
(c_{\alpha}c_{\beta}-2s_{\alpha}s_{\beta})
(s_{W}^{2}-s_{\beta}^{2})
\qty[C_{hVV2}^{SVV}(G^{\pm}, W, W)+C_{hVV2}^{VVS}(W, W, G^{\pm})]
\notag \\ &\quad
+g^{3}m_{W}
(c_{\alpha}s_{\beta}+2s_{\alpha}c_{\beta})
s_{\beta}c_{\beta}
\qty[C_{hVV2}^{SVV}(H^{\pm}, W, W)+C_{hVV2}^{VVS}(W, W, H^{\pm})]
\notag \\ &\quad
-gg_{Z}^{2}m_{W}
(c_{\alpha}c_{\beta}-2s_{\alpha}s_{\beta})
(c_{2W}+s_{\beta}^{2})
(s_{W}^{2}-s_{\beta}^{2})
\qty[C_{hVV2}^{VSS}(W, G^{\pm}, G^{\pm})+C_{hVV2}^{SSV}(G^{\pm}, G^{\pm}, W)]
\notag \\ &\quad
-gg_{Z}^{2}m_{W}
(c_{\alpha}s_{\beta}+2s_{\alpha}c_{\beta})
(c_{2W}+c_{\beta}^{2})
s_{\beta}c_{\beta}
\qty[C_{hVV2}^{VSS}(W, H^{\pm}, H^{\pm})+C_{hVV2}^{SSV}(H^{\pm}, H^{\pm}, W)]
\notag \\ &\quad
+gg_{Z}^{2}m_{W}
(c_{\alpha}c_{\beta}-2s_{\alpha}s_{\beta})
s_{\beta}^{2}c_{\beta}^{2}
\qty[C_{hVV2}^{VSS}(W, H^{\pm}, G^{\pm})+C_{hVV2}^{SSV}(G^{\pm}, H^{\pm}, W)]
\notag \\ &\quad
+gg_{Z}^{2}m_{W}
(c_{\alpha}s_{\beta}+2s_{\alpha}c_{\beta})
s_{\beta}c_{\beta}(s_{W}^{2}-s_{\beta}^{2})
\qty[C_{hVV2}^{VSS}(W, G^{\pm}, H^{\pm})+C_{hVV2}^{SSV}(H^{\pm}, G^{\pm}, W)]
\notag \\ &\quad
-2g^{3}m_{W}c_{W}^{2}
(c_{\alpha}c_{\beta}-2s_{\alpha}s_{\beta})
C_{1223}(c^{\pm}, c^{\pm}, c^{\pm})
\notag \\ &\quad
+\frac{g_{Z}^{3}}{2}m_{Z}
c_{\alpha}^{3}
\qty[C_{hVV2}^{VSS}(Z, h, G^{0})+C_{hVV2}^{SSV}(G^{0}, h, Z)]
\notag\\ &\quad
+\frac{g_{Z}^{3}}{2}m_{Z}
c_{\alpha}s_{\alpha}^{2}
\qty[C_{hVV2}^{VSS}(Z, H, G^{0})+C_{hVV2}^{SSV}(G^{0}, H, Z)]
\notag\\ &\quad
-4g_{Z}^{2}\bigg[
\frac{1}{2}g_{H^{+}H^{-}h}^{}
(c_{2W}+c_{\beta}^{2})^{2}
C_{1223}(H^{\pm},H^{\pm},H^{\pm})
\notag\\ &\quad
+\frac{1}{2}g_{H^{+}H^{-}h}^{}
c_{\beta}^{2}s_{\beta}^{2}
C_{1223}(H^{\pm},G^{\pm},H^{\pm})
+\frac{1}{2}g_{G^{+}G^{-}h}^{}
c_{\beta}^{2}s_{\beta}^{2}
C_{1223}(G^{\pm},H^{\pm},G^{\pm})
\notag\\ &\quad
+\frac{1}{2}g_{H^{+}G^{-}h}^{}
c_{\beta}s_{\beta}(c_{2W}+c_{\beta}^{2})
\qty[C_{1223}(H^{\pm},H^{\pm},G^{\pm})+C_{1223}(G^{\pm},H^{\pm},H^{\pm})]
\notag\\ &\quad
+\frac{1}{2}g_{H^{+}G^{-}h}^{}
c_{\beta}s_{\beta}(c_{2W}+s_{\beta}^{2})
\qty[C_{1223}(H^{\pm},G^{\pm},G^{\pm})+C_{1223}(G^{\pm},G^{\pm},H^{\pm})]
\notag\\ &\quad
+\frac{1}{2}g_{G^{+}G^{-}h}^{}
(c_{2W}+s_{\beta}^{2})^{2} 
C_{1223}(G^{\pm},G^{\pm},G^{\pm})
\notag\\ &\quad
+\frac{1}{2}g_{G^{0}G^{0}h}^{}
s_{\alpha}^{2}
C_{1223}(G^{0},H,G^{0})
+\frac{1}{2}g_{G^{0}G^{0}h}^{}
c_{\alpha}^{2}
C_{1223}(G^{0},h,G^{0})
\notag\\ &\quad
+\frac{1}{2}g_{HHh}^{}
s_{\alpha}^{2}
C_{1223}(H,G^{0},H)
+\frac{3}{2}g_{hhh}^{}
c_{\alpha}^{2}
C_{1223}(h,G^{0},h)
\notag\\ &\quad
-\frac{1}{2}g_{Hhh}^{}
s_{\alpha}
c_{\alpha}
\qty[C_{1223}(H,G^{0},h)+C_{1223}(h,G^{0},H)]
\bigg], \\
&16\pi^{2}\Gamma_{hZZ}^{3, {\rm 1PI}}(p_{1}^{2},p_{2}^{2},q^{2})_{B}=0,
\end{align}
where the loop functions are defined as~\cite{Kanemura:2015mxa}.
\begin{align}
C_{hVV1}^{VVV}(X,Y,Z)
&=
\Big[18C_{24}+p_{1}^{2}(2C_{21}+3C_{11}+C_{0})+p_{2}^{2}(2C_{22}+C_{12})
\notag \\ &\quad
+p_{1}\cdot p_{2}(4C_{23}+3C_{12}+C_{11}-4C_{0})\Big](X,Y,Z)-3, \\
C_{hVV2}^{VVV}(X,Y,Z)&= \left(10C_{23}+9C_{12}+C_{11}+5C_0\right)(X,Y,Z), \\
C_{hVV1}^{SVV}(X,Y,Z)
&=
\Big[3C_{24}+p_1^2(C_{21}-C_0)+p_2^2(C_{22}-2C_{12}+C_0)
\notag \\ &\qquad
+2p_1\cdot p_2 (C_{23}-C_{11})\Big](X,Y,Z)-\frac{1}{2}, \\
C_{hVV2}^{SVV}(X,Y,Z)&= \left(4C_{11}-3C_{12}-C_{23}\right)(X,Y,Z), \\
C_{hVV1}^{VVS}(X,Y,Z)
&=
\Big[3C_{24}+p_1^2(C_{21}+4C_{11}+4C_0)+p_2^2(C_{22}+2C_{12})
\notag \\ &\qquad
+2p_1\cdot p_2 (C_{23}+2C_{12}+C_{11}+2C_0)\Big](X,Y,Z)-\frac{1}{2}, \\
C_{hVV2}^{VVS}(X,Y,Z)&= \left(2C_{11}-5C_{12}-2C_0-C_{23}\right)(X,Y,Z), \\
C_{hVV2}^{VSS}(X,Y,Z)&= (C_{23}+C_{12}+2C_{11}+2C_0)(X,Y,Z), \\
C_{hVV2}^{SSV}(X,Y,Z)&= (C_{23}-C_{12})(X,Y,Z), \\
C_{1223}(X,Y,Z)&= (C_{12}+C_{23}) (X,Y,Z). 
\end{align}

The fermionic contributions to the 1PI $hWW$ vertices are obtained as
\begin{align}
\Gamma^{i, \mathrm{1PI}}_{hWW}(p_{1}^{2},p_{2}^{2},q^{2})_{F}
=
\zeta_{h}\Gamma^{i, \mathrm{1PI}}_{hWW}(p_{1}^{2},p_{2}^{2},q^{2})_{F,\, \SM}
\qquad (i=1,2,3),
\end{align}
where
\begin{align}
&16\pi^{2}\Gamma^{1, \mathrm{1PI}}_{hWW}(p_{1}^{2},p_{2}^{2},q^{2})_{F,\, \SM}
\notag \\
&=
\sum_{f, f'}N_{c}^{f}\frac{g^{2} m_{f}^{2}}{v}\Bigg[
\frac{1}{2}B_{0}(p_{1}^{2},f,f')+\frac{1}{2}B_{0}(p_{2}^{2},f,f')+B_{0}(q^{2},f,f)
-4C_{24}(p_{1}^{2},p_{2}^{2},q^{2},f,f',f)
\notag\\ &\quad
+\qty(m_{f}^{2}+m_{f'}^{2}-\frac{p_{1}^{2}}{2}-\frac{p_{2}^{2}}{2})C_{0}(f,f',f)
\Bigg]
+(m_{f} \leftrightarrow m_{f'}), \\
&\qty(m_{W}^{2})^{-1}16\pi^{2}\Gamma^{2, \mathrm{1PI}}_{hWW}(p_{1}^{2},p_{2}^{2},q^{2})_{F,\, \SM}
\notag \\
&=
-\sum_{f, f'}N_{c}^{f}\frac{g^{2} m_{f}^{2}}{v}
\qty(4C_{23}+3C_{12}+C_{11}+C_{0})(f,f',f)
+(m_{f}\leftrightarrow m_{f'}), \\
&\qty(m_{W}^{2})^{-1}16\pi^{2}\Gamma^{3,\mathrm{1PI}}_{hWW}(p_{1}^{2},p_{2}^{2},q^{2})_{F,\, \SM}
\notag \\
&=
-\sum_{f, f'}N_{c}^{f}\frac{g^{2} m_{f}^{2}}{v}\qty(C_{11}+C_{12}+C_{0})(f,f',f)
+(m_{f}\leftrightarrow m_{f'}), \\
&16\pi^{2}\Gamma^{1,\text{1PI}}_{hWW}(p_{1}^{2},p_{2}^{2},q^{2})_{B}
\notag \\
&=
g^{3}c_{W}m_{Z}c_{\alpha}
C_{hVV1}^{VVV}(Z,W,Z)
+g^{3}c_{W}^{2}m_{W}
(c_{\alpha}c_{\beta}-2s_{\alpha}s_{\beta})
C_{hVV1}^{VVV}(W,Z,W)
\notag\\ &\quad
+g^{3}s_{W}^{2}m_{W}
(c_{\alpha}c_{\beta}-2s_{\alpha}s_{\beta})
C_{hVV1}^{VVV}(W,\gamma,W)
\notag\\ &\quad
-\frac{1}{2}g^{3}m_{W}
(c_{\alpha}c_{\beta}-2s_{\alpha}s_{\beta})
(s_{W}^{2}-s_{\beta}^{2})
\qty[C_{hVV1}^{SVV}(G^{\pm},Z,W)+C_{hVV1}^{VVS}(W,Z,G^{\pm})]
\notag\\ &\quad
+\frac{1}{2}g^{3}s_{W}^{2}m_{W}
(c_{\alpha}c_{\beta}-2s_{\alpha}s_{\beta})
\qty[C_{hVV1}^{SVV}(G^{\pm},\gamma,W)+C_{hVV1}^{VVS}(W,\gamma,G^{\pm})]
\notag\\ &\quad
-\frac{1}{2}g^{3}m_{W}
(c_{\alpha}s_{\beta}+2s_{\alpha}c_{\beta})
s_{\beta}c_{\beta}
\qty[C_{hVV1}^{SVV}(H^{\pm},Z,W)+C_{hVV1}^{VVS}(W,Z,H^{\pm})]
\notag\\ &\quad
-g_{Z}^{3}m_{Z}m_{W}^{2}(s_{W}^{2}-s_{\beta}^{2})^{2}
c_{\alpha}
C_{0}(Z,G^{\pm},Z)
-g_{Z}^{3}m_{Z}m_{W}^{2}s_{\beta}^{2}c_{\beta}^{2}
c_{\alpha}
C_{0}(Z,H^{\pm},Z)
\notag\\ &\quad
-g^{3}m_{W}^{3}
(c_{\alpha}c_{\beta}-2s_{\alpha}s_{\beta})^{3}
C_{0}(W,h,W)
\notag\\ &\quad
-g^{3}m_{W}^{3}
(c_{\alpha}c_{\beta}-2s_{\alpha}s_{\beta})
(s_{\alpha}c_{\beta}+2c_{\alpha}s_{\beta})^{2}
C_{0}(W,H,W)
\notag\\ &\quad
+6g_{hhh}^{}
g^{2}m_{W}^{2}
(c_{\alpha}c_{\beta}-2s_{\alpha}s_{\beta})^{2}
C_{0}(h,W,h)
\notag\\ &\quad
+2g_{HHh}^{}
g^{2}m_{W}^{2}
(s_{\alpha}c_{\beta}+2c_{\alpha}s_{\beta})^{2}
C_{0}(H,W,H)
\notag\\ &\quad
-2g_{Hhh}^{}
g^{2}m_{W}^{2}
(c_{\alpha}c_{\beta}-2s_{\alpha}s_{\beta})
(s_{\alpha}c_{\beta}+2c_{\alpha}s_{\beta})
\qty[C_{0}(h,W,H)+C_{0}(H,W,h)]
\notag\\ &\quad
+g_{G^{+}G^{-}h}^{}
g_{Z}^{2}m_{W}^{2}
(s_{W}^{2}-s_{\beta}^{2})^{2}
C_{0}(G^{\pm},Z,G^{\pm})
+g_{H^{+}H^{-}h}^{}
g_{Z}^{2}m_{W}^{2}
s_{\beta}^{2}c_{\beta}^{2}
C_{0}(H^{\pm},Z,H^{\pm})
\notag\\ &\quad
+g_{G^{+}G^{-}h}^{}
g^{2}s_{W}^{2}m_{W}^{2}
C_{0}(G^{\pm},\gamma,G^{\pm})
\notag\\ &\quad
-g_{H^{+}G^{-}h}^{}
g_{Z}^{2}m_{W}^{2}
s_{\beta}c_{\beta}(s_{W}^{2}-s_{\beta}^{2})
\qty[C_{0}(H^{\pm},Z,G^{\pm})+C_{0}(G^{\pm},Z,H^{\pm})]
\notag\\ &\quad
+\frac{1}{2}g^{3}m_{W}
(c_{\alpha}c_{\beta}-2s_{\alpha}s_{\beta})^{3}
\qty[C_{24}(W,h,G^{\pm})+C_{24}(G^{\pm},h,W)]
\notag\\ &\quad
+\frac{1}{2}g^{3}m_{W}
(c_{\alpha}c_{\beta}-2s_{\alpha}s_{\beta})
(s_{\alpha}c_{\beta}+2c_{\alpha}s_{\beta})^{2}
\qty[C_{24}(W,H,G^{\pm})+C_{24}(G^{\pm},H,W)]
\notag\\ &\quad
+\frac{1}{2}g^{3}m_{W}
(c_{\alpha}c_{\beta}-2s_{\alpha}s_{\beta})
(c_{\alpha}s_{\beta}+2s_{\alpha}c_{\beta})^{2}
\qty[C_{24}(W,h,H^{\pm})+C_{24}(H^{\pm},h,W)]
\notag\\ &\quad
+\frac{1}{2}g^{3}m_{W}
(c_{\alpha}s_{\beta}+2s_{\alpha}c_{\beta})
(s_{\alpha}c_{\beta}+2c_{\alpha}s_{\beta})
(s_{\alpha}s_{\beta}-2c_{\alpha}c_{\beta})
\qty[C_{24}(W,H,H^{\pm})+C_{24}(H^{\pm},H,W)]
\notag\\ &\quad
+\frac{1}{2}gg_{Z}^{2}m_{W}
(s_{W}^{2}-s_{\beta}^{2})
c_{\beta}
c_{\alpha}
\qty[C_{24}(G^{0},G^{\pm},Z)+C_{24}(Z,G^{\pm},G^{0})]
\notag\\ &\quad
+\frac{1}{2}gg_{Z}^{2}m_{W}
s_{\beta}^{2}c_{\beta}
c_{\alpha}
\qty[C_{24}(G^{0},H^{\pm},Z)+C_{24}(Z,H^{\pm},G^{0})]
\notag\\ &\quad
-g^{3}c_{W}m_{Z}
c_{\alpha}
C_{24}(c_{Z},c^{\pm},c_{Z})
-g^{3}c_{W}^{2}m_{W}
(c_{\alpha}c_{\beta}-2s_{\alpha}s_{\beta})
C_{24}(c^{\pm},c_{Z},c^{\pm})
\notag\\ &\quad
-g^{3}s_{W}^{2}m_{W}
(c_{\alpha}c_{\beta}-2s_{\alpha}s_{\beta})
C_{24}(c^{\pm},c_{\gamma},c^{\pm})
\notag\\ &\quad
-g^{3}m_{W}
(c_{\alpha}c_{\beta}-2s_{\alpha}s_{\beta})
\qty[3B_{0}(q^{2},W,W)-2]
-g^{3}c_{W}m_{Z}
c_{\alpha}
\qty[3B_{0}(q^{2},Z,Z)-2]
\notag\\ &\quad
-\frac{1}{2}g^{3}(1+3s_{\alpha}^{2})m_{W}
(c_{\alpha}c_{\beta}-2s_{\alpha}s_{\beta})
\qty[B_{0}(p_{1}^{2}; W, h)+B_{0}(p_{2}^{2},W, h)]
\notag\\ &\quad
+\frac{3}{2}g^{3}s_{\alpha}c_{\alpha}m_{W}
(s_{\alpha}c_{\beta}+2c_{\alpha}s_{\beta})
\qty[B_{0}(p_{1}^{2}; W,H)+B_{0}(p_{2}^{2}; W,H)]
\notag\\ &\quad
-\frac{1}{2}gg_{Z}^{2}m_{W}
\qty[c_{\alpha}c_{\beta}s_{W}^{2}+2c_{W}^{2}s_{\alpha}s_{\beta}]
(s_{W}^{2}-s_{\beta}^{2})
\qty[B_{0}(p_{1}^{2}; Z,G^{\pm})+B_{0}(p_{2}^{2}; Z,G^{\pm})]
\notag\\ &\quad
-\frac{1}{2}gg_{Z}^{2}m_{W}
\qty[c_{\alpha}s_{\beta}s_{W}^{2}-2c_{W}^{2}s_{\alpha}c_{\beta}]
s_{\beta}c_{\beta}
\qty[B_{0}(p_{1}^{2}; Z,H^{\pm})+B_{0}(p_{2}^{2}; Z,H^{\pm})]
\notag\\ &\quad
-\frac{1}{2}g^{3}s_{W}^{2}m_{W}
(c_{\alpha}c_{\beta}-2s_{\alpha}s_{\beta})
\qty[B_{0}(p_{1}^{2}; \gamma,G^{\pm})+B_{0}(p_{2}^{2}; \gamma,G^{\pm})]
\notag\\ &\quad
-4g^{2}\bigg\{
\frac{1}{2}g_{G^{0}G^{0}h}^{}\qty[
c_{\beta}^{2}
C_{24}(G^{0},G^{\pm},G^{0})
+s_{\beta}^{2}
C_{24}(G^{0},H^{\pm},G^{0})]
\notag\\ &\quad
+\frac{3}{2}g_{hhh}^{}\qty[
\qty(c_{\beta}c_{\alpha}-2s_{\beta}s_{\alpha})^{2}
C_{24}(h,G^{\pm},h)
+\qty(s_{\beta}c_{\alpha}+2c_{\beta}s_{\alpha})^{2}
C_{24}(h,H^{\pm},h)]
\notag\\ &\quad
+\frac{1}{2}g_{HHh}^{}\qty[
\qty(c_{\beta}s_{\alpha}+2s_{\beta}c_{\alpha})^{2}
C_{24}(H,G^{\pm},H)
+\qty(s_{\beta}s_{\alpha}-2c_{\beta}c_{\alpha})^{2}
C_{24}(H,H^{\pm},H)]
\notag\\ &\quad
-\frac{1}{2}g_{Hhh}^{}
\qty(c_{\beta}c_{\alpha}-2s_{\beta}s_{\alpha})
\qty(c_{\beta}s_{\alpha}+2s_{\beta}c_{\alpha})
\qty[C_{24}(h,G^{\pm},H)+C_{24}(H,G^{\pm},h)]
\notag\\ &\quad
-\frac{1}{2}g_{Hhh}^{}
\qty(s_{\beta}c_{\alpha}+2c_{\beta}s_{\alpha})
\qty(s_{\beta}s_{\alpha}-2c_{\beta}c_{\alpha})
\qty[C_{24}(h,H^{\pm},H)+C_{24}(H,H^{\pm},h)]
\notag\\ &\quad
+\frac{1}{4}g_{H^{+}H^{-}h}^{}
s_{\beta}^{2}
C_{24}(H^{\pm},G^{0},H^{\pm})
\notag\\ &\quad
+\frac{1}{4}g_{H^{+}H^{-}h}^{}\qty[
\qty(s_{\beta}c_{\alpha}+2c_{\beta}s_{\alpha})^{2}
C_{24}(H^{\pm},h,H^{\pm})
+\qty(s_{\beta}s_{\alpha}-2c_{\beta}c_{\alpha})^{2}
C_{24}(H^{\pm},H,H^{\pm})]
\notag\\ &\quad
+\frac{1}{4}g_{G^{+}G^{-}h}^{}
c_{\beta}^{2}
C_{24}(G^{\pm},G^{0},G^{\pm})
\notag\\ &\quad
+\frac{1}{4}g_{G^{+}G^{-}h}^{}\qty[
\qty(c_{\beta}c_{\alpha}-2s_{\beta}s_{\alpha})^{2}
C_{24}(G^{\pm},h,G^{\pm})
+\qty(c_{\beta}s_{\alpha}+2s_{\beta}c_{\alpha})^{2}
C_{24}(G^{\pm},H,G^{\pm})]
\notag\\ &\quad
-\frac{1}{4}g_{H^{+}G^{-}h}^{}
s_{\beta}c_{\beta}
\qty[C_{24}(G^{\pm},G^{0},H^{\pm})+C_{24}(H^{\pm},G^{0},G^{\pm})]
\notag\\ &\quad
-\frac{1}{4}g_{H^{+}G^{-}h}^{}
\qty(s_{\beta}c_{\alpha}+2c_{\beta}s_{\alpha})
\qty(c_{\beta}c_{\alpha}-2s_{\beta}s_{\alpha})
\qty[C_{24}(G^{\pm},h,H^{\pm})+C_{24}(H^{\pm},h,G^{\pm})]
\notag\\ &\quad
-\frac{1}{4}g_{H^{+}G^{-}h}^{}
\qty(s_{\beta}s_{\alpha}-2c_{\beta}c_{\alpha})
\qty(c_{\beta}s_{\alpha}+2s_{\beta}c_{\alpha})
\qty[C_{24}(G^{\pm},H,H^{\pm})+C_{24}(H^{\pm},H,G^{\pm})]
\bigg\}
\notag\\ &\quad
+g^{2}\bigg[
\frac{1}{2} g_{G^{+}G^{-}h}^{} (1+s_{\beta}^{2})
B_{0}(q^{2};G^{\pm},G^{\pm})
+\frac{1}{2} g_{H^{+}H^{-}h}^{} (1+c_{\beta}^{2})
B_{0}(q^{2};H^{\pm},H^{\pm})
\notag\\ &\quad
+g_{H^{+}G^{-}h}^{} s_{\beta}c_{\beta}
B_{0}(q^{2};H^{\pm},G^{\pm})
+\frac{1}{2}g_{G^{0}G^{0}h}^{}
B_{0}(q^{2};G^{0},G^{0})
\notag\\ &\quad
+\frac{3}{2}g_{hhh}^{} (1+3s_{\alpha}^{2})
B_{0}(q^{2};h,h)
+\frac{1}{2}g_{HHh}^{} (1+3c_{\alpha}^{2})
B_{0}(q^{2};H,H)
\notag\\ &\quad
+3g_{Hhh}^{} s_{\alpha}c_{\alpha}
B_{0}(q^{2};H,h)\bigg], \\
&(m_{W}^{2})^{-1}16\pi^{2}\Gamma_{hWW}^{2, \mathrm{1PI}}(p_{1}^{2},p_{2}^{2},q^{2})_{B}
\notag \\
&=
g^{3}c_{W}m_{Z}
c_{\alpha}
C_{hVV2}^{VVV}(Z,W,Z)
+g^{3}c_{W}^{2}m_{W}
(c_{\alpha}c_{\beta}-2s_{\alpha}s_{\beta})
C_{hVV2}^{VVV}(W,Z,W)
\notag\\ &\quad
+g^{3}s_{W}^{2}m_{W}
(c_{\alpha}c_{\beta}-2s_{\alpha}s_{\beta})
C_{hVV2}^{VVV}(W,\gamma,W)
\notag\\ &\quad
-\frac{1}{2}g^{3}m_{W}
(c_{\alpha}c_{\beta}-2s_{\alpha}s_{\beta})
(s_{W}^{2}-s_{\beta}^{2})
\qty[C_{hVV2}^{SVV}(G^{\pm},Z,W)+C_{hVV2}^{VVS}(W,Z,G^{\pm})]
\notag\\ &\quad
+\frac{1}{2}g^{3}s_{W}^{2}m_{W}
(c_{\alpha}c_{\beta}-2s_{\alpha}s_{\beta})
\qty[C_{hVV2}^{SVV}(G^{\pm},\gamma,W)+C_{hVV2}^{VVS}(W,\gamma,G^{\pm})]
\notag\\ &\quad
-\frac{1}{2}g^{3}m_{W}
(c_{\alpha}s_{\beta}+2s_{\alpha}c_{\beta})
s_{\beta}c_{\beta}
\qty[C_{hVV2}^{SVV}(H^{\pm},Z,W)+C_{hVV2}^{VVS}(W,Z,H^{\pm})]
\notag\\ &\quad
+\frac{1}{2}g^{3}m_{W}
(c_{\alpha}c_{\beta}-2s_{\alpha}s_{\beta})^{3}
\qty[C_{hVV2}^{VSS}(W,h,G^{\pm})+C_{hVV2}^{SSV}(G^{\pm},h,W)]
\notag\\ &\quad
+\frac{1}{2}g^{3}m_{W}
(c_{\alpha}c_{\beta}-2s_{\alpha}s_{\beta})
(s_{\alpha}c_{\beta}+2c_{\alpha}s_{\beta})^{2}
\qty[C_{hVV2}^{VSS}(W,H,G^{\pm})+C_{hVV2}^{SSV}(G^{\pm},H,W)]
\notag\\ &\quad
+\frac{1}{2}g^{3}m_{W}
(c_{\alpha}c_{\beta}-2s_{\alpha}s_{\beta})
(c_{\alpha}s_{\beta}+2s_{\alpha}c_{\beta})^{2}
\qty[C_{hVV2}^{VSS}(W,h,H^{\pm})+C_{hVV2}^{SSV}(H^{\pm},h,W)]
\notag\\ &\quad
+\frac{1}{2}g^{3}m_{W}
(c_{\alpha}s_{\beta}+2s_{\alpha}c_{\beta})
(s_{\alpha}c_{\beta}+2c_{\alpha}s_{\beta})
(s_{\alpha}s_{\beta}-2c_{\alpha}c_{\beta})
\qty[C_{hVV2}^{VSS}(W,H,H^{\pm})+C_{hVV2}^{SSV}(H^{\pm},H,W)]
\notag\\ &\quad
+\frac{1}{2}gg_{Z}^{2}m_{W}
(s_{W}^{2}-s_{\beta}^{2})
c_{\beta}c_{\alpha}
\qty[C_{hVV2}^{VSS}(Z,G^{\pm},G^{0})+C_{hVV2}^{SSV}(G^{0},G^{\pm},Z)]
\notag\\ &\quad
+\frac{1}{2}gg_{Z}^{2}m_{W}
s_{\beta}^{2}c_{\beta}c_{\alpha}
\qty[C_{hVV2}^{VSS}(Z,H^{\pm},G^{0})+C_{hVV2}^{SSV}(G^{0},H^{\pm},Z)]
\notag\\ &\quad
-g^{3}c_{W}m_{Z}
c_{\alpha}
C_{1223}(c_{Z},c^{\pm},c_{Z})
-g^{3}c_{W}^{2}m_{W}
(c_{\alpha}c_{\beta}-2s_{\alpha}s_{\beta})
C_{1223}(c^{\pm},c_{Z},c^{\pm})
\notag\\ &\quad
-g^{3}s_{W}^{2}m_{W}
(c_{\alpha}c_{\beta}-2s_{\alpha}s_{\beta})
C_{1223}(c^{\pm},c_{\gamma},c^{\pm})
\notag\\ &\quad
-4g^{2}\bigg\{
\frac{1}{2}g_{G^{0}G^{0}h}^{}
\qty[
c_{\beta}^{2}C_{1223}(G^{0},G^{\pm},G^{0})
+s_{\beta}^{2}C_{1223}(G^{0},H^{\pm},G^{0})
]
\notag\\ &\quad
+\frac{3}{2}g_{hhh}^{}
\qty[
\qty(c_{\beta}c_{\alpha}-2s_{\beta}s_{\alpha})^{2}
C_{1223}(h,G^{\pm},h)
+\qty(s_{\beta}c_{\alpha}+2c_{\beta}s_{\alpha})^{2}
C_{1223}(h,H^{\pm},h)
]
\notag\\ &\quad
+\frac{1}{2}g_{HHh}^{}
\qty[
\qty(c_{\beta}s_{\alpha}+2s_{\beta}c_{\alpha})^{2}
C_{1223}(H,G^{\pm},H)
+\qty(s_{\beta}s_{\alpha}-2c_{\beta}c_{\alpha})^{2}
C_{1223}(H,H^{\pm},H)
]
\notag\\ &\quad
-\frac{1}{2}g_{Hhh}^{}
\qty(c_{\beta}c_{\alpha}-2s_{\beta}s_{\alpha})
\qty(c_{\beta}s_{\alpha}+2s_{\beta}c_{\alpha})
\qty[C_{1223}(h,G^{\pm},H)+C_{1223}(H,G^{\pm},h)]
\notag\\ &\quad
-\frac{1}{2}g_{Hhh}^{}
\qty(s_{\beta}c_{\alpha}+2c_{\beta}s_{\alpha})
\qty(s_{\beta}s_{\alpha}-2c_{\beta}c_{\alpha})
\qty[C_{1223}(h,H^{\pm},H)+C_{1223}(H,H^{\pm},h)]
\notag\\ &\quad
+\frac{1}{4}g_{H^{+}H^{-}h}^{}
s_{\beta}^{2}
C_{1223}(H^{\pm},G^{0},H^{\pm})
\notag\\ &\quad
+\frac{1}{4}g_{H^{+}H^{-}h}^{}
\qty[
\qty(s_{\beta}c_{\alpha}+2c_{\beta}s_{\alpha})^{2}
C_{1223}(H^{\pm},h,H^{\pm})
+\qty(s_{\beta}s_{\alpha}-2c_{\beta}c_{\alpha})^{2}
C_{1223}(H^{\pm},H,H^{\pm})
]
\notag\\ &\quad
+\frac{1}{4}g_{G^{+}G^{-}h}^{}
c_{\beta}^{2}
C_{1223}(G^{\pm},G^{0},G^{\pm})
\notag\\ &\quad
+\frac{1}{4}g_{G^{+}G^{-}h}^{}
\qty[
\qty(c_{\beta}c_{\alpha}-2s_{\beta}s_{\alpha})^{2}
C_{1223}(G^{\pm},h,G^{\pm})
+\qty(c_{\beta}s_{\alpha}+2s_{\beta}c_{\alpha})^{2}
C_{1223}(G^{\pm},H,G^{\pm})
]
\notag\\ &\quad
-\frac{1}{4}g_{H^{+}G^{-}h}^{}
s_{\beta}c_{\beta}
\qty[C_{1223}(G^{\pm},G^{0},H^{\pm})+C_{1223}(H^{\pm},G^{0},G^{\pm})]
\notag\\ &\quad
-\frac{1}{4}g_{H^{+}G^{-}h}^{}
\qty(s_{\beta}c_{\alpha}+2c_{\beta}s_{\alpha})
\qty(c_{\beta}c_{\alpha}-2s_{\beta}s_{\alpha})
\qty[C_{1223}(G^{\pm},h,H^{\pm})+C_{1223}(H^{\pm},h,G^{\pm})]
\notag\\ &\quad
-\frac{1}{4}g_{H^{+}G^{-}h}^{}
\qty(s_{\beta}s_{\alpha}-2c_{\beta}c_{\alpha})
\qty(c_{\beta}s_{\alpha}+2s_{\beta}c_{\alpha})
\qty[C_{1223}(G^{\pm},H,H^{\pm})+C_{1223}(H^{\pm},H,G^{\pm})]
\bigg\}, \\
&16\pi^{2}\Gamma_{hWW}^{3,\mathrm{1PI}}(p_{1}^{2},p_{2}^{2},q^{2})_{B} = 0.
\end{align}

\subsubsection{\texorpdfstring{$hf\bar{f}$}{} vertex} \label{app: hff_1PI}

The $hf\bar{f}$ vertex can be decomposed into the eight form factors as shown in Eq.~\eqref{eq:hff-form}.
The 1PI contributions are given by
\begin{align}
&\qty(\frac{m_{f}}{v})^{-1}16\pi^{2}\Gamma^{S, \mathrm{1PI}}_{hf\bar{f}}
\notag \\ &
=
-4\zeta_{h}\left[
e^{2}Q_{f}^{2}C_{hff}^{FVF}(f,\gamma,f)
+g_{Z}^{2}(v_{f}^{2}-a_{f}^{2})C_{hff}^{FVF}(f,Z,f)\right]
\notag \\ &\quad
+\frac{m_{f}^{2}}{v^{2}}\zeta_{h}\qty[
\zeta_{h}^{2}C_{hff}^{FSF}(f,h,f)
+\zeta_{H}^{2}C_{hff}^{FSF}(f,H,f)
-\zeta_{G^{0}}^{2}C_{hff}^{FSF}(f,G^{0},f)
]
\notag \\ &\quad
-\frac{2m_{f'}^{2}}{v^{2}}\zeta_{h}\qty[
\zeta_{G^{\pm}}^{2}C_{hff}^{FSF}(f',G^{\pm},f')
+\zeta_{H^{\pm}}^{2}C_{hff}^{FSF}(f',H^{\pm},f')
]
\notag \\ &\quad
-4g_{Z}^{3}m_{Z}v(v_{f}^{2}-a_{f}^{2})c_{\alpha}
C_0(Z,f,Z)
\notag \\ &\quad
-\frac{m_{f}^{2}}{v}\Big\{
6g_{hhh}^{} \zeta_{h}^{2} C_{0}(h,f,h)
+2g_{HHh}^{} \zeta_{H}^{2} C_{0}(H,f,H)
+2g_{Hhh}^{} \zeta_{h} \zeta_{H} [C_{0}(h,f,H)+C_{0}(H,f,h)]
\notag \\ &\qquad
-2g_{G^{0}G^{0}h}^{} \zeta_{G^{0}}^{2} C_{0}(G^{0},f,G^{0})
\Big\}
\notag \\ &\quad
+\frac{2m_{f'}^{2}}{v}\Big\{
g_{G^{+}G^{-}h}^{} \zeta_{G^{\pm}}^{2} C_{0}(G^{\pm},f',G^{\pm})
+g_{H^{+}H^{-}h}^{} \zeta_{H^{\pm}}^{2} C_{0}(H^{\pm},f',H^{\pm})
\notag\\ &\qquad
+g_{G^{-}H^{+}h}^{} \zeta_{G^{\pm}}\zeta_{H^{\pm}}
[C_{0}(G^{\pm},f',H^{\pm})+C_{0}(H^{\pm},f',G^{\pm})]
\Big\}
\notag \\ &\quad
-\frac{g^{2}}{4}\zeta_{G^{\pm}}(c_{\beta}c_{\alpha}-2s_{\beta}s_{\alpha})\qty[
C_{hff}^{VFS}(W,f',G^{\pm})+C_{hff}^{SFV}(G^{\pm},f',W)
]
\notag \\ &\quad
+\frac{g^{2}}{4}\zeta_{H^{\pm}}(s_{\beta}c_{\alpha}+2c_{\beta}s_{\alpha})\qty[
C_{hff}^{VFS}(W,f',H^{\pm})+C_{hff}^{SFV}(H^{\pm},f',W)
]
\notag \\ &\quad
-\frac{g_{Z}^{2}}{8}\zeta_{G^{0}}c_{\alpha}\qty[
C_{hff}^{VFS}(Z,f,G^{0})+C_{hff}^{SFV}(G^{0},f,Z)
], \\
&\qty(\frac{m_{f}}{v})^{-1}16\pi^{2}\Gamma^{P, \mathrm{1PI}}_{hf\bar{f}}
\notag \\ &
=
\frac{g^{2}}{4}\zeta_{G^{\pm}}(c_{\beta}c_{\alpha}-2s_{\beta}s_{\alpha})\qty[
C_{hff}^{VFS}(W,f',G^{\pm})-C_{hff}^{SFV}(G^{\pm},f',W)
]
\notag \\ &\quad
-\frac{g^{2}}{4}\zeta_{H^{\pm}}(s_{\beta}c_{\alpha}+2c_{\beta}s_{\alpha})\qty[
C_{hff}^{VFS}(W,f',H^{\pm})-C_{hff}^{SFV}(H^{\pm},f',W)
]
\notag \\ &\quad
+2g_{Z}^{2}v_{f}a_{f}\zeta_{G^{0}}c_{\alpha}\qty[
C_{hff}^{VFS}(Z,f,G^{0})-C_{hff}^{SFV}(G^{0},f,Z)
], \\
&(16\pi^{2})\Gamma^{V_{1}, \mathrm{1PI}}_{hf\bar{f}}
\notag \\
&=
-\frac{2m_{f}^{2}}{v}\zeta_{h}\qty[
e^{2}Q_{f}^{2}(C_{0}+2C_{11})(f,\gamma,f)
+g_{Z}^{2}(v_{f}^{2}+a_{f}^{2})(C_{0}+2C_{11})(f,Z,f)
]
\notag \\ &\quad
-\frac{g^{2}}{2}\frac{m_{f'}^{2}}{v}\zeta_{h}(C_{0}+2C_{11})(f',W,f')
\notag \\ &\quad
-\frac{m_{f}^{4}}{v^{3}}\zeta_{h}\Big[
\zeta_{h}^{2}(C_{0}+2C_{11})(f,h,f)
+\zeta_{H}^{2}(C_{0}+2C_{11})(f,H,f)
\notag \\ &\qquad
+\zeta_{G^{0}}^{2}(C_{0}+2C_{11})(f,G^{0},f)
\Big]
\notag \\ &\quad
-\frac{m_{f'}^{2}}{v^{3}}(m_{f}^{2}+m_{f'}^{2})\zeta_{h}\qty[
\zeta_{G^{\pm}}^{2}(C_{0}+2C_{11})(f',G^{\pm},f')
+\zeta_{H^{\pm}}^{2}(C_{0}+2C_{11})(f',H^{\pm},f')
]
\notag \\ &\quad
+2g_{Z}^{3}m_{Z}(v_{f}^{2}+a_{f}^{2})c_{\alpha}
(C_{0}+C_{11})(Z,f,Z)
+\frac{g^{3}m_{W}}{2}(c_{\beta}c_{\alpha}-2s_{\beta}s_{\alpha})
(C_{0}+C_{11})(W,f',W)
\notag \\ &\quad
-\frac{m_{f}^{2}}{v^{2}}\Big\{
6g_{hhh}^{} \zeta_{h}^{2} (C_{0}+C_{11})(h,f,h)
+2g_{HHh}^{} \zeta_{H}^{2} (C_{0}+C_{11})(H,f,H)
\notag \\ &\qquad
+2g_{Hhh}^{} \zeta_{h} \zeta_{H}
\left[(C_{0}+C_{11})(H,f,h)+(C_{0}+C_{11})(h,f,H)\right]
\notag \\ &\qquad
+2g_{G^{0}G^{0}h}^{} \zeta_{G^{0}}^{2} (C_{0}+C_{11})(G^{0},f,G^{0})
\Big\}
\notag \\ &\quad
-\frac{m_{f}^{2}+m_{f'}^{2}}{v^{2}}\Big\{
g_{G^{+}G^{-}h}^{} \zeta_{G^{\pm}}^{2}
(C_{0}+C_{11})(G^{\pm},f',G^{\pm})
+g_{H^{+}H^{-}h}^{} \zeta_{H^{\pm}}^{2}
(C_{0}+C_{11})(H^{\pm},f',H^{\pm})
\notag \\ &\qquad
+g_{H^{+}G^{-}h}^{} \zeta_{G^{\pm}} \zeta_{H^{\pm}}
\left[(C_{0}+C_{11})(G^{\pm},f',H^{\pm})+(C_{0}+C_{11})(H^{\pm},f',G^{\pm})\right]
\Big\}
\notag \\ &\quad
+\frac{g^{2}}{4}\frac{m_{f'}^{2}}{v}\Big\{
\zeta_{G^{\pm}}(c_{\beta}c_{\alpha}-2s_{\beta}s_{\alpha})
\left[(2C_{0}+C_{11})(W,f',G^{\pm})+(-C_{0}+C_{11})(G^{\pm},f',W)\right]
\notag \\ &\qquad
-\zeta_{H^{\pm}}(s_{\beta}c_{\alpha}+2c_{\beta}s_{\alpha})
\left[(2C_{0}+C_{11})(W,f',H^{\pm})+(-C_{0}+C_{11})(H^{\pm},f',W)\right]
\Big\}
\notag \\ &\quad
+\frac{g_{Z}^{2}}{8}\frac{m_{f}^{2}}{v}
\zeta_{G^{0}}c_{\alpha}
\left[(2C_{0}+C_{11})(Z,f,G^{0})+(-C_{0}+C_{11})(G^{0},f,Z)\right], \\
&(16\pi^{2})\Gamma^{V_{2}, \mathrm{1PI}}_{hf\bar{f}}
\notag \\ &
=
-\frac{2m_{f}^{2}}{v}\zeta_{h}\left[
e^{2}Q_{f}^{2}(C_{0}+2C_{12})(f,\gamma,f)
+g_{Z}^{2}(v_{f}^{2}+a_{f}^{2})(C_{0}+2C_{12})(f,Z,f)
\right]
\notag \\ &\quad
-\frac{g^{2}}{2}\frac{m_{f'}^{2}}{v}\zeta_{h}(C_{0}+2C_{12})(f',W,f')
\notag \\ &\quad
-\frac{m_{f}^{4}}{v^{3}}\zeta_{h}\Big[
\zeta_{h}^{2}(C_{0}+2C_{12})(f,h,f)
+\zeta_{H}^{2}(C_{0}+2C_{12})(f,H,f)
+\zeta_{G^{0}}^{2}(C_{0}+2C_{12})(f,G^{0},f)
\Big]
\notag \\ &\quad
-\frac{m_{f'}^{2}}{v^{3}}(m_{f}^{2}+m_{f'}^{2})\zeta_{h}\qty[
\zeta_{G^{\pm}}^{2}(C_{0}+2C_{12})(f',G^{\pm},f')
+\zeta_{H^{\pm}}^{2}(C_{0}+2C_{12})(f',H^{\pm},f')
]
\notag \\ &\quad
+2g_{Z}^{3}m_{Z}(v_{f}^{2}+a_{f}^{2})c_{\alpha}
C_{12}(Z,f,Z)
+\frac{g^{3}m_{W}}{2}(c_{\beta}c_{\alpha}-2s_{\beta}s_{\alpha})
C_{12}(W,f',W)
\notag \\ &\quad
-\frac{m_{f}^{2}}{v^{2}}\Big\{
6g_{hhh}^{} \zeta_{h}^{2} C_{12}(h,f,h)
+2g_{HHh}^{} \zeta_{H}^{2} C_{12}(H,f,H)
+2g_{Hhh}^{} \zeta_{h} \zeta_{H}
\left[C_{12}(H,f,h)+C_{12}(h,f,H)\right]
\notag \\ &\qquad
+2g_{G^{0}G^{0}h}^{} \zeta_{G^{0}}^{2} C_{12}(G^{0},f,G^{0})
\Big\}
\notag \\ &\quad
-\frac{m_{f}^{2}+m_{f'}^{2}}{v^{2}}\Big\{
g_{G^{+}G^{-}h}^{} \zeta_{G^{\pm}}^{2}
C_{12}(G^{\pm},f',G^{\pm})
+g_{H^{+}H^{-}h}^{} \zeta_{H^{\pm}}^{2}
C_{12}(H^{\pm},f',H^{\pm})
\notag \\ &\qquad
+g_{H^{+}G^{-}h}^{} \zeta_{G^{\pm}} \zeta_{H^{\pm}}
\left[C_{12}(G^{\pm},f',H^{\pm})+C_{12}(H^{\pm},f',G^{\pm})\right]
\Big\}
\notag \\ &\quad
+\frac{g^{2}}{4}\frac{m_{f'}^{2}}{v}\Big\{
\zeta_{G^{\pm}}(c_{\beta}c_{\alpha}-2s_{\beta}s_{\alpha})
\left[(2C_{0}+C_{12})(W,f',G^{\pm})+(-C_{0}+C_{12})(G^{\pm},f',W)\right]
\notag \\ &\qquad
-\zeta_{H^{\pm}}(s_{\beta}c_{\alpha}+2c_{\beta}s_{\alpha})
\left[(2C_{0}+C_{12})(W,f',H^{\pm})+(-C_{0}+C_{12})(H^{\pm},f',W)\right]
\Big\}
\notag \\ &\quad
+\frac{g_{Z}^{2}}{8}\frac{m_{f}^{2}}{v}
\zeta_{G^{0}}c_{\alpha}
\left[(2C_{0}+C_{12})(Z,f,G^{0})+(-C_{0}+C_{12})(G^{0},f,Z)\right], \\
&(16\pi^{2})\Gamma^{A_{1}, \mathrm{1PI}}_{hf\bar{f}}
\notag \\ &
=
4g_{Z}^{2}v_{f}a_{f}\frac{m_{f}^{2}}{v}\zeta_{h}(C_{0}+2C_{11})(f,Z,f)
+\frac{g^{2}}{2}\frac{m_{f'}^{2}}{v}\zeta_h(C_{0}+2C_{11})(f',W,f')
\notag \\ &\quad
-\frac{m_{f'}^{2}}{v^{3}}(m_{f}^{2}-m_{f'}^{2})\zeta_{h}\left[
\zeta_{G^{\pm}}^{2}(C_{0}+2C_{11})(f',G^{\pm},f')
+\zeta_{H^{\pm}}^{2}(C_{0}+2C_{11})(f',H^{\pm},f')
\right]
\notag\\ &\quad
-4g_{Z}^{3}m_{Z}v_{f}a_{f}c_{\alpha}
(C_{0}+C_{11})(Z,f,Z)
-\frac{g^{3}m_{W}}{2}(c_{\beta}c_{\alpha}-2s_{\beta}s_{\alpha})
(C_{0}+C_{11})(W,f',W)
\notag\\ &\quad
-\frac{m_{f}^{2}-m_{f'}^{2}}{v^2}\Big\{
\zeta_{G^{\pm}}^{2} g_{G^{+}G^{-}h}^{} (C_{0}+C_{11})(G^{\pm},f',G^{\pm})
+\zeta_{H^{\pm}}^{2} g_{H^{+}H^{-}h}^{} (C_{0}+C_{11})(H^{\pm},f',H^{\pm})
\notag\\ &\quad
+\zeta_{G^{\pm}} \zeta_{H^{\pm}} g_{H^{+}G^{-}h}^{}
\left[(C_{0}+C_{11})(G^{\pm},f',H^{\pm})+(C_{0}+C_{11})(H^{\pm},f',G^{\pm})\right]
\Big\}
\notag\\ &\quad
-\frac{g^{2}}{4}\frac{m_{f'}^2}{v}\Big\{
\zeta_{G^{\pm}} (c_{\beta}c_{\alpha}-2s_{\beta}s_{\alpha})
\left[(2C_{0}+C_{11})(W,f',G^{\pm})+(-C_{0}+C_{11})(G^{\pm},f',W)\right]
\notag\\ &\qquad
-\zeta_{H^{\pm}} (s_{\beta}c_{\alpha}+2c_{\beta}s_{\alpha})
\left[(2C_{0}+C_{11})(W,f',H^{\pm})+(-C_{0}+C_{11})(H^{\pm},f',W)\right]
\Big\}
\notag\\ &\quad
-2g_{Z}^{2}v_{f}a_{f}\frac{m_{f}^{2}}{v}
\zeta_{G^{0}}c_{\alpha}
\left[(2C_{0}+C_{11})(Z,f,G^{0})+(-C_{0}+C_{11})(G^{0},f,Z)\right], \\
&(16\pi^{2})\Gamma^{A_{2}, \mathrm{1PI}}_{hf\bar{f}}
\notag \\ &
=
4g_{Z}^{2}v_{f}a_{f}\frac{m_{f}^{2}}{v}\zeta_{h}(C_{0}+2C_{12})(f,Z,f)
+\frac{g^{2}}{2}\frac{m_{f'}^{2}}{v}\zeta_h(C_{0}+2C_{12})(f',W,f')
\notag \\ &\quad
-\frac{m_{f'}^{2}}{v^{3}}(m_{f}^{2}-m_{f'}^{2})\zeta_{h}\left[
\zeta_{G^{\pm}}^{2}(C_{0}+2C_{12})(f',G^{\pm},f')
+\zeta_{H^{\pm}}^{2}(C_{0}+2C_{12})(f',H^{\pm},f')
\right]
\notag\\ &\quad
-4g_{Z}^{3}m_{Z}v_{f}a_{f}c_{\alpha}
C_{12}(Z,f,Z)
-\frac{g^{3}m_{W}}{2}(c_{\beta}c_{\alpha}-2s_{\beta}s_{\alpha})
C_{12}(W,f',W)
\notag\\ &\quad
-\frac{m_{f}^{2}-m_{f'}^{2}}{v^2}\Big\{
\zeta_{G^{\pm}}^{2} g_{G^{+}G^{-}h}^{} C_{12}(G^{\pm},f',G^{\pm})
+\zeta_{H^{\pm}}^{2} g_{H^{+}H^{-}h}^{} C_{12}(H^{\pm},f',H^{\pm})
\notag\\ &\qquad
+\zeta_{G^{\pm}} \zeta_{H^{\pm}} g_{H^{+}G^{-}h}^{}
\left[C_{12}(G^{\pm},f',H^{\pm})+C_{12}(H^{\pm},f',G^{\pm})\right]
\Big\}
\notag\\ &\quad
-\frac{g^{2}}{4}\frac{m_{f'}^2}{v}\Big[
\zeta_{G^{\pm}} (c_{\beta}c_{\alpha}-2s_{\beta}s_{\alpha})
\left[(2C_{0}+C_{12})(W,f',G^{\pm})+(-C_{0}+C_{12})(G^{\pm},f',W)\right]
\notag\\ &\qquad
-\zeta_{H^{\pm}} (s_{\beta}c_{\alpha}+2c_{\beta}s_{\alpha})
\left[(2C_{0}+C_{12})(W,f',H^{\pm})+(-C_{0}+C_{12})(H^{\pm},f',W)\right]
\Big\}
\notag\\ &\quad
-2g_{Z}^{2}v_{f}a_{f}\frac{m_{f}^{2}}{v}\zeta_{G^{0}}c_{\alpha}
\left[(2C_{0}+C_{12})(Z,f,G^{0})+(-C_{0}+C_{12})(G^{0},f,Z)\right], \\
&\qty(\frac{m_{f}}{v})^{-1}16\pi^{2}\Gamma^{T, \mathrm{1PI}}_{hf\bar{f}}
\notag \\ &
=
-\frac{m_{f}^{2}}{v^{2}}\zeta_{h}\Big[
\zeta_{h}^{2}(C_{11}-C_{12})(f,h,f)+\zeta_{H}^{2}(C_{11}-C_{12})(f,H,f)
-\zeta_{G^{0}}^{2}(C_{11}-C_{12})(f,G^{0},f)
\Big]
\notag\\ &\quad
+\frac{2m_{f'}^{2}}{v^{2}}\zeta_{h}\Big[
\zeta_{G^{\pm}}^{2}(C_{11}-C_{12})(f',G^{\pm},f')
+\zeta_{H^{\pm}}^{2}(C_{11}-C_{12})(f',H^{\pm},f')
\Big]
\notag\\ &\quad
-\frac{g^{2}}{4}\Big\{
\zeta_{G^{\pm}}(c_{\beta}c_{\alpha}-2s_{\beta}s_{\alpha})
\left[(2C_{0}+2C_{11}-C_{12})(W,f',G^{\pm})
+(C_{0}+C_{11}-2C_{12})(G^{\pm},f',W)\right]
\notag\\ &\qquad
-\zeta_{H^{\pm}}(s_{\beta}c_{\alpha}+2c_{\beta}s_{\alpha})
\left[(2C_{0}+2C_{11}-C_{12})(W,f',H^{\pm})
+(C_{0}+C_{11}-2C_{12})(H^{\pm},f',W)\right]
\Big\}
\notag\\ &\quad
-\frac{g_{Z}^{2}}{8}\zeta_{G^{0}}c_{\alpha}
\left[(2C_{0}+2C_{11}-C_{12})(Z,f,G^{0})
+(C_{0}+C_{11}-2C_{12})(G^{0},f,Z)\right], \\
&\qty(\frac{m_{f}}{v})^{-1}16\pi^{2}\Gamma^{PT, \mathrm{1PI}}_{hf\bar{f}}
\notag \\ &
=
\frac{g^{2}}{4}\Big\{
\zeta_{G^{\pm}}(c_{\beta}c_{\alpha}-2s_{\beta}s_{\alpha})
\left[(2C_{0}+2C_{11}-C_{12})(W,f',G^{\pm})
-(C_{0}+C_{11}-2C_{12})(G^{\pm},f',W)\right]
\notag\\ &\qquad
-\zeta_{H^{\pm}}(s_{\beta}c_{\alpha}+2c_{\beta}s_{\alpha})
\left[(2C_{0}+2C_{11}-C_{12})(W,f',H^{\pm})
-(C_{0}+C_{11}-2C_{12})(H^{\pm},f',W)\right]
\Big\}
\notag\\ &\quad
+2g_{Z}^{2}v_{f}a_{f}\zeta_{G^{0}}c_{\alpha}
\left[(2C_{0}+2C_{11}-C_{12})(Z,f,G^{0})
-(C_{0}+C_{11}-2C_{12})(G^{0},f,Z)\right],
\end{align}
where the loop functions are defined as~\cite{Kanemura:2015mxa, Aiko:2026qzi}
\begin{align}
&C_{hff}^{FVF}(X,Y,Z)
\notag\\
&=
[m_{f}^{2}C_{0}+p_{1}^{2}(C_{11}+C_{21})+p_{2}^{2}(C_{12}+C_{22})
+p_{1}\cdot p_{2}(2C_{23}-C_{0})+4C_{24}](X,Y,Z)-1,\\
&C_{hff}^{FSF}(X,Y,Z)
\notag \\
&=
[m_{f}^{2}C_{0}+p_{1}^{2}(C_{11}+C_{21})+p_{2}^{2}(C_{12}+C_{22})
+2p_{1}\cdot p_{2}(C_{11}+C_{23})+4C_{24}](X,Y,Z)-\frac{1}{2}, \\
&C_{hff}^{VFS}(X,Y,Z)
\notag\\
&=
[p_{1}^{2}(2C_{0}+3C_{11}+C_{21})+p_{2}^{2}(2C_{12}+C_{22})
+2p_{1}\cdot p_{2}(2C_{12}+C_{23})+4C_{24}](X,Y,Z)-\frac{1}{2}, \\
&C_{hff}^{SFV}(X,Y,Z)
\notag\\
&=
[p_{1}^{2}(-C_{0}+C_{21})+p_{2}^{2}(-C_{12}+C_{22})
\notag\\ &\qquad
+2p_{1}\cdot p_{2}(C_{23}+C_{12}-C_{0}-C_{11})+4C_{24}](X,Y,Z)-\frac{1}{2}. 
\end{align}

\subsubsection{\texorpdfstring{$h\mathcal{V}\mathcal{V'}$}{} vertex}
\label{app: hVV'_1PI}

The 1PI $hgg$ vertices are obtained as
\begin{align}
\Gamma_{hg^{a}g^{b}}^{i}(p_{1}^{2}, p_{2}^{2}, q^{2})
=
\zeta_{h}\Gamma_{hg^{a}g^{b}}^{i}(p_{1}^{2}, p_{2}^{2}, q^{2})_{\SM}
\qquad (i=1,2),
\end{align}
where
\begin{align}
&16\pi^{2}\Gamma_{hg^{a}g^{b}}^{1}(p_{1}^{2}, p_{2}^{2}, q^{2})_{\SM}
\notag \\
&=
-\sum_{q}\frac{4g_{s}^{2}m_{q}^{2}}{v}\qty[
8C_{24}(q, q, q)-2B_{0}(q^{2}; q, q)+(p_{1}^{2}+p_{2}^{2}-q^{2})C_{0}(q, q, q)
]\delta^{ab}, \\
&16\pi^{2}\Gamma_{hg^{a}g^{b}}^{2}(p_{1}^{2}, p_{2}^{2}, q^{2})_{\SM}
=
-\sum_{q}\frac{8g_{s}^{2}m_{q}^{2}}{v}q^{2}\qty[
C_{0}+4C_{1223}](q, q, q)\delta^{ab}.
\end{align}

The 1PI $h\gamma\gamma$ vertex is decomposed into the fermion, SM boson and charged scalar contributions.
The fermion contributions are obtained as
\begin{align}
\Gamma_{h\gamma\gamma}^{1}(p_{1}^{2}, p_{2}^{2}, q^{2})_{F}
=
\zeta_{h}\Gamma_{h\gamma\gamma}^{1}(p_{1}^{2}, p_{2}^{2}, q^{2})_{F,\, \SM}
\quad (i=1,2),
\end{align}
where
\begin{align}
&16\pi^{2}\Gamma_{h\gamma\gamma}^{1}(p_{1}^{2}, p_{2}^{2}, q^{2})_{F,\, \SM}
\notag \\
&=
-\sum_{f}N_{c}^{f}\frac{4e^{2}Q_{f}^{2}m_{f}^{2}}{v}\qty[
8C_{24}(f, f, f)-2B_{0}(q^{2}; f, f)+(p_{1}^{2}+p_{2}^{2}-q^{2})C_{0}(f, f, f)
], \\
&16\pi^{2}\Gamma_{h\gamma\gamma}^{2}(p_{1}^{2}, p_{2}^{2}, q^{2})_{F,\, \SM}
=
-\sum_{f}N_{c}^{f}\frac{8e^{2}Q_{f}^{2}m_{f}^{2}}{v}\qty[
C_{0}+4C_{1223}](f, f, f).
\end{align}
The SM boson and charged scalar contributions are obtained as
\begin{align}
&16\pi^{2}\Gamma_{h\gamma\gamma}^{1}(p_{1}^{2}, p_{2}^{2}, q^{2})_{V}
\notag \\
&=
\frac{2e^{2}m_{W}^{2}}{v}(c_{\beta}c_{\alpha}-2s_{\beta}s_{\alpha})
\notag \\ &\quad \times
\Big\{
(5p_{1}^{2}+5p_{2}^{2}-7q^{2})C_{0}(W, W, W)
-6[B_{0}(q^{2}; W, W)-4C_{24}(W, W, W)]
\notag \\ &\quad
-\frac{m_{h}^{2}}{m_{W}^{2}}[B_{0}(q^{2}; W, W)-4C_{24}(W, W, W)]-m_{h}^{2}C_{0}(W, W, W)
\Big\}, \\
&16\pi^{2}\Gamma_{h\gamma\gamma}^{2}(p_{1}^{2}, p_{2}^{2}, q^{2})_{V}
\notag \\
&=
\frac{8e^{2}m_{W}^{2}}{v}(c_{\beta}c_{\alpha}-2s_{\beta}s_{\alpha})\qty[
4C_{0}+6C_{1223}+\frac{m_{h}^{2}}{m_{W}^{2}}C_{1223}](W, W, W), \\
&16\pi^{2}\Gamma_{h\gamma\gamma}^{1}(p_{1}^{2}, p_{2}^{2}, q^{2})_{S}
=
-2e^{2}g_{H^{+}H^{-}h}^{} \left[
4C_{24}(H^{\pm}, H^{\pm}, H^{\pm})-B_{0}(q^{2}; H^{\pm}, H^{\pm})
\right], \\
&16\pi^{2}\Gamma_{h\gamma\gamma}^{2}(p_{1}^{2}, p_{2}^{2}, q^{2})_{S}
=
-8e^{2}g_{H^{+}H^{-}h}C_{1223}(H^{\pm}, H^{\pm}, H^{\pm}).
\end{align}
The fermion contributions to the 1PI $hZ\gamma$ vertices are obtained as
\begin{align}
\Gamma_{hZ\gamma}^{i}(p_{1}^{2}, p_{2}^{2}, q^{2})_{F}
=
\zeta_{h}\Gamma_{hZ\gamma}^{i}(p_{1}^{2}, p_{2}^{2}, q^{2})_{F,\, \SM}
\quad (i=1,2),
\end{align}
where
\begin{align}
&16\pi^{2}\Gamma_{hZ\gamma}^{1}(p_{1}^{2}, p_{2}^{2}, q^{2})_{F,\, \SM}
\notag \\
&=
-\sum_{f}N_{c}^{f}v_{f}Q_{f}\frac{4eg_{Z}m_{f}^{2}}{v}\big[
8C_{24}(f, f, f)-2B_{0}(q^{2}; f, f)
\notag \\ &\quad
+(p_{1}^{2}+p_{2}^{2}-q^{2})C_{0}(f, f, f)
\big], \\
&16\pi^{2}\Gamma_{hZ\gamma}^{2}(p_{1}^{2}, p_{2}^{2}, q^{2})_{F,\, \SM}
=
-\sum_{f}N_{c}^{f}v_{f}Q_{f}\frac{8eg_{Z}m_{f}^{2}}{v}\qty[
C_{0}+4C_{1223}](f, f, f).
\end{align}
The bosonic contributions are obtained as
\begin{align}
&16\pi^{2}\Gamma_{hZ\gamma}^{1}(p_{1}^{2}, p_{2}^{2}, q^{2})_{B}
\notag \\
&=
eg^{2}m_{W}c_{W}(c_{\alpha}c_{\beta}-2s_{\alpha}s_{\beta})
\qty[2C^{VVV}_{hVV1}(W, W, W)-6B_{0}(q^{2}; W, W)+4-2C_{24}(c^{\pm}, c^{\pm}, c^{\pm})]
\notag \\ &\quad
+egg_{Z}m_{W}(s_{W}^{2}-s_{\beta}^{2})
(c_{\alpha}c_{\beta}-2s_{\alpha}s_{\beta})
\notag \\ &\qquad \times
\qty[C^{SVV}_{hVV1}(G^{\pm}, W, W)+2m_{W}^{2}C_{0}(W, G^{\pm}, W)-2C_{24}(W, G^{\pm}, G^{\pm})+B_{0}(p_{1}^{2}; G^{\pm}, W)]
\notag \\ &\quad
-eg^{2}m_{W}c_{W}(c_{\alpha}c_{\beta}-2s_{\alpha}s_{\beta})
C^{VVS}_{hVV1}(W, W, G^{\pm})
\notag \\ &\quad
-2eg_{Z}m_{W}^{2}(s_{W}^{2}-s_{\beta}^{2})g_{G^{+}G^{-}h}
C_{0}(G^{\pm}, W, G^{\pm})
\notag \\ &\quad
+egg_{Z}m_{W}(c_{2W}+s_{\beta}^{2})
(c_{\alpha}c_{\beta}-2s_{\alpha}s_{\beta})
C_{24}(G^{\pm}, G^{\pm}, W)
\notag \\ &\quad
+egg_{Z}m_{W}\qty(c_{\alpha}c_{\beta}s_{W}^{2}+2s_{\alpha}s_{\beta}c_{W}^{2})
B_{0}(p_{2}^{2}; G^{\pm}, W)
\notag \\ &\quad
-4eg_{Z}(c_{2W}+s_{\beta}^{2})g_{G^{+}G^{-}h}
\qty[C_{24}(G^{\pm}, G^{\pm}, G^{\pm})-\frac{1}{4}B_{0}(q^{2}; G^{\pm}, G^{\pm})]
\notag \\ &\quad
+egg_{Z}m_{W}s_{\beta}c_{\beta}
(c_{\alpha}s_{\beta}+2s_{\alpha}c_{\beta})
\qty[C^{SVV}_{hVV1}(H^{\pm}, W, W)-2C_{24}(W, H^{\pm}, H^{\pm})]
\notag \\ &\quad
+2eg_{Z}m_{W}^{2}s_{\beta}c_{\beta}g_{H^{+}G^{-}h}
C_{0}(H^{\pm}, W, G^{\pm})
\notag \\ &\quad
-egg_{Z}m_{W}s_{\beta}c_{\beta}
(c_{\alpha}s_{\beta}+2s_{\alpha}c_{\beta})
\qty[C_{24}(H^{\pm}, G^{\pm}, W)-B_{0}(p_{1}^{2}; H^{\pm}, W)]
\notag \\ &\quad
-4eg_{Z}(c_{2W}+c_{\beta}^{2})g_{H^{+}H^{-}h}
\qty[C_{24}(H^{\pm}, H^{\pm}, H^{\pm})-\frac{1}{4}B_{0}(q^{2}; H^{\pm}, H^{\pm})]
\notag \\ &\quad
-4eg_{Z}s_{\beta}c_{\beta}g_{H^{+}G^{-}h}
\qty[C_{24}(H^{\pm}, G^{\pm}, G^{\pm})+C_{24}(G^{\pm}, H^{\pm}, H^{\pm})-\frac{1}{2}B_{0}(q^{2}; H^{\pm}, G^{\pm})], \\
&16\pi^{2}\Gamma_{hZ\gamma}^{2}(p_{1}^{2}, p_{2}^{2}, q^{2})_{B}
\notag \\
&=
eg^{2}m_{W}c_{W}(c_{\alpha}c_{\beta}-2s_{\alpha}s_{\beta})
\qty[2C^{VVV}_{hVV2}(W, W, W)-2C_{1223}(c^{\pm}, c^{\pm}, c^{\pm})-C^{VVS}_{hVV2}(W, W, G^{\pm})]
\notag \\ &\quad
+egg_{Z}m_{W}(s_{W}^{2}-s_{\beta}^{2})
(c_{\alpha}c_{\beta}-2s_{\alpha}s_{\beta})
\qty[C^{SVV}_{hVV2}(G^{\pm}, W, W)-2C^{VSS}_{hVV2}(W, G^{\pm}, G^{\pm})]
\notag \\ &\quad
+egg_{Z}m_{W}(c_{2W}+s_{\beta}^{2})
(c_{\alpha}c_{\beta}-2s_{\alpha}s_{\beta})
C^{SSV}_{hVV2}(G^{\pm}, G^{\pm}, W)
\notag \\ &\quad
-4eg_{Z}(c_{2W}+s_{\beta}^{2})g_{G^{+}G^{-}h}
C_{1223}(G^{\pm}, G^{\pm}, G^{\pm})
\notag \\ &\quad
+egg_{Z}m_{W}s_{\beta}c_{\beta}
(c_{\alpha}s_{\beta}+2s_{\alpha}c_{\beta})
\notag \\ &\qquad \times
\qty[C^{SVV}_{hVV2}(H^{\pm}, W, W)-C^{SSV}_{hVV2}(H^{\pm}, G^{\pm}, W)-2C^{VSS}_{hVV2}(W, H^{\pm}, H^{\pm})]
\notag \\ &\quad
-4eg_{Z}(c_{2W}+c_{\beta}^{2})g_{H^{+}H^{-}h}
C_{1223}(H^{\pm}, H^{\pm}, H^{\pm})
\notag \\ &\quad
-4eg_{Z}s_{\beta}c_{\beta}g_{H^{+}G^{-}h}
\qty[C_{1223}(H^{\pm}, G^{\pm}, G^{\pm})+C_{1223}(G^{\pm}, H^{\pm}, H^{\pm})].
\end{align}

\subsection{Four-point functions}
\label{app: 4PFs}

\subsubsection{Box diagrams for \texorpdfstring{$h\to Zf\bar{f}$}{}}

There are six SM-like diagrams and two diagrams with the charged Higgs boson.
The Feynman diagrams are given in Ref.~\cite{Aiko:2026qzi}, and we use the same diagram numbering as in that reference.
For the diagrams without a $Z$ boson, the NLO contributions are obtained as
\begin{align}
(16\pi^{2})2\Re\qty[\sum \mathcal{M}_{\mathrm{tree}}^{*}\mathcal{M}_{BZi}]
&=
4(v_{f}+a_{f})C_{\mathrm{tree}}^{Z}C^{BZi}B^{BZi}.
\end{align}
On the other hand, for the $Z$-boson mediated diagram, we obtain 
\begin{align}
(16\pi^{2})2\Re\qty[\sum \mathcal{M}_{\mathrm{tree}}^{*}\mathcal{M}_{BZ6}]
&=
8(v_{f}^{4}+6v_{f}^{2}a_{f}^{2}+a_{f}^{4})C^{\mathrm{tree}}C^{BZ6}B^{BZ6}.
\end{align}
The tree-level factor is obtained as $C_{\mathrm{tree}}^{Z} = g_{Z}\Gamma_{hZZ}^{1, \mathrm{tree}}/(s-m_{Z}^{2})$.
The loop functions, $B^{BZi}$, are the same as those in the CHTM and are given in Ref.~\cite{Aiko:2026qzi}.
The one-loop factor $C^{BZi}$ are obtained as
\begin{align}
C^{BZ1} = C^{BZ2}
&=
-2a_{f}g^{4}m_{W}c_{W}(c_{\beta}c_{\alpha}-2s_{\beta}s_{\alpha}), \\
C^{BZ3} = C^{BZ4}
&=
a_{f}g^{3}g_{Z}m_{W}(s_{W}^{2}-s_{\beta}^{2})(c_{\beta}c_{\alpha}-2s_{\beta}s_{\alpha}), \\
C^{BZ5}
&=
-(v_{f'}+a_{f'})g^{3}g_{Z}m_{W}(c_{\beta}c_{\alpha}-2s_{\beta}s_{\alpha}), \\
C^{BZ6}
&=
-2g_{Z}^{4}m_{Z}c_{\alpha}, \\
C^{BZ7} = C^{BZ8}
&=
a_{f}g^{3}g_{Z}m_{W}s_{\beta}c_{\beta}(s_{\beta}c_{\alpha}+2c_{\beta}s_{\alpha}).
\end{align}

\subsubsection{Box diagrams for \texorpdfstring{$h\to Wf\bar{f}'$}{}}

There are eleven SM-like diagrams and two diagrams with the charged Higgs boson.
The Feynman diagrams are given in Ref.~\cite{Aiko:2026qzi}, and we use the same diagram numbering as in that reference.
The NLO contributions are obtained as
\begin{align}
(16\pi^{2})2\Re\qty[\sum \mathcal{M}_{\mathrm{tree}}^{*}\mathcal{M}_{BWi}]
&=
4C_{\mathrm{tree}}^{W}C^{BWi}\Re\qty[B^{BWi}].
\end{align}
The tree-level factor is given by $C_{\mathrm{tree}}^{W} = g_{W}\Gamma_{hWW}^{1, \mathrm{tree}}/(s-m_{W}^{2})$.
The loop functions, $B^{BWi}$, are the same as those in the CHTM and are given in Ref.~\cite{Aiko:2026qzi}.
The one-loop factors $C^{BWi}$ are obtained as
\begin{align}
C^{BW1}
&=
-4a_{f'}(v_{f'}+a_{f'})\frac{g^{2}g_{Z}^{2}}{\sqrt{2}}c_{W}m_{Z}c_{\alpha}, \\
C^{BW3}
&=
-4a_{f}(v_{f}+a_{f})\frac{g^{3}g_{Z}}{\sqrt{2}}c_{W}m_{W}(c_{\beta}c_{\alpha}-2s_{\beta}s_{\alpha}), \\
C^{BW5}
&=
-4a_{f}eQ_{f}\frac{g^{3}}{\sqrt{2}}s_{W}m_{W}(c_{\beta}c_{\alpha}-2s_{\beta}s_{\alpha}), \\
C^{BW7}
&=
-2a_{f}(v_{f}+a_{f})\frac{g^{2}g_{Z}^{2}}{\sqrt{2}}(s_{W}^{2}-s_{\beta}^{2})m_{W}(c_{\beta}c_{\alpha}-2s_{\beta}s_{\alpha}), \\
C^{BW9}
&=
2a_{f}Q_{f}\frac{e^{2}g^{2}}{\sqrt{2}}m_{W}(c_{\beta}c_{\alpha}-2s_{\beta}s_{\alpha}), \\
C^{BW10}
&=
2a_{f'}Q_{f'}\frac{e^{2}g^{2}}{\sqrt{2}}m_{W}(c_{\beta}c_{\alpha}-2s_{\beta}s_{\alpha}), \\
C^{BW11}
&=
-2(v_{f}+a_{f})(v_{f'}+a_{f'})\frac{g_{Z}^{3}g}{\sqrt{2}}m_{Z}c_{\alpha}, \\
C^{BW12}
&=
-2a_{f}(v_{f}+a_{f})\frac{g^{2}g_{Z}^{2}}{\sqrt{2}}s_{\beta}c_{\beta}m_{W}(c_{\alpha}s_{\beta}+2s_{\alpha}c_{\beta}).
\end{align}
The one-loop factors $C^{BW2}$, $C^{BW4}$, $C^{BW6}$, $C^{BW8}$, and $C^{BW13}$ can be obtained from $C^{BW1}$, $C^{BW3}$, $C^{BW5}$, $C^{BW7}$, and $C^{BW12}$, respectively, by interchanging $f$ and $f'$.

\section{Gauge dependence of scalar self-energies}\label{sec:pinch}
\begin{figure}
    \centering
    \includegraphics[width=1\linewidth]{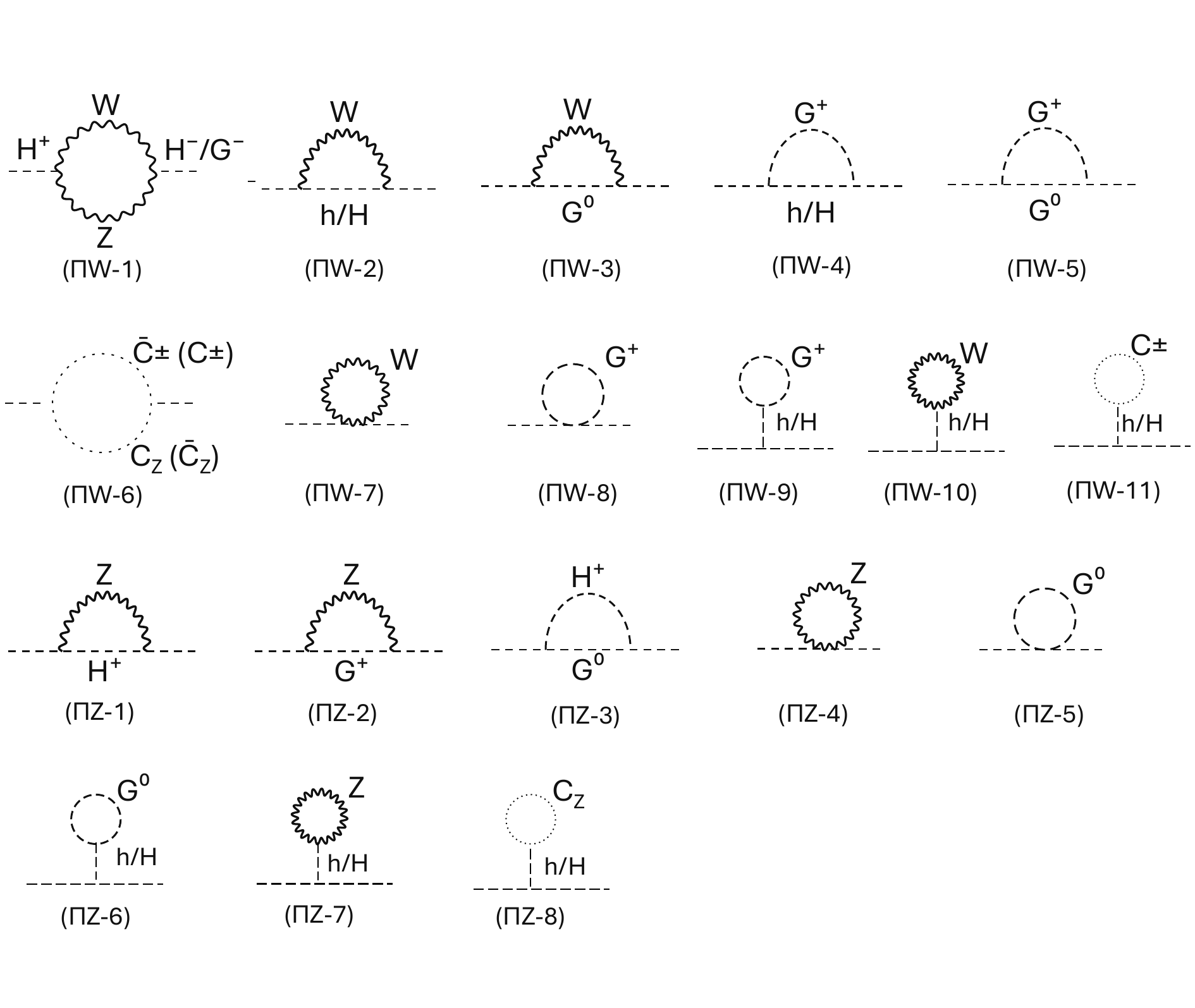}    
 \caption{Feynman diagrams contributing to the gauge-dependent
parts of the charged scalar self-energies.}
    \label{fig:Pi_xi}
\end{figure}
In this appendix, we briefly review the derivation of
the pinch terms for the scalar self-energies, 
which are given in Eqs. (\ref{eq:PT_CPeven}) and  (\ref{eq:PT_Charged}). 
The pinch terms for the self-energies of the CP-even scalar bosons are calculated in the same way as in the CHTM~\cite{Aiko:2026qzi}, except that contributions from loops involving the CP-odd scalar $A$ are absent. 
We therefore omit the discussion of the CP-even sector and focus on the pinch terms for the charged scalar two-point functions.

\subsection{Cancellation of the gauge dependence} 
In this subsection, we describe the cancellation of the gauge dependence 
associated with the gauge-fixing parameters $\xi_W$ and $\xi_Z$ in the charged scalar self-energies. 
Since the $\xi_\gamma$ dependence has the same structure
as in the 2HDM~\cite{Kanemura:2017wtm}, we omit its detailed discussion. 
In order to analytically confirm that the gauge dependence of the $H^+H^-$ and $H^+G^-$ self-energies is completely canceled by the pinched contributions,  
we define the gauge-dependent part of a quantity
$\mathcal{O}$ as
\begin{align}
\mathcal{O}|_\textrm{G.D}
\equiv
\mathcal{O}\big|_{R_\xi}
-
\mathcal{O}\big|_{\xi_V=1}, 
\end{align}
where the first and second terms on the right-hand side
are calculated in the $R_\xi$ gauge and the
't Hooft--Feynman gauge, respectively.
The Feynman diagrams contributing to the gauge-dependent
parts of the charged scalar self-energies are shown
in Fig.~\ref{fig:Pi_xi}.
The gauge-dependent parts of the $H^+H^-$ and
$H^+G^-$ self-energies are calculated as follows:
\begin{align}
 &\left.\Biggl(\sum_{i=1}^{11}\Pi_{H^+H^-}^{W-i}
 +\sum_{i=1}^{8}\Pi_{H^+H^-}^{Z-i}\Biggr)
 \right|_{\mathrm{G.D.}}\notag\\
 &=\frac{p^2-m_{H^\pm}^2}{64\pi^2}
 \Biggl[-2g^2(1-\xi_W)(1+c_\beta^2)B_0(0;W,G^+)
 \nonumber\\
 &\quad-g_Z^2(1-\xi_Z)
 \left(c_{H^+H^-Z}^2+s_\beta^2c_\beta^2\right)B_0(0;Z,G^0)
 \nonumber\\
 &\quad+g^2(1-\xi_W)\sum_{S=h,H,G^0}c_{H^\pm SW^\mp}^2
 \left\{p^2+m_{H^\pm}^2-2\delta_{S}m_S^2\right\}
 C_0(p^2;W,G^+,S)
 \nonumber\\
 &\quad+g_Z^2(1-\xi_Z)\sum_{\substack{(\phi,X)=(H^+,H^+),\\\qquad(G^+,W)}}c_{H^+\phi^-Z}^2
 \left\{p^2+m_{H^\pm}^2-2\delta_{\phi}m_{H^\pm}^2\right\}
 C_0(p^2;Z,G^0,X)\Biggr],\label{eq:HHself}\\
 &\left.\Biggl(\sum_{i=1}^{11}\Pi_{H^+G^-}^{W-i}
 +\sum_{i=1}^{8}\Pi_{H^+G^-}^{Z-i}\Biggr)
 \right|_{\mathrm{G.D.}}
 \nonumber\\
 &={}\frac{1}
 {64\pi^2}
 \Biggl[-g^2s_\beta c_\beta(1-\xi_W)
 \left\{2p^2-m_{H^\pm}^2+m_W^2(1+\xi_Z)\right\}B_0(0;W,G^+)
 \nonumber\\
 &\quad-g_Z^2s_\beta c_\beta(1-\xi_Z)
 \left\{\left(c_{2W}+\frac12\right)(2p^2-m_{H^\pm}^2)
 -c_W^2m_W^2(1+\xi_W)\right\}B_0(0;Z,G^0)
 \nonumber\\
 &\quad+2g^2s_\beta c_\beta m_W^2
 \left\{(1-\xi_W)-(1-\xi_Z)\right\}B_0(p^2;W,Z)
 \nonumber\\
 &\quad+g^2(1-\xi_W)\sum_{S=h,H,G^0}c_{H^\pm SW^\mp}c_{G^\pm SW^\mp}
 \nonumber\\
 &\qquad\times\left\{p^4-(2p^2-m_{H^\pm}^2)m_S^2\delta_{S}\right\}
 C_0(p^2;W,G^+,S)
 \nonumber\\
 &\quad+g_Z^2(1-\xi_Z)\sum_{\substack{(\phi,X)=(H^+,H^+),\\\qquad(G^+,W)}}c_{H^+\phi^-Z}c_{G^+\phi^-Z}
 \left\{p^4-\delta_{\phi}m_{H^\pm}^2(2p^2-m_{H^\pm}^2)\right\}
 C_0(p^2;Z,G^0,X)
 \nonumber\\
 &\quad-g^2s_\beta c_\beta\frac{m_W^2}{m_Z^2}(1-\xi_W)
 \left\{(p^2-m_Z^2-m_{G^+}^2)^2-4m_Z^2m_{G^+}^2\right\}
 C_0(p^2;W,G^+,Z)
 \nonumber\\
 &\quad+g^2s_\beta c_\beta\frac{m_W^2}{m_Z^2}(1-\xi_W)
 \left\{(p^2-m_{G^+}^2)^2-m_{G^0}^4+2p^2m_{G^0}^2\right\}
 C_0(p^2;W,G^+,G^0)
 \nonumber\\
 &\quad+g^2s_\beta c_\beta(1-\xi_Z)
 \left\{m_W^4+m_{G^+}^4-2p^2(m_W^2+m_{G^+}^2)-2m_W^2m_{G^0}^2\right\}
 C_0(p^2;Z,G^0,W)\Biggr], 
 \label{eq:HGself}
\end{align}
with $\delta_S=1\, (0)$ for $S=h,H$ ($S=G^0$), $\delta_\phi=1\, (0)$ for $\phi=H^+$ ($\phi=G^+$). We also obtain $\Pi_{G^+H^-}^{V-i} =\Pi_{H^+G^-}^{V-i}$.
Moreover, we here use the useful relations among the Passarino–Veltman functions, i.e., $B_0(0;X,Y)=  (A(X)- A(Y))/(m_X^2 - m_Y^2)$ and 
$C_0(p^2,X,Y,Z) \equiv (B_0(p^2;X,Z)- B_0(p^2;Y,Z))/(m_X^2 - m_Y^2)$.

\begin{figure}
    \centering
    \includegraphics[width=1\linewidth]{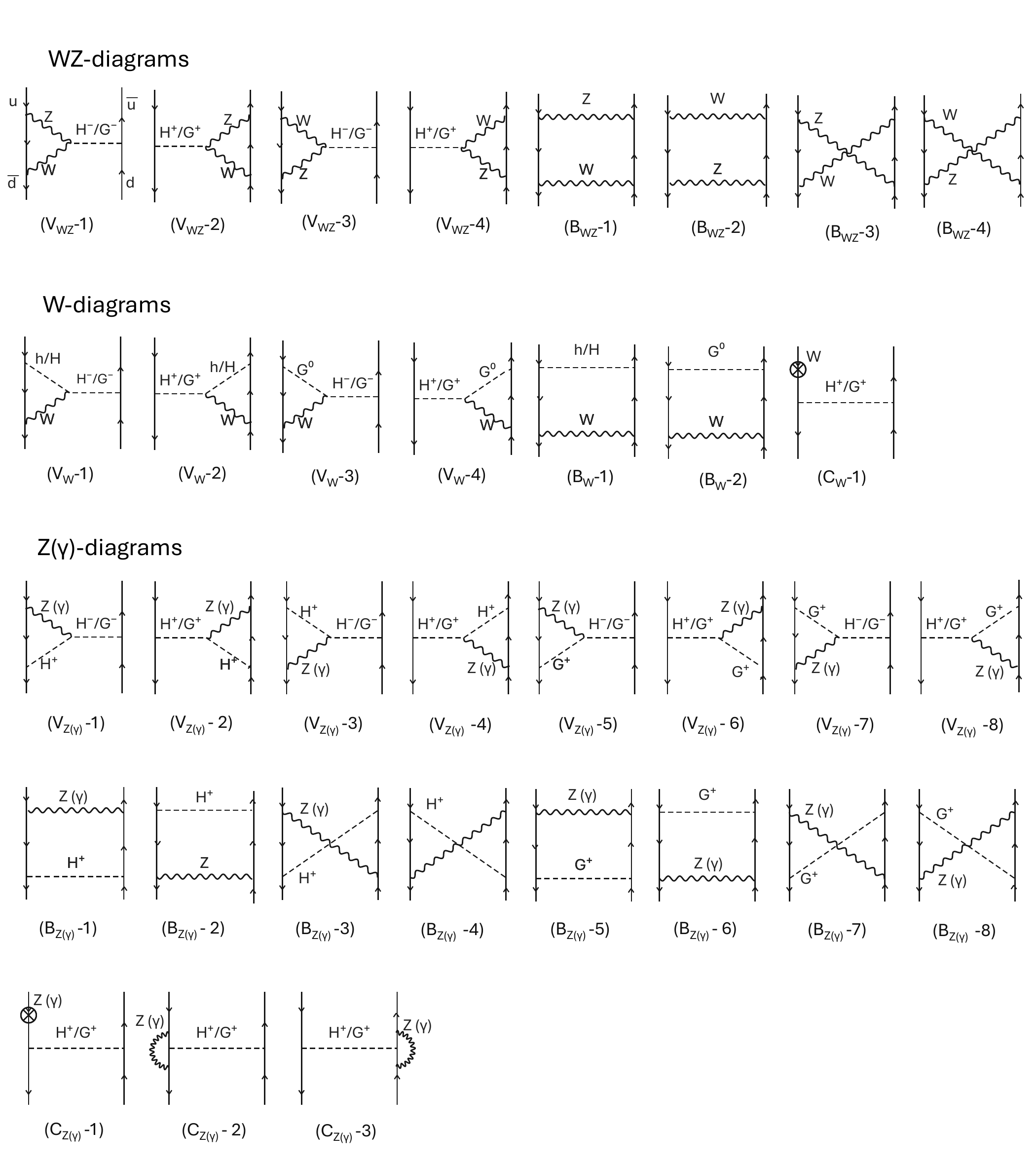}    
 \caption{Feynman diagrams contributing to the pinch terms
extracted from the $u\bar{d}\to u\bar{d}$ scattering amplitude.}
    \label{fig:M_PT}
\end{figure}
To identify the contributions that cancel the gauge dependence of these self-energies, we consider the scattering process $u\bar{d}\to u\bar{d}$. 
We here introduce the reduced amplitude $\overline{\mathcal{M}}$, defined as
\begin{align}
   \mathcal{M} =\overline{\mathcal{M}} \left(2\frac{m_u^2}{v^2}\right)(\overline{d}P_R u)\times(\overline{u}P_L^{} d). \label{eq:red_M}
\end{align} 
For the self-energies of the charged scalars, 
the pinch contributions can be extracted from the reduced amplitude which arises from diagrams shown in Fig.~\ref{fig:M_PT}.

In the processes in which 
one or more vector bosons are attached to a fermion line, 
the pinch terms are calculated by canceling the internal fermion propagators.  
The resulting contributions are referred to as pinched amplitudes. 
In particular, diagrams containing two vector bosons ($W$ and $Z$) attached to a fermion line generate the following pinch contributions: 
\begin{align}
\overline{d}\big[
(v_u - a_u)m_u P_R +(v_u + a_u)P_R\,\slashed{k}  
\big]u, 
 \label{eq:kkT} 
\end{align}
where $k$ indicates the loop momentum, and irrelevant overall factors are omitted. 
The vector and axial-vector coupling factors of $u$ are denoted by $v_u^{}$ and $a_u^{}$, respectively.
There are also contributions from diagrams in which
the order of the $Z$ and $W$ boson attachments
to the fermion line is interchanged. 
The first and second term correspond to the ``scalar-like'' pinch contribution, and the ``vector-like'' pinch contribution, respectively.  
In the RHTM, the $ H^\pm W^\mp Z$ 
coupling exists at the tree level, giving rise to
the $WZ$ diagrams shown in Fig.~\ref{fig:M_PT},
such as (V$_{WZ}$-1)--(V$_{WZ}$-4). 
These diagrams generate both scalar-like and
vector-like pinch contributions, both of which
contribute to the pinch term for the $H^+G^-$ 
mixing self-energy. 
In contrast, in the 2HDM, the $WZ$ diagrams do not
contribute to the pinch terms for the $H^+H^-$
and $H^+G^-$ self-energies. 
\footnote{In the 2HDM, although there are the $V_{WZ}$-diagrams involving
a propagating $G^\pm$ boson and $B_{WZ}$-diagrams, 
their pinch terms do not contribute 
to the cancellation of the gauge dependence
in the $H^+H^-$ and $H^+G^-$ self-energies.} 
Thus, these contributions represent a distinctive
feature of models with a tree-level
$H^\pm W^\mp Z$ coupling, such as the RHTM. 
Diagrams containing a single vector boson attached to a fermion line,
i.e., the $W$-, $Z$-, and $\gamma$-diagrams in Fig.~\ref{fig:M_PT},
generate only scalar-like pinch contributions.

We here give explicit expressions for gauge-dependent parts of each vertex, box, and fermion wave-function contributions shown in Fig.~\ref{fig:M_PT}. 
The gauge-dependent parts of the scalar-like pinch contributions are given by
\begin{align}
 \left.\overline{\mathcal M}_{\mathrm{ver}}[g_V,V,S,\phi]\right|_{\mathrm{G.D.}}
 &=\frac{g_V^2}{32\pi^2}\frac{1-\xi_V}{p^2-m_\phi^2}
 \left[B_0(0;V,G_V)-(p^2-m_S^2)C_0(p^2;V,G_V,S)\right],
 \label{eq:PT-ver}\\
 \left.\overline{\mathcal M}_{\mathrm{box}}[g_V,V,S]\right|_{\mathrm{G.D.}}
 &=\frac{g_V^2}{64\pi^2}(1-\xi_V)C_0(p^2;V,G_V,S),
 \label{eq:PT-box}\\
 \left.\overline{\mathcal M}_{\mathrm{ver}+\delta}[g_V,V,\phi]\right|_{\mathrm{G.D.}}
 &=-\frac{g_V^2}{64\pi^2}\frac{1-\xi_V}{p^2-m_\phi^2}B_0(0;V,G_V).
 \label{eq:PT-fermion}
\end{align}
For $W$- and $Z$-diagrams in Fig.~\ref{fig:M_PT}, we obtain
\begin{align}
 \left.\sum_{i=1}^{4}\overline{\mathcal M}^\textrm{sc}_{VW-i}\right|_{\mathrm{G.D.}}
 ={}&\sum_{\phi^+,S}\zeta_{\phi^+}^{}\zeta_Sc_{\phi^+ SW^\mp}
 \left.\overline{\mathcal M}_{\mathrm{ver}}[g,W,S,\phi^+]\right|_{\mathrm{G.D.}}
 \nonumber\\
 &+\frac{g^2}{32\pi^2}\sum_{\phi^+}
 \frac{\zeta_{\phi^+}\zeta_{G^0}c_{\phi^\pm G^0W^\mp}}{p^2-m_{\phi^+}^2}
 \left[B_0(p^2;W,Z)-B_0(p^2;W,G^0)\right],
 \label{eq:PT-Wver}\\
 \left.\sum_{i=1}^{2}\overline{\mathcal M}^\textrm{sc}_{BW-i}\right|_{\mathrm{G.D.}}
 ={}&\sum_S\zeta_S^2
 \left.\overline{\mathcal M}_{\mathrm{box}}[g,W,S]\right|_{\mathrm{G.D.}},
 \label{eq:PT-Wbox}\\
 \left.\overline{\mathcal M}^\textrm{sc}_{CW-1}\right|_{\mathrm{G.D.}}
 ={}&2\sum_{\phi^+}\zeta_{\phi^+}^2
 \left.\overline{\mathcal M}_{\mathrm{ver}+\delta}[g,W,\phi^+]\right|_{\mathrm{G.D.}},
 \label{eq:PT-Wfermion}\\
 \left.\sum_{i=1}^{8}\overline{\mathcal M}^\textrm{sc}_{VZ-i}\right|_{\mathrm{G.D.}}
 ={}&c_{2W}\sum_{\phi^+,\chi^+}\zeta_{\phi^+}\zeta_{\chi^+} c_{\phi^+\chi^-Z}
 \left.\overline{\mathcal M}_{\mathrm{ver}}[g_Z,Z,\chi^+,\phi^+]\right|_{\mathrm{G.D.}}
 \nonumber\\
 &+\frac{g_Z^2c_{2W}}{32\pi^2}\sum_{\phi^+}
 \frac{\zeta_{\phi^+}\zeta_{G^+}c_{\phi^\pm G^\mp Z}}{p^2-m_{\phi^+}^2}
 \left[B_0(p^2;W,Z)-B_0(p^2;G^+,Z)\right],
 \label{eq:PT-Zver}\\
 \left.\sum_{i=1}^{8}\overline{\mathcal M}^\textrm{sc}_{BZ-i}\right|_{\mathrm{G.D.}}
 ={}&c_{2W}^2\sum_\phi\zeta_{\phi^+}^2
 \left.\overline{\mathcal M}_{\mathrm{box}}[g_Z,Z,\phi^+]\right|_{\mathrm{G.D.}},
 \label{eq:PT-Zbox}\\
 \left.\sum_{i=1}^{3}\overline{\mathcal M}^\textrm{sc}_{CZ-i}\right|_{\mathrm{G.D.}}
 ={}&c_{2W}^2\sum_{\phi^+}\zeta_{\phi^+}^2
 \left.\overline{\mathcal M}_{\mathrm{ver}+\delta}[g_Z,Z,\phi^+]\right|_{\mathrm{G.D.}}.
 \label{eq:PT-Zfermion}
\end{align}

As mentioned above, (V$_{WZ}$-1)--(V$_{WZ}$-4) and (B$_{WZ}$-1)--(B$_{WZ}$-4) generate not only scalar-like pinch contributions but also additional vector-like pinch contributions. 
The scalar-like contributions are given by
\begin{align}
 &\left.\sum_{i=1}^{4}\overline{\mathcal M}_{VWZ-i}^{\mathrm{sc}}\right|_{\mathrm{G.D.}}
 \nonumber\\
 &={}-\frac{g_Z^2c_{2W}}{32\pi^2}
\sum_{\phi=H^+,G^+}
\frac{\zeta_\phi c_{\phi^\pm W^\mp Z}}{p^2-m_\phi^2}
 \Biggl[\frac{m_W^2}{m_Z^2}(1-\xi_W)
 \Bigl\{(p^2+m_Z^2-m_{G^+}^2)C_0(p^2;W,G^+,Z)
 \nonumber\\
 &\quad+(m_{G^+}^2+m_{G^0}^2-p^2)C_0(p^2;W,G^+,G^0)\Bigr\}
 \nonumber\\
 &\quad+(1-\xi_Z)(m_W^2+m_{G^+}^2)C_0(p^2;Z,G^0,W)\Biggr],
 \label{eq:PT-WZver}\\
 &\left.\sum_{i=1}^{4}\overline{\mathcal M}_{BWZ-i}^{\mathrm{sc}}\right|_{\mathrm{G.D.}}
 \nonumber\\
 &={}\frac{g_Z^2c_{2W}^2}{64\pi^2}\frac{m_W^2}{m_Z^2}(1-\xi_W)
 \Bigl[C_0(p^2;W,G^+,Z)-C_0(p^2;W,G^+,G^0)\Bigr],
 \label{eq:PT-WZbox}
\end{align}
where $c_{\phi^\pm W^\mp Z}$ is defined by $c_{H^\pm W^\mp Z} = s_\beta c_\beta$, $c_{G^\pm W^\mp Z} = s_\beta^2 - s_W^2$.

Using the coupling sum rules in
Eqs.~\eqref{eq:sumrule-W-s}--\eqref{eq:sumrule-Z-G},
we organize the pinch contributions into terms proportional to
$\zeta_{H^+}^2$, $\zeta_{H^+}\zeta_{G^+}$, and $\zeta_{G^+}^2$,
which are assigned to the $H^+H^-$, $H^\pm G^\mp$, and
$G^+G^-$ self-energies, respectively,
following Ref.~\cite{Aiko:2026qzi}.
For some scalar-like contributions and the vector-like contributions,
we further separate the $H^+G^-$ and $H^+W^-$ contributions.
For this purpose, we define
\begin{align}
 \Lambda_{G^+}
 &=i\frac{\sqrt{2}m_u}{v}\bar uP_Ld,
 &
 \Lambda_W^\mu
 &=i\frac{g}{\sqrt{2}}\bar u\gamma^\mu P_Ld,
 \label{eq:charged-fermion-vertices}
\end{align}
and use the identity
\begin{align}
 \frac{\Lambda_{G^+}}{p^2}
 =\frac{\Lambda_{G^+}}{p^2-m_{G^+}^2}
 -\frac{i m_W}{p^2}
 \Lambda_W^\mu(\Delta_W)_{\mu\nu}p^\nu,
 \label{eq:charged-GW-decomposition}
\end{align}
where $\Delta_W$ is the $W$ propagator.
The first and second terms on the right-hand side are assigned
to $H^+G^-$ and $H^+W^-$ mixing, respectively.
We denote the resulting contributions to the reduced scattering
amplitude by $\overline{\mathcal M}^{\mathrm{PT}}_{XY}$,
where $XY$ specifies the self-energy channel.

The $H^+G^-$ contributions obtained from the vector-like parts
of diagrams (V$_{WZ}$-1)--(V$_{WZ}$-4) are
\begin{align}
 &\left.
 \sum_{i=1}^{4}
 \overline{\mathcal M}^{\mathrm{PT},\mathrm{vec}}_{H^+G^-,\,VWZ-i}
 \right|_{\mathrm{G.D.}}
 \nonumber\\
 &=
 \frac{g^2\zeta_{H^+}\zeta_{G^+}s_\beta c_\beta}
 {64\pi^2(p^2-m_{H^\pm}^2)(p^2-m_{G^+}^2)}
 \Biggl[
 m_W^2(1-\xi_W)(1+\xi_Z)B_0(0;W,G^+)
 \nonumber\\
 &\quad
 -m_W^2(1-\xi_Z)(1+\xi_W)B_0(0;Z,G^0)
 \nonumber\\
 &\quad
 -2m_W^2\bigl\{(1-\xi_W)-(1-\xi_Z)\bigr\}
 B_0(p^2;W,Z)
 \nonumber\\
 &\quad
 +\frac{m_W^2}{m_Z^2}(1-\xi_W)
 \Bigl[
 \bigl\{(p^2-m_Z^2-m_{G^+}^2)^2
       -4m_Z^2m_{G^+}^2\bigr\}
 C_0(p^2;W,G^+,Z)
 \nonumber\\
 &\qquad
 -\bigl\{(p^2-m_{G^+}^2)^2-m_{G^0}^4\bigr\}
 C_0(p^2;W,G^+,G^0)
 \Bigr]
 \nonumber\\
 &\quad
 -(1-\xi_Z)
 \bigl\{
 m_W^4+m_{G^+}^4
 -2p^2(2m_W^2+m_{G^+}^2)
 -2m_W^2m_{G^0}^2
 \bigr\}
 C_0(p^2;Z,G^0,W)
 \Biggr].
 \label{eq:charged-WZ-vector-vertex}
\end{align}
Similarly, the vector-like parts of diagrams
(B$_{WZ}$-1)--(B$_{WZ}$-4) give
\begin{align}
 &\left.
 \sum_{i=1}^{4}
 \overline{\mathcal M}^{\mathrm{PT},\mathrm{vec}}_{H^+G^-,\,BWZ-i}
 \right|_{\mathrm{G.D.}}
 \nonumber\\
 &=
 \frac{g^2\zeta_{H^+}\zeta_{G^+}s_\beta c_\beta}
 {64\pi^2(p^2-m_{G^+}^2)}
 \Biggl[
 \frac{m_W^2}{m_Z^2}(1-\xi_W)
 \Bigl\{
 -(p^2+m_Z^2-m_{G^+}^2)
 C_0(p^2;W,G^+,Z)
 \nonumber\\
 &\qquad
 +(p^2+m_{G^0}^2-m_{G^+}^2)
 C_0(p^2;W,G^+,G^0)
 \Bigr\}
 \nonumber\\
 &\quad
 +(1-\xi_Z)(m_W^2-m_{G^+}^2)
 C_0(p^2;Z,G^0,W)
 \Biggr].
 \label{eq:charged-WZ-vector-box}
\end{align}

We have analytically confirmed that 
the gauge-dependent contributions cancel separately in the
$H^+H^-$ and $H^+G^-$ channels:
\begin{align}
 &\Biggl[\frac{\zeta_{H^+}^2}{(p^2-m_{H^\pm}^2)^2}\left(\sum_{i=1}^{11}\Pi_{H^+H^-}^{W-i}
 +\sum_{i=1}^{8}\Pi_{H^+H^-}^{Z-i}\right)
 +\overline{\mathcal M}^{\mathrm{PT, sc}}_{H^+H^-,\textrm{all}}
 \Biggr]\Bigg|_{\mathrm{G.D.}}=0,
 \label{eq:PT-HHcancel}\\
 &\Biggl[\frac{\zeta_{H^+}^{}\zeta_{G^+}^{}}{(p^2-m_{H^\pm}^2)(p^2-m_{G^\pm}^2)}\left(\sum_{i=1}^{11}\Pi_{H^+G^-}^{W-i}
 +\sum_{i=1}^{8}\Pi_{H^+G^-}^{Z-i}\right)\notag\\
 &\quad+\overline{\mathcal M}^{\mathrm{PT,sc}}_{H^+G^-,\textrm{all}}
 +\overline{\mathcal M}^{\mathrm{PT,vec}}_{H^+G^-,\textrm{all}}
 \Biggr]\Bigg|_{\mathrm{G.D.}}=0. 
 \label{eq:PT-HGcancel}
\end{align}
The gauge dependence in the $G^+H^-$ self-energy also cancels
in exactly the same way as in Eq.~(\ref{eq:PT-HGcancel}).

\subsection{Pinch terms of the self-energies of charged scalars}
In the 't Hooft--Feynman gauge, the only nonvanishing pinch contributions arise from the vertex correction diagrams (V$_W$-1)--(V$_W$-4), (V$_Z$-1)--(V$_Z$-8), and (V$_\gamma$-1)--(V$_\gamma$-8). 
The corresponding pinched amplitudes are calculated as follows:
\begin{align}
\sum_{i=1-4}\overline{\mathcal{M}}_{V_W\text{-}i} &=
-\frac{g^2}{32 \pi^2}\sum_{s=h,H}\left[\left(\frac{\zeta_{H^+}^{}\zeta_{s}^{}c_{H^\pm s W^{\mp}}^{}}{p^2 - m_{H^\pm}^2} +\frac{\zeta_{G^+}^{}\zeta_{s}^{}c_{G^\pm s W^{\mp}}^{}}{p^2 - m_{G^\pm}^2}\right)
B_0(p^2;W,s)\right]\notag\\
&-\frac{g^2}{32 \pi^2}\left(\frac{\zeta_{H^+}^{}\zeta_{G^0}^{}c_{H^\pm G^0 W^{\mp}}^{}}{p^2 - m_{H^\pm}^2} +\frac{\zeta_{G^+}^{}\zeta_{G^0}^{}c_{G^\pm G^0 W^{\mp}}^{}}{p^2 - m_{G^\pm}^2}\right)
B_0(p^2;W,Z),\label{eq:M_VW}\\
\sum_{i=1-8}\overline{\mathcal{M}}_{V_Z\text{-}i}&=
-\frac{g_Z^2c_{2W}^{}}{32 \pi^2}\left(\frac{\zeta_{H^+}^{2}c_{H^+ H^- Z}^{}}{p^2 - m_{H^\pm}^2} +\frac{\zeta_{H^+}^{}\zeta_{G^+}^{}c_{H^\pm G^\mp Z^{}}}{p^2 - m_{G^\pm}^2}\right)
B_0(p^2;Z,H^\pm)\notag\\
&-\frac{g_Z^2c_{2W}^{}}{32 \pi^2}\left(\frac{\zeta_{H^+}^{}\zeta_{G^+}^{}c_{H^\pm G^\mp Z^{}}}{p^2 - m_{H^\pm}^2} +\frac{\zeta_{G^+}^{2}c_{G^+ G^- Z^{}}}{p^2 - m_{G^\pm}^2}\right)
B_0(p^2;Z,W),\label{eq:M_VZ}\\
\sum_{i=1-8}\overline{\mathcal{M}}_{V_\gamma\text{-}i}
&=-\frac{e^2}{8 \pi^2}\left(\frac{\zeta_{H^+}^2}{p^2 - m_{H^\pm}^2} B_0(p^2;\gamma,H^\pm)+\frac{\zeta_{G^+}^2}{p^2 - m_{G^\pm}^2}B_0(p^2;\gamma,G^\pm)\right). \label{eq:M_VG}
\end{align} 
The pinch terms of these self-energies are related to
these pinched amplitudes as
\begin{align}
&\sum_{i=1-4}\overline{\mathcal{M}}_{V_W\text{-}i} + \sum_{i=1-8}\overline{\mathcal{M}}_{V_Z\text{-}i}+\sum_{i=1-8}\overline{\mathcal{M}}_{V_\gamma\text{-}i}\to
\frac{\zeta_{H^+}^2}{(p^2-m_{H^\pm}^2)^2}\Pi_{H^+H^-}^\textrm{PT}(p^2)\notag\\
&\quad +\frac{\zeta_{G^+}^2}{(p^2-m_{G^\pm}^2)^2}\Pi_{G^+G^-}^\textrm{PT}(p^2) 
+\frac{\zeta_{H^+}^{}\zeta_{G^+}^{}}{(p^2-m_{H^{\pm}}^2)(p^2-m_{G^\pm}^2)}\left(\Pi_{H^+G^-}^\textrm{PT}(p^2) +\Pi_{G^+H^-}^\textrm{PT}(p^2)\right), \label{eq:decompose2}
\end{align}
where $\Pi_{H^+G^-}^\textrm{PT}(p^2) =\Pi_{G^+H^-}^\textrm{PT}(p^2)$. 
Here, the arrow instead of an equality sign indicates that the contributions associated with vector--scalar mixing self-energies, such as the $H^+W^-$ mixing self-energy, have been removed from the pinched amplitudes. 
We isolate the contributions relevant to the scalar self-energies following the procedure described in Eq.~(\ref{eq:charged-GW-decomposition}).
Therefore, we decompose these pinched amplitudes into the pinch terms as follows;
\begin{align}
\Pi_{H^+H^-}^\mathrm{PT}(p^2) 
&= -\frac{g^2}{32 \pi^2}(p^2-m_{H^\pm}^2)\left[\sum_{s=h,H}\left\{c_{H^\pm s W^{\mp}}^{2}
B_0(p^2;W,s)\right\} + c_{H^\pm G^0 W^{\mp}}^{2}
B_0(p^2;W,Z)\right]\notag\\
&-\frac{g_Z^2}{32 \pi^2}(p^2-m_{H^\pm}^2)\left[c_{H^+ H^- Z}^{2} 
B_0(p^2;Z,H^\pm)+c_{H^\pm G^\mp Z}^{2} 
B_0(p^2;Z,W)\right]\notag\\
&-\frac{e^2}{8 \pi^2}(p^2-m_{H^\pm}^2)B_0(p^2;\gamma,H^\pm). 
\end{align}
The pinch term of the $H^+G^-$ self-energy is given in Eq.~(\ref{eq:PT_Charged}).   
In the decomposition, we use the following sum rules among the coupling constants; 
\begin{align}
 \zeta_s
 &= \zeta_{H^+}c_{H^\pm sW^\mp}
  + \zeta_{G^+}c_{G^\pm sW^\mp},
 \label{eq:sumrule-W-s}\\
 \zeta_{G^0}
 &= \zeta_{H^+}c_{H^\pm G^0W^\mp}
  + \zeta_{G^+}c_{G^\pm G^0W^\mp},
 \label{eq:sumrule-W-G0}\\
 c_{2W}\zeta_{H^+}
 &= \zeta_{H^+}c_{H^+H^-Z}
  + \zeta_{G^+}c_{H^\pm G^\mp Z},
 \label{eq:sumrule-Z-H}\\
 c_{2W}\zeta_{G^+}
 &= \zeta_{H^+}c_{H^\pm G^\mp Z}
  + \zeta_{G^+}c_{G^+G^-Z}.
 \label{eq:sumrule-Z-G}
\end{align}

\bibliographystyle{utphys28mod}
\bibliography{reference}

\end{document}